\documentclass[letterpaper,english,iop]{emulateapj}
\usepackage[LGR,T1]{fontenc}
\usepackage{color}
\usepackage{babel}
\usepackage{array}
\usepackage{booktabs}
\usepackage{textcomp}
\usepackage{multirow}
\usepackage{graphicx}
\usepackage{wasysym}
\usepackage[unicode=true]
 {hyperref}
\usepackage{breakurl}

\makeatletter

\DeclareRobustCommand{\greektext}{%
  \fontencoding{LGR}\selectfont\def\encodingdefault{LGR}}
\DeclareRobustCommand{\textgreek}[1]{\leavevmode{\greektext #1}}
\ProvideTextCommand{\~}{LGR}[1]{\char126#1}

\providecommand{\tabularnewline}{\\}

\shorttitle{EXTRAPLANAR UV \& H$\alpha$ EMISSIONS}
\shortauthors{JO ET AL.}

\newenvironment{continuedfigure*}{%
\addtocounter{figure}{-1}%
\begin{figure}}{%
\end{figure}}
\makeatother

\begin{document}

\title{Comparison of the extraplanar H$\alpha$ and UV emissions \\in the
halos of nearby edge\textendash on spiral galaxies }

\author{Young-Soo Jo\altaffilmark{1,2}, Kwang-il Seon\altaffilmark{1,3},
Jong-Ho Shinn\altaffilmark{1}, Yujin Yang\altaffilmark{1}, Dukhang
Lee\altaffilmark{1,4}, and Kyoung-Wook Min\altaffilmark{2}}

\altaffiltext{1}{Korea Astronomy and Space Science Institute (KASI), 776 Daedeokdae-ro, Yuseong-gu, Daejeon, 34055, Korea; stspeak@gmail.com}
\altaffiltext{2}{Korea Advanced Institute of Science and Technology (KAIST), 291, Daehak-ro, Yuseong-gu, Daejeon, 34141, Korea}
\altaffiltext{3}{Astronomy and Space Science Major, Korea University of Science and Technology, Daejeon, 34113, Korea}
\altaffiltext{4}{York University, 4700 Keele Street, Toronto, Ontario, M3J 1P3, Canada}
\begin{abstract}
We compare vertical profiles of the extraplanar H$\alpha$ emission
to those of the UV emission for 38 nearby edge-on late-type galaxies.
It is found that detection of the ``diffuse'' extraplanar dust (eDust),
traced by the vertically extended, scattered UV starlight, always
coincides with the presence of the extraplanar H\textgreek{a} emission.
A strong correlation between the scale heights of the extraplanar
H$\alpha$ and UV emissions is also found; the scale height at H$\alpha$
is found to be $\sim0.74$ of the scale height at FUV. Our results
may indicate the multiphase nature of the diffuse ionized gas and
dust in the galactic halos. The existence of eDust in galaxies where
the extraplanar H$\alpha$ emission is detected suggests that a
larger portion of the extraplanar H$\alpha$ emission than that predicted
in previous studies may be caused by H$\alpha$ photons that originate
from \ion{H}{2} regions in the galactic plane and are subsequently
scattered by the eDust. This possibility raise a in studying the eDIG.
We also find that the scale heights of the extraplanar emissions normalized
to the galaxy size correlate well with the star formation rate surface
density of the galaxies. The properties of eDust in our galaxies is
on a continuation line of that found through previous observations
of the extraplanar polycyclic aromatic hydrocarbons emission in more
active galaxies known to have galactic winds.
\end{abstract}


\section{INTRODUCTION}

Study of the feedback between stars and interstellar
medium (ISM) is essential to understand the formation and evolution
of galaxies. Although the ISM in late-type galaxies
including the Milky Way Galaxy are mostly concentrated in
the galactic plane, it has been found that a considerable amount
of materials exists at high altitudes above the midplane \citep[e.g.,][]{2009RvMP...81..969H,2012ARA&A..50..491P,2014ApJ...785L..18S,2014ApJ...789..131H}.
The extraplanar material in the galactic halo traces infalling star-formation
fuel and feedback from a galaxy's disk and is therefore a crucial
component of galactic evolution that probes disk-halo interaction
\citep{2005ASPC..331..155D,2012ARA&A..50..491P}.

Narrow band and spectroscopic observations of the H$\alpha$ emission
of nearby late-type edge-on galaxies have revealed the existence of
extraplanar H$\alpha$ emission from many galaxies with sufficient
star formation rates (SFRs). The extraplanar H$\alpha$
emission is generally believed to originate from the extraplanar
diffuse ionized gas (hereafter, eDIG) \citep{1996ApJ...462..712R,2000A&A...359..433R,2003A&A...406..493R,2003A&A...406..505R,2004AJ....128..674R,2003ApJS..148..383M,2003ApJ...592...79M,2016MNRAS.457.1257H}.
The eDIG is believed to be maintained by photoionization by ionizing
photons (Lyman continuum; Lyc) that are mainly produced by O-type
stars in the galactic plane \citep{1984ApJ...282..191R,1994ApJ...428..647D,1996AJ....111.2265F,2002A&A...386..801Z,2004MNRAS.353.1126W,2009RvMP...81..969H,2015MNRAS.447..559B}.

However, it is still not clear whether Lyc leaked
out of \ion{H}{2} regions in the galactic plane is the major ionization
source of the high latitude and interarm gas \citep[e.g.,][]{2009ApJ...703.1159S,2011ApJ...743..188S,2011ApJ...727...35D,2011MNRAS.415.2182F,2017A&A...599A.141J}.
\citet{2009ApJ...703.1159S} pointed out that an incredibly low absorption
coefficient of Lyc is required to explain the diffuse H$\alpha$ emission
of a face-on galaxy M 51 with the standard photoionization model.
One alternative explanation of the extraplanar H$\alpha$ emission
was proposed by \citet{2011ApJ...727...35D}. The diffuse H$\alpha$
emission in their model originates from gas that was photoionized
in the past, but which is currently cooling and recombining; the ionizing
radiation should last only for a very short time ($\sim10^{5}$ yrs),
compared to O star lifetimes ($\sim3\times10^{6}$ yrs), before the
photoionization switches off and the gas begins to cool. Therefore,
the ionizing radiation should be mostly provided by runaway OB stars
with velocities of $\gtrsim100$ km s$^{-1}$. 

The processes that expel gas from the galactic disk may also act on
the interstellar dust grains. It has been suggested that dust could
be elevated by radiation pressure from the galactic disk into the
halo \citep{1987Natur.327..214G,1991ApJ...366..443F,1991ApJ...381..137F}
and/or by hydrodynamic motions due to supernovae and stellar winds
\citep{1997AJ....114.2463H}. The extraplanar dust (hereafter, eDust)
in galactic halos (or thick disks) can be utilized to study the feedback
process in the cold phase of the ISM. High-resolution optical images
of nearby edge-on galaxies have revealed an extensive filamentary
structure of the eDust seen in absorption against the background stellar
light of the bulge and thick stellar disk \citep{1997AJ....114.2463H,1999AJ....117.2077H,2000AJ....119..644H,2004AJ....128..662T}.
Many studies in the mid-infrared (MIR) and far-IR (FIR) wavelengths
have provided evidence of eDust \citep{2000A&AS..145...83A,2000A&A...356..795A,2006A&A...445..123I,2007ApJ...668..918B,2009ApJ...698L.125K,2013A&A...556A..54V,2016A&A...586A...8B}.
Far-ultraviolet (FUV) and/or near-UV (NUV) observations of edge-on
galaxies have revealed the stellar continuum scattered into the line
of sight by the eDust \citep{2014ApJ...785L..18S,2014ApJ...789..131H,2015ApJ...815..133S,2016ApJ...833...58H}.
The FUV emission in the galactic outflows of starburst galaxies were
also attributed to starlight scattered by dust in the outflow \citep{2005ApJ...619L..99H}.
The UV continuum emission mostly originates from OB stars in the galactic
thin disk and hence the UV reflection halo caused by the eDust can
be used to estimate the amount of eDust. Based on this idea, \citet{2014ApJ...785L..18S}
and \citet{2015ApJ...815..133S} quantitatively derived the amount
of eDust by comparing dust radiative transfer models with the observed
UV images of edge-on galaxies. \citet{2016A&A...587A..86B} showed
that the spectral energy density of NGC 3628 is well reproduced using
the dust and stellar geometries obtained by \citet{2015ApJ...815..133S}.

\citet{2003A&A...406..505R} found a correlation between the presence/non-presence
of eDust and eDIG in spiral galaxies; absorbing dusty features at
high altitudes are usually found in the galaxies where the eDIG is
also detected. The simultaneous existence of the eDust and eDIG indicates
the multiphase nature of the ISM in galactic halos. In this regard,
\citet{2000AJ....119..644H} and \citet{2013AJ....145...62R} compared
the absorbing, filamentary dust structures in high-resolution optical
(\emph{BVI} bands) images to the H$\alpha$ emission of edge-on galaxies,
but found that the filamentary morphologies of the dust absorption
have no counterpart in the smoothly distributed H$\alpha$ emission.
They concluded that the diffuse eDIG and filamentary eDust trace physically
distinct phases of the thick disk ISM. However, it should be noted
that the absorbing dust features trace only opaque clouds with optical
depths of $\apprge1$, as noted in \citet{2014ApJ...785L..18S}, and
the H$\alpha$ emission would be extinguished by the absorbing dust
clouds. Thus, the absence of a spatial correspondence between the
absorbing dust and H$\alpha$ emission does not necessarily imply
distinct phases. The total amount of eDust, which is missing in the
high-resolution optical studies, is better traced by observing the
scattered starlight in UV wavelengths, as studied by \citet{2014ApJ...785L..18S}
and \citet{2015ApJ...815..133S}. Therefore, the multiphase nature
of eDust and eDIG can be best studied by comparing the UV halo and
the extraplanar H$\alpha$ emission.

Moreover, we note that the diffuse eDust that scatters the UV starlight
from the midplane \citep{2014ApJ...789..131H,2014ApJ...785L..18S,2015ApJ...815..133S,2016ApJ...833...58H}
and produces the UV reflection halo has a potential to scatter the
H$\alpha$ photons originating from \ion{H}{2} regions in the galactic
plane. \citet{1996ApJ...467L..69F} investigated the amount of H$\alpha$
photons that originates in \ion{H}{2} regions and scattered by dust
at high altitude and found that $\sim10$\% of the extraplanar H$\alpha$
emission at $z\sim600$ pc can be attributed to the scattered light
(see also \citet{1999ApJ...525..799W} for a similar model in the
Milky Way Galaxy). However, they took into account only a thin dust
disk with a scale height of $\sim0.2$ kpc to calculate the fraction
of the scattered H$\alpha$ component. The presence of eDust in galaxies
where the extraplanar H$\alpha$ emission is detected will raise the
fraction of scattered H$\alpha$ photons compared to the estimation
of \citet{1996ApJ...467L..69F} and thus decrease the amount of in-situ
photoionized gas in the halo.

The above two concerns regarding the multiphase nature of the extraplanar
ISM and the possibility of H$\alpha$ to be scattered by the eDust
motivated the present study of comparing the extraplanar H$\alpha$
and UV emissions in the halos of nearby edge-on late\textendash type
galaxies. In this study, we compare the vertical profiles of H$\alpha$
emission to those of UV emission in the nearby edge-on galaxies. In
Section 2, we describe the sample galaxies. We examine correlation
relations between the vertical profiles of the extraplanar H$\alpha$
and UV emissions in Section 3. Correlation of the extraplanar emissions
with the star formation rate (SFR) is also investigated. Section 4
presents a summary and discussion.

\section{Data}

We analyzed the narrow band H$\alpha$ and UV images of the nearby
late-type edge-on galaxies which were taken from previous H$\alpha$
and UV galaxy surveys. The H$\alpha$ images were obtained from the
following databases: the Spitzer Local Volume Legacy \citep[LVL;][]{2008ApJS..178..247K,2009ApJ...703..517D}
survey, the Spitzer Infrared Nearby Galaxies Survey \citep[SINGS;][]{2003PASP..115..928K,2010ApJS..190..233M},
the Survey for Ionization in Neutral Gas Galaxies \citep[SINGG;][]{2006ApJS..165..307M},
the H$\alpha$ Galaxy Survey \citep[H$\alpha$GS;][]{2004A&A...414...23J},
the H$\alpha$ narrow band imaging survey of galaxies (H$\alpha$3;
\citealt{2003A&A...400..451G,2012A&A...545A..16G}), and the H$\alpha$
survey (henceforth Rossa) of \citet{2003A&A...406..493R,2003A&A...406..505R}.
The H$\alpha$ images are available in the relevant websites of LVL\footnote{\href{http://irsa.ipac.caltech.edu/data/SPITZER/LVL/summary.html}{http://irsa.ipac.caltech.edu/data/SPITZER/LVL/summary.html}},
SINGS\footnote{\href{http://irsa.ipac.caltech.edu/data/SPITZER/SINGS/}{http://irsa.ipac.caltech.edu/data/SPITZER/SINGS/}},
SINGG\footnote{\href{http://sungg.pha.jhu.edu/PubData/Portal/index.html}{http://sungg.pha.jhu.edu/PubData/Portal/index.html}},
H$\alpha$GS\footnote{\href{http://www.astro.ljmu.ac.uk/HaGS/fits/index.html}{http://www.astro.ljmu.ac.uk/HaGS/fits/index.html}},
H$\alpha$3\footnote{\href{http://goldmine.mib.infn.it/}{http://goldmine.mib.infn.it/}},
and Rossa\footnote{\href{https://ned.ipac.caltech.edu/cgi-bin/objsearch?refcode=2003A\%26A...406..505R&search_type=Search}{https://ned.ipac.caltech.edu/cgi-bin/objsearch?refcode=2003A\%{}26A...406..505R\&{}search\_{}type=Search}}.
In order to compare the H$\alpha$ images with the FUV and NUV images,
we also have retrieved the GALEX archival data\footnote{\href{http://galex.stsci.edu/GR6/}{http://galex.stsci.edu/GR6/}}
(Galaxy Evolution Explorer; \citealt{2005ApJ...619L...1M,2005ApJ...619L...7M,2007ApJS..173..682M})
of the sample galaxies. In total, 38 edge-on late-type galaxies with
a distance of less than 30 Mpc were selected. Visual inspection was
performed to exclude galaxies with noticeable spiral or asymmetry
patterns. The data with the highest signal-to-noise ratio were adopted
if the galaxy was observed several times in different databases.

Table \ref{table1} shows the sample galaxies and their basic information
mostly taken from the NASA/IPAC Extragalactic Database (NED)\footnote{\href{https://ned.ipac.caltech.edu/}{https://ned.ipac.caltech.edu/}}.
The H$\alpha$ data for 12 galaxies were obtained from LVL \citep{2008ApJS..178..247K,2009ApJ...703..517D},
12 galaxies from SINGG \citep{2006ApJS..165..307M}, 2 galaxies from
SINGS\citep{2003PASP..115..928K,2010ApJS..190..233M}, 4 galaxies
from H$\alpha$GS \citep{2004A&A...414...23J}, 6 galaxies from H$\alpha$3
\citep{2003A&A...400..451G,2012A&A...545A..16G}, and 2 galaxies from
Rossa \citep{2003A&A...406..493R,2003A&A...406..505R}.

The entries in Table \ref{table1} are organized as follows: column
(1) \textendash{} the running index number, column (2) \textendash{}
galaxy name taken from the NED\textquoteright s preferred object name,
column (3) \textendash{} galaxy morphology, column (4) \textendash{}
galactic longitude, column (5) \textendash{} galactic latitude, column
(6) - average value of the redshift-independent distances taken from
NED, column (7) \textendash{} major axis diameter in arcmin, column
(8) - major axis diameter in kpc, column (9) \textendash{} integrated
H$\alpha$ luminosity taken from the reference papers, column (10)
\textendash{} star formation rate SFR$_{{\rm H}\alpha}$ estimated
using the H$\alpha$ luminosity (column 9), column (11) \textendash{}
star formation rate SFR$_{{\rm FIR}}$ estimated using the far-infrared
(FIR) luminosity $L_{{\rm FIR}}$, and column (12) \textendash{} references
for the H$\alpha$ data. The star formation rates, shown in columns
(10) and (11), were estimated using the following relations \citep{1998ApJ...498..541K}:
\begin{equation}
{\rm SFR}_{{\rm H}\alpha}=\frac{L_{{\rm H}\alpha}}{1.26\times10^{41}\ {\rm erg}\ {\rm s}^{-1}}M{}_{\odot}\ {\rm yr}^{-1},\label{eq:1}
\end{equation}
\begin{equation}
{\rm SFR}_{{\rm FIR}}=\frac{L_{{\rm FIR}}}{2.2\times10^{43}\ {\rm erg}\ {\rm s}^{-1}}M{}_{\odot}\ {\rm yr}^{-1},\label{eq:2}
\end{equation}
where the FIR luminosity $L_{{\rm FIR}}$ was calculated using the
relation of \citet{1988ApJS...68...91R}, $L_{{\rm FIR}}=1.51\times10^{39}d_{{\rm Mpc}}^{2}(2.58f_{60}+f_{100})$
erg s$^{-1}$. Here, $f_{60}$ and $f_{100}$ are the fluxes at 60
and 100 $\mu$m, respectively, in Jy obtained from the IRAS catalog
\citep{1990IRASF.C......0M}. The FIR luminosities for seven galaxies
(UGCA 442, UGCA 193, ESO 347-G 017, IC 4951, UGC 08313, NGC 5229,
and NGC 5023) that were not provided in the IRAS catalog were calculated
using the monochromatic luminosity $L_{{\rm FIR}}=4\pi d^{2}(c/\lambda)f_{70}=5.39\times10^{39}d_{{\rm Mpc}}^{2}f_{70}$
erg s$^{-1}$, where $f_{70}$ is the flux at 70 \textgreek{m}m in
Jy obtained from the MIPS catalog (Multiband Imaging Photometer for
Spitzer; \citealt{2009ApJ...703..517D}). The SFR$_{{\rm FIR}}$ for
three galaxies (NGC 7412A, ESO 249-G 035, NGC 3365) that have no FIR
luminosity were estimated from the empirical relation SFR$_{{\rm FIR}}$
$=1.31\times{\rm SFR}_{{\rm H}\alpha}^{1.54}$. This
relation was derived by using the correlation between ${\rm SFR}_{{\rm FIR}}$
and ${\rm SFR}_{{\rm H}\alpha}$ of our galaxies, of which the FIR
luminosities are available. Thus, the relation would be suitable for
edge-on galaxies in a statistical sense, although H$\alpha$ emission
is not a good tracer of SFR for edge-on galaxies.

The H$\alpha$ and UV images of 38 galaxies were processed
in the following order. First, point-like sources as well as extended
sources except the target galaxy were masked out using Source Extractor
\citep{1996A&AS..117..393B}. Most point sources in the H$\alpha$
images were removed through the continuum subtraction process. Second,
the masked image was rotated about the center of the galaxy to align
the major axis of the galaxy with the horizontal axis of the image.
Third, the rotated image was cropped to a rectangular shape, putting
the center of the galaxy at the center of the rectangle. The final
images and vertical profiles in the H$\alpha$, FUV, and NUV wavelengths
for the 38 galaxies are shown in Figure \ref{fig1}.

An important factor affecting vertical profiles of the extraplanar
emission is an extended wing of the point spread function (PSF) of
a telescope \citep{2014A&A...567A..97S,2015A&A...577A.106S}. \citet{2015ApJ...815..133S}
took this effect into account in dust radiative transfer models and
\citet{2016ApJ...833...58H} subtracted the contamination by the PSF-wing
from the observed images. The effect of the extended
wing is severe only when studying surface brightnesses that are much
lower than $\sim10^{-2}$ ($\sim5$ mag) of peak intensity at the
galactic plane. As can be seen in Figure \ref{fig1}, we mainly focus
on higher intensity levels. To test the PSF-wing effect, we examined
whether the scale heights of extraplanar H$\alpha$, FUV, and NUV
emissions, derived in Section 3.2, systematically increase with distance
to the galaxies. But, no systematic trend was found. Therefore, we
conclude that this effect is not significant for galaxies in our sample.
The contamination by the extended wing of PSF may marginally change
the scale heights of the extraplanar emission measured in the present
study. However, this effect does not significantly alter the results
presented in this paper.

We also note that, even at small deviations from $90^{\circ}$,
the projected disk may appear as vertical emission. Some of the galaxies
in Figure \ref{fig1} do not appear to be completely edge-on (e.g.,
NGC 5951, NGC 493, NGC 803, NGC 4020, NGC 3365, IC 2000, and NGC 5356).
In addition, some of the galaxies show disturbed disks in optical
and Near-IR images (e.g., NGC 5107, NGC 4631, NGC 3432, and NGC 3628).
Therefore, in Figures \ref{fig2} to \ref{fig5}, we used different
colors to denote these galaxies to see if they stand out in any particular
way: blue for the less-inclined galaxies and yellow for the disturbed
disks. In the figures, it can be immediately recognized that our results
are not affected by the galaxies.

\section{Results}

\subsection{Morphology}

As shown in Figure \ref{fig1}, the morphology of
the H$\alpha$ emission, especially in the galactic plane, is in
general more compact than those shown in the FUV and NUV images, although
deeper and higher-resolution observations are required for a more
detailed comparison. In other words, there is less contrast in the
UV light, compared to the H$\alpha$ emission. We also note that the
radial and vertical extents appear to be smaller at H$\alpha$.

The trend is consistent with the anti-correlation between the FUV
to H$\alpha$ intensity ratio and H$\alpha$ intensity found in 10
face-on spiral galaxies \citep{2000ApJ...541..597H,2001ApJ...559..878H}
and two starburst galaxies \citep{2005ApJ...619L..99H}. A similar
trend was also found in the Milky Way Galaxy \citep{2011ApJ...743..188S}.
Moving from a bright region into diffuse regions, the FUV to H$\alpha$
intensity ratio increases and thus the H$\alpha$ intensity decreases
faster than the decrease at FUV. This property is equivalent to the
more compact morphology at H$\alpha$. This is due to the fact that
\ion{H}{2} regions are more spatially clumped than stars that emit
the FUV and NUV continuum. The H$\alpha$ emission originates mainly
from the \ion{H}{2} regions around OB associations, while the FUV
and NUV emissions arise not only from OB associations but also from
late field OB and A stars, which are more spatially extended than
OB associations. Further discussion of the morphology is given in
Section \ref{sec:Conclusions}.

\subsection{Vertical Profile}

\begin{figure*}[tp]
\begin{centering}
\medskip{}
\par\end{centering}
\begin{centering}
\includegraphics[clip,scale=0.7]{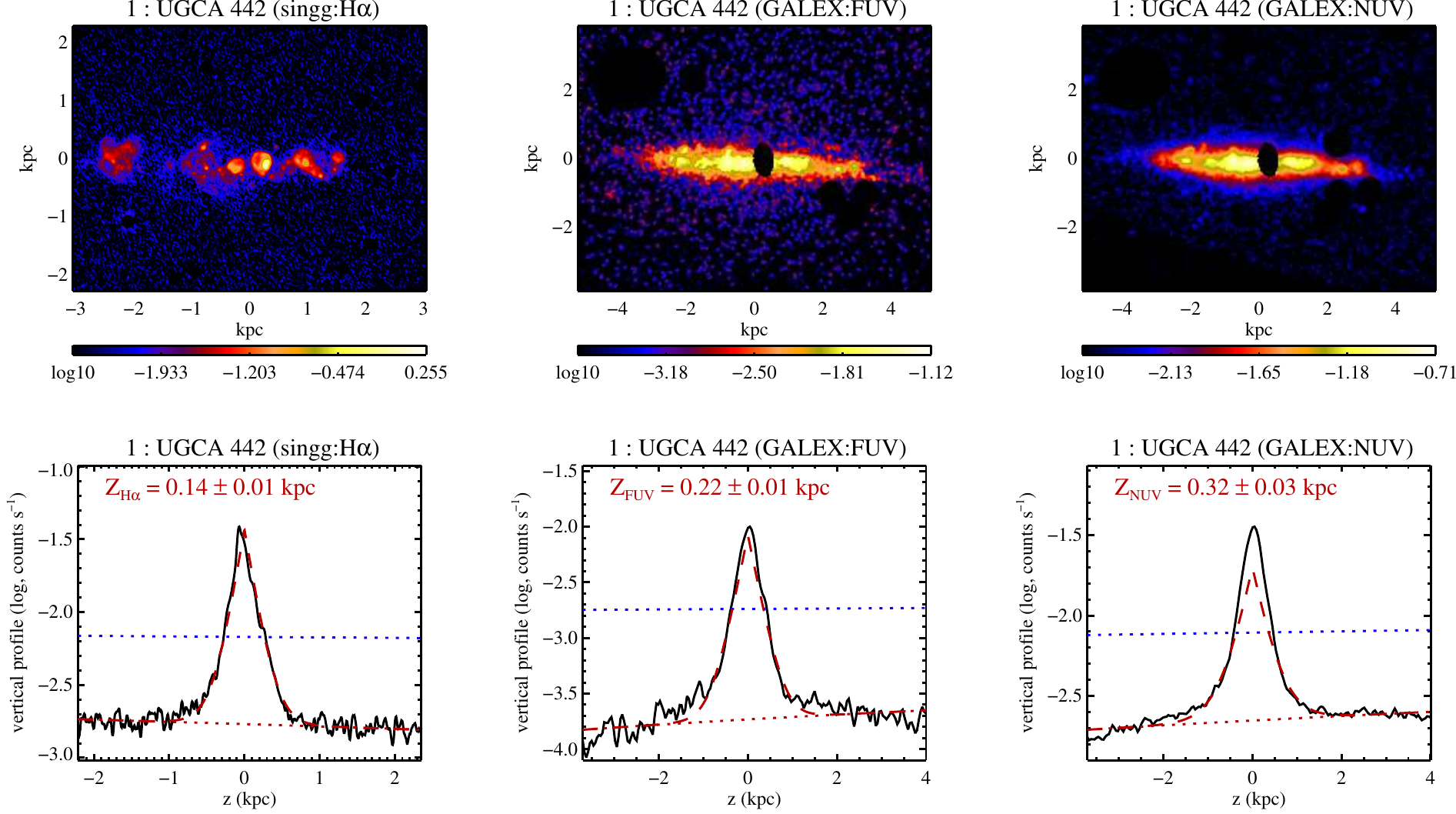}
\par\end{centering}
\begin{centering}
\includegraphics[clip,scale=0.7]{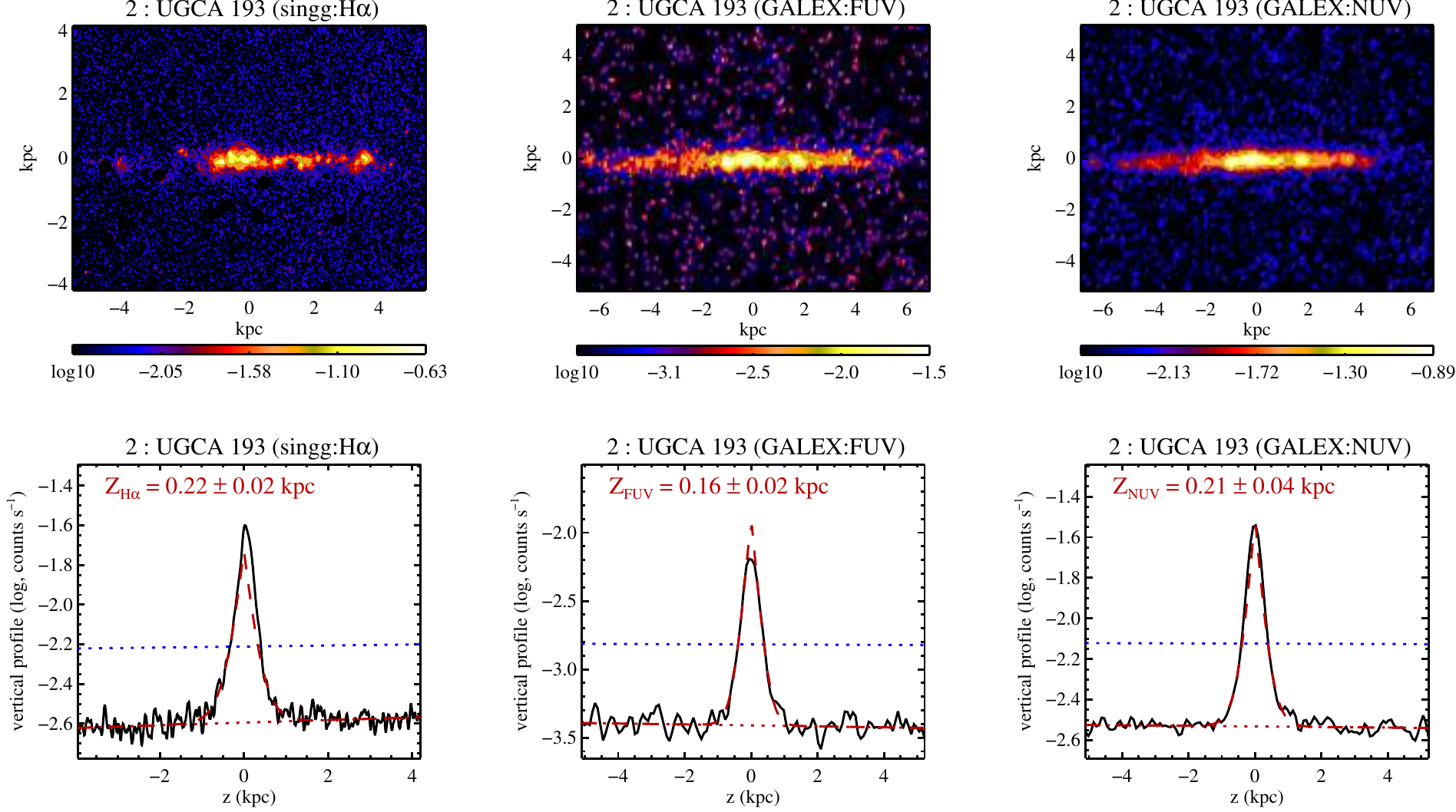}
\par\end{centering}
\begin{centering}
\medskip{}
\par\end{centering}
\caption{\label{fig1}(left top) H$\alpha$ emission map, (middle top) GALEX
FUV map, (right top) GALEX NUV map, (left bottom) H$\alpha$ vertical
profile, (middle bottom) FUV vertical profile, and (right bottom)
NUV vertical profile for each galaxy. In the vertical profiles, the
red dotted lines indicate the background levels. The red dashed lines
denote the best fit exponential function for the vertical profiles.
The horizontal blue dotted lines are the line to distinguish the bright
galactic plane region from the diffuse extraplanar region of the galaxies. }
\end{figure*}

The vertical profiles of the H$\alpha$, FUV, and NUV emissions for
the 38 edge-on galaxies were obtained by horizontally averaging each
image and then the profiles, denoted by black solid lines in Figure
\ref{fig1}, were fitted with an exponential function. The adopted
exponential function to fit the extraplanar emission is 
\begin{equation}
y=a_{0}\exp\left(-\frac{\left|x-a_{1}\right|}{a_{2}}\right)+a_{3}+a_{4}x,\label{eq:3}
\end{equation}
where the first term on the right-hand side is an exponential function
representing the vertical profile of the extraplanar emission and
the remaining terms are a linear function representing the background
of the profile. The parameters of the exponential function in Equation
(\ref{eq:3}) are the scale height ($a_{2}$), the peak intensity
($a_{0}$), and the location of the galactic center ($a_{1}$), respectively.
The resulting best-fit exponential functions are represented by red
dashed lines in Figure \ref{fig1}. The background levels are denoted
by red dotted lines in Figure \ref{fig1}. The scale height found
for each image is shown at the top left corner of each figure together
with its 1$\sigma$ error in units of kpc. The best-fit parameters
are shown in Table \ref{table2}.

The thin disk component of dust with a scale height of $\sim0.2$
kpc was assumed to be mostly confined in a region with intensity brighter
than a certain threshold marked by blue dotted lines in Figure \ref{fig1}.
The region above the threshold level was excluded from the fit to
minimize contamination by the thin disk component in estimating the
extraplanar component. The threshold for most galaxies was set to
be $e^{-2}$ times the difference between the peak intensity and the
background. For 10 galaxies (ESO 249-G 035, NGC 7412A, NGC 1311, NGC
0784, UGC 08313, IC 2233, NGC 3510, NGC 4313, IC 5176, and NGC 3628)
that have relatively poor signal-to-noise ratios, we set the threshold
as $e^{-1}$ times the difference. The threshold was empirically chosen
to discriminate the extraplanar component from the thin disk component.
It should be noted that the radial scale length of the thin dust disk
tends to be larger than that of the stellar disk \citep{1997A&A...325..135X,1998A&A...331..894X,2014MNRAS.441..869D,2014ApJ...785L..18S,2015ApJ...815..133S}.
Therefore, the observed peak intensity is an attenuated value by the
thin dust disk and the adopted threshold level is much lower than
the value estimated by multiplying $e^{-2}$ or $e^{-1}$ to the intrinsic
peak intensity. This implies that we are analyzing regions far enough
from the galactic plane. The best-fit scale height would be higher
than the value expected from the galactic thin disk, if there is an
additional, thick component in the halo as detected in \citet{2014ApJ...785L..18S},
\citet{2015ApJ...815..133S}, \citet{2014ApJ...789..131H}, and \citet{2016ApJ...833...58H}.
We also adopted lower thresholds corresponding to more outer regions
from the midplane and repeated the analysis, but the following results
were not significantly altered.

We compare the scale heights estimated from the H$\alpha$, FUV, and
NUV data in Figure \ref{fig2}. Figure \ref{fig2}(a) shows a strong
correlation between the scale heights of the FUV ($Z_{{\rm FUV}}$)
and NUV (Z$_{{\rm NUV}}$) emissions. The scale height of the H$\alpha$
emission (Z$_{{\rm H}\alpha}$) is compared to that of the FUV emission
(Z$_{{\rm FUV}}$) in Figure \ref{fig2}(c). In Figures \ref{fig2}(b)
and (d), the scale heights normalized to the size of major axis (D$_{25}$)
are compared. The black dashed diagonal lines in Figure \ref{fig2}
indicate one-to-one correspondence lines. There are strong correlations
not only between Z$_{{\rm FUV}}$ and Z$_{{\rm NUV}}$ but also between
the normalized values Z$_{{\rm FUV}}$/$D_{25}$ and Z$_{{\rm NUV}}$/$D_{25}$.
The correlation coefficient between Z$_{{\rm FUV}}$ and Z$_{{\rm NUV}}$
(Z$_{{\rm FUV}}$/D$_{25}$ and Z$_{{\rm NUV}}$/D$_{25}$) is obtained
as 0.96 (0.88) in Figure \ref{fig2}(a) (Figure \ref{fig2}(b)). In
Figure \ref{fig2}(c) (Figure \ref{fig2}(d)), the correlation coefficient
between Z$_{{\rm H}\alpha}$ and Z$_{{\rm FUV}}$ (Z$_{{\rm H}\alpha}$/D$_{25}$
and Z$_{{\rm FUV}}$/D$_{25}$) is found to be 0.89 (0.67). The correlations
shown in Figures \ref{fig2}(c) and (d) are less significant than
the cases of Z$_{{\rm FUV}}$ and Z$_{{\rm NUV}}$, but still strong.
In Figures \ref{fig2}(a) (Figure \ref{fig2}(b)), the blue dot-dashed
line denotes the linear line representing a direct proportional relation
between Z$_{{\rm FUV}}$ and Z$_{{\rm NUV}}$ (Z$_{{\rm FUV}}$/D$_{25}$
and Z$_{{\rm NUV}}$/D$_{25}$). The proportional relations are also
shown in the figures. As shown in the figure, the scale height of
the NUV emission is found to be in general the same as (but slightly
lower than) that of the FUV emission. The scale height of the H$\alpha$
emission tends to be smaller than that of the FUV emission. In Figures
\ref{fig2}(c) and (d), the blue triple-dot-dashed lines denote the
lines corresponding to $Z_{{\rm H}\alpha}=0.5Z_{{\rm FUV}}$. The
H$\alpha$ intensity scales with the emission measure (defined as
the square of the number density of electrons integrated over the
volume of ionized gas); the scale height of electron (ionized gas)
may be twice as large as the H$\alpha$ scale heights. If the ionized
gas traced by H$\alpha$ emission is well mixed with dust, then the
H$\alpha$ scale height will be half of the dust scale height.

The sample galaxies were divided into two groups for convenience,
as shown in Table \ref{table2}: Group A with a scale height less
than 0.4 kpc in both UV wavelength bands and Group B with a scale
height greater than 0.4 kpc in both UV bands. The scale height threshold
dividing Groups A and B was determined based on the observations that
the scale height of OB stars that are the main source of the UV continuum
is $\lesssim0.2$ kpc and the scale height of thin dust disk tends
to be $\sim0.2$ kpc (ranging from $\sim0.1$ kpc to $\sim0.4$ kpc)
\citep{1997A&A...325..135X,1998A&A...331..894X,1999A&A...344..868X,2004A&A...425..109A,2014MNRAS.441..869D}.
The galaxies that have relatively small inclinations
or show disturbed disks in optical images belong to Group B. In Figures
\ref{fig2} to \ref{fig5}, Groups A is denoted by black diamonds.
Group B is denoted by red, blue, and yellow squares. The blue and
yellow squares indicate the less-inclined galaxies and disturbed disks,
respectively. The red squares denote remaining galaxies in Group B.
The black dashed, vertical, and horizontal lines in Figures \ref{fig2}(a)
and (c) indicate lines corresponding to Z$_{{\rm FUV}}$ = 0.4 kpc
and Z$_{{\rm NUV}}$ = 0.4 kpc, respectively. The symbol size in Figures
\ref{fig2} to \ref{fig5} is proportional to the logarithm of the
galaxy size ($\log{\rm D}_{25}$).

The average scale heights at H$\alpha$ and FUV for Group A are Z$_{{\rm H}\alpha}$
$=0.21\pm0.5$ kpc and Z$_{{\rm FUV}}$ $=0.23\pm0.6$ kpc, respectively.
These values are consistent with the scale height of the thin dust
disk as well as OB associations. This indicates that the Group A galaxies
have no (or negligible) additional geometrically thick dust component
and we are detecting the exponential tail of the thin disk. On the
other hand, the larger scale heights found in Group B imply the presence
of an additional component in the galaxies of Group B. The scale height
Z$_{{\rm H}\alpha}$ of Group A ranges from $\sim$ 0.1 to 0.3 kpc
except IC 5052. Most galaxies in Group B, except seven galaxies, have
a scale height at H\textgreek{a} larger than 0.4 kpc. We also note
that the galaxies with small $Z_{{\rm H}\alpha}(<0.4$ kpc) in Group
B have relatively small Z$_{{\rm FUV}}$. In other words, the galaxies
with the additional extraplanar H$\alpha$ emission appear to have
the extraplanar FUV emission as well. The trend is consistent with
the results of \citet{2000AJ....119..644H} and \citet{2003A&A...406..505R},
in that they also found a similar trend by comparing the clumpy features
of eDust and the H$\alpha$ emission. However, it should be noted
that Z$_{{\rm H}\alpha}$ is in general smaller than Z$_{{\rm FUV}}$.
The average ratio of Z$_{{\rm H}\alpha}$ to Z$_{{\rm FUV}}$ is $0.74\pm0.30$
for the galaxies in Group B. This point will be discussed in Section
\ref{sec:Conclusions}.

The normalized scale heights (Z$_{{\rm FUV}}$/D$_{25}$ and Z$_{{\rm H}\alpha}$/D$_{25}$)
in Figure \ref{fig2}(b) and (d) range from 0.01 to 0.1. In the figure,
averages of the normalized scale heights of Group B appear to be slightly
higher than those of Group A, although the differences are not large.
The absence of a substantial difference in the normalized scale height
between the two groups suggests that the scale height tends to increase
with the galaxy size. Nonetheless, the finding that the normalized
scale height does not approach to a single value indicates that the
scale height depends on other properties (e.g., star formation rate)
of galaxies as well. This issue will be discussed in the next section.

\subsection{Comparison with Star Formation Rate}

Most phenomena in spiral galaxies are closely associated with the
star formation activity. We therefore compare the scale heights of
FUV and H$\alpha$ emissions with the star formation rates derived
from the FIR luminosity (SFR$_{{\rm FIR}}$) of host galaxies in Figure
\ref{fig3}. It is clear that both the scale heights Z$_{{\rm FUV}}$
and Z$_{{\rm H}\alpha}$ strongly correlate with SFR$_{{\rm FIR}}$.
The correlation coefficients in Figures \ref{fig3}(a) and (b) are
0.92 and 0.87, respectively. It is found that the scale heights are
well described by a power law function of SFR$_{{\rm FIR}}$. The
best-fit power law function is overplotted as a black dashed line
in the figure. The equation describing the best-fit power law is also
shown at the top left corner. In Figure \ref{fig3}, the condition
for the detection of the extraplanar emission is SFR$_{{\rm FIR}}$
$\gtrsim0.03M_{\odot}$ yr$^{-1}$.

For a given SFR, the scale height tends to increase with the galaxy
size. For instance, the three galaxies NGC 5107, NGC 4313, and NGC
3365 with SFR $\sim0.1M_{\odot}$ yr$^{-1}$ show the trend clearly.
Figure \ref{fig4} shows a strong correlation between the scale height
and the galaxy size, although the correlation is slightly weaker than
the correlation between the scale height and SFR shown in Figure \ref{fig3}.
Figure \ref{fig4} also shows that the scale height increases with
SFR for a given galaxy size, as in Figures \ref{fig2} and \ref{fig3}.
The stronger correlation of the scale height with SFR$_{{\rm FIR}}$
than with the galaxy size is ascribed to the wider dynamic range of
SFR.

Figures \ref{fig5}(a) and (b) compare the scale heights
normalized by D$_{25}$ with the surface density of SFR ($\Sigma_{{\rm SFR,FIR}}\equiv{\rm SFR}_{{\rm FIR}}/\pi D_{25}^{2}$
) of the host galaxies. The (non-normalized) scale
heights are compared with the surface density of SFR in Figures \ref{fig5}(c)
and (d). Both the normalized and non-normalized scale heights of
the FUV and H$\alpha$ emissions show a good correlation with $\Sigma_{{\rm SFR,FIR}}$.
The correlation of the SFR surface density is stronger with the H$\alpha$
scale height than with the FUV scale height. It is interesting that
the normalized scale height of Group A appears to decrease as the
galaxy size, indicated by the symbol size, increases. On the other
hands, there is no clear tendency between the normalized scale height
and the galaxy size for the Group B galaxies.

\citet{2013ApJ...774..126M} found a correlation between the SFR surface
density and the normalized scale height of the extraplanar emission
by polycyclic aromatic hydrocarbons (PAHs) for 16 local, active galaxies
known to have galactic winds. \citet{2015ApJ...815..133S} modeled
the extraplanar dust of six nearby galaxies using a radiative transfer
simulation for FUV images, and compared the obtained scale height
of eDust with the SFR surface density for three targets (NGC 891,
NGC 3628, and UGC 11794), which apparently show the extraplanar FUV
emission. Since the galaxies analyzed in \citet{2013ApJ...774..126M}
are more active than the galaxies in this paper, it would be interesting
to combine their results with ours. Figure \ref{fig6} shows the SFR
surface density as a function of the normalized scale height for our
galaxies in Group B together with the results of \citet{2013ApJ...774..126M}
and \citet{2015ApJ...815..133S}. The blue pluses denote 16 galaxies
of \citet{2013ApJ...774..126M} and the black asterisks indicate three
galaxies of \citet{2015ApJ...815..133S}. The red triangles and orange
crosses indicate the results obtained from the extraplanar FUV and
H$\alpha$ emissions, respectively, of 21 galaxies (Group B galaxies)
in the present study. Note that two of the galaxies (NGC 891 and NGC
3628) in \citet{2015ApJ...815..133S} are also included in the present
study. It is clear that the correlation relation between the normalized
scale height of the PAH emission and the SFR surface density is consistent
with the relation estimated using the extraplanar FUV and H$\alpha$
emissions. This implies that the correlation relation between the
normalized scale height and the SFR surface density holds for a very
wide range of SF activity.

\section{Summary and Discussion}

\label{sec:Conclusions}

We measured the vertical scale heights of the extraplanar H$\alpha$,
FUV, and NUV emissions for 38 nearby edge-on galaxies. The scale height
at NUV is found to be similar to or slightly higher than the scale
height at FUV. This might be due to the fact that the NUV continuum
source could be a slightly later-type and thus has a slightly higher
scale height than the source of the FUV continuum. It is also found
that galaxies with the extraplanar H$\alpha$ emission always show
the extraplanar FUV emission as well. The scale height of the H$\alpha$
emission strongly correlates with the scale height of the UV emission.
The scale height at H$\alpha$ is found to be in general lower than
that at UV.

\citet{2003A&A...406..505R} found a correlation between the presence/non-presence
of eDust and eDIG in edge-on galaxies. \citet{2000AJ....119..644H}
and \citet{2013AJ....145...62R} found no spatial correspondence between
the smoothly distributed H\textgreek{a} emission and the opaque eDust
filamentary structure. However, it should be noted that these studies
are based on the observations of opaque dust clumps at high altitudes.
The present study has an advantage over these studies in that the
UV reflection halos can probe the diffuse eDust, which was not detectable
in the studies of \citet{2003A&A...406..505R}, \citet{2000AJ....119..644H},
and \citet{2013AJ....145...62R}.

The scale heights of the extraplanar UV and H$\alpha$ emissions are
found to correlate with the star formation rate (SFR$_{{\rm FIR}}$),
the galaxy size (D$_{25}$), and the SFR surface density ($\Sigma_{{\rm SFR,FIR}}$).
The scale heights at the extraplanar emissions correlate more strongly
with the SFR than with the size of galaxy. The SFR surface density
correlates with the normalized scale heights measured at UV and H$\alpha$.
This result is in good agreement with the relation found for other
galaxies in which the extraplanar PAH emission was detected by \citet{2013ApJ...774..126M}.
We note that the galaxies observed in \citet{2013ApJ...774..126M}
are known to have galactic winds and thus are more active than our
galaxies. The correlation suggests that the extraplanar ISM is closely
associated with the galactic SF activities, such as stellar radiation
pressure and supernovae feedback. The validity of the single correlation
relation over a wide type of galaxies, ranging from normal star forming
galaxies with no apparent galactic winds to starburst galaxies with
strong winds, indicates that the properties of the extraplanar ISM
do not abruptly change at a critical level as the SF activity increases.

The strong correlation between the presence/non-presence of the FUV
and H$\alpha$ emissions and between their vertical profiles suggests
two possibilities for the origin of the diffuse H$\alpha$ emission.
First, the extraplanar FUV and H$\alpha$ emissions trace the eDust
and ionized gas (eDIG), respectively, and the correlation is caused
by the multiphase nature of the ISM in galactic halo. The eDIG in
the traditional scenario is believed to be produced by ionizing photons
transported through transparent pathways carved out by superbubbles
or chimneys \citep{1999ApJ...513..142M,2002ApJ...576..745C,2004ApJS..151..193S,2004ApJ...606..829S,2005ARA&A..43..769V}.
Second, a substantial portion of the extraplanar H$\alpha$ emission
is caused by dust scattering of the photons originating from \ion{H}{2}
regions in the galactic disk.

Detailed model calculations for photoionization and dust radiative
transfer in the galactic scale taking into account not only the global
ISM structure but also small structures must be carried out in order
to pin down the main origin of the extraplanar H$\alpha$ emission.
\citet{2010ApJ...721.1397W} investigated models for the photoionization
of the DIG in galaxies using hydrodynamic simulations of a supernova-driven
ISM. However, the emission measure distributions in their simulations
were found to be wider than those derived from H$\alpha$ observations,
implying the adopted ISM models are too porous to represent the realistic
density structure of the ISM. The emission measure distribution or
the H$\alpha$ intensity distribution is directly coupled to the density
distribution or porosity of the ISM, as noted in \citet{2009ApJ...703.1159S}.
Therefore, more extensive studies on photoionization models of the
DIG are required to reproduce not only the high altitude H$\alpha$
emission but also the observed distribution of emission measure.

Recently, \citet{2017MNRAS.466.3217Z} investigated the DIG using
a sample of 365 face-on galaxies and concluded that ionization by
evolved stars in the galactic halo with LI(N)ER-like emission is likely
a major ionization source for DIG. An alternative scenario was proposed
by \citet{2011ApJ...727...35D}. In their model, the diffuse H$\alpha$
emission is likely powered by runaway OB stars. We note that a time
dependent photoionization model is needed to investigate the cooling
and recombining gas model proposed by \citet{2011ApJ...727...35D},
which is even more challenging than most of the photoionization models
assuming a steady ionization state. \citet{2000AJ....119..644H} suggested
a scenario that may be relevant to the proposal of \citet{2011ApJ...727...35D}.
They found an anti-correlation between the FUV to H$\alpha$ intensity
ratio and the H$\alpha$ intensity in face-on galaxies and proposed
that photoionization of the diffuse ISM is maintained by late-type
field OB stars. \citet{2011ApJ...743..188S} and \citet{2012ApJ...758..109S}
also investigated this possibility. In the present study, it was found
that the morphology at H$\alpha$ is more compact and clumpier than
that at FUV, implying an anti-correlation between the FUV to H$\alpha$
intensity ratio and the H$\alpha$ intensity.

The FUV halo emission could be either starlight from
the stellar halo or a reflection nebula produced by scattering of
FUV photons that escape the disk. The recent studies suggesting that
evolved hot stars may contribute to the ionization of DIG \citep[e.g.,][]{2017MNRAS.466.3217Z}
might imply that some of the extraplanar FUV emission is also stellar in
origin. One way to examine this possibility may be to compare the
various correlations presented in this study with the stellar, vertical
light distributions. \citet{2014ApJ...789..131H} examined which scenario
is more consistent with the data from the perspective of UV\textminus r
color in the halo, and SED fitting using dust models. They found that
UV\textminus r colors and SEDs in the halos are more consistent with
being a reflection nebula. Using dust radiative transfer models, \citet{2014ApJ...785L..18S}
and \citet{2015ApJ...815..133S} could successfully explain the vertically
extended FUV and NUV emissions as being due to dust-scattered starlight.
The amount and scale height of eDust that were calculated from the
radiative transfer models were found to be consistent with the observed
vertical profile of FIR emission in NGC 891 \citep{2016A&A...586A...8B}.
Polarization maps in the optical wavelengths can also trace large-scale
galactic dust distributions. In edge-on galaxies, for instance NGC
891 and NGC 4565, extended optical polarization features were found
in the halo regions above the galactic midplane \citep{1990IAUS..140..245S,1996MNRAS.278..519S,1996A&A...308..713F}.
If only a thin dust layer with a scale height of $\sim0.2$ kpc is
assumed, the polarization arising from scattering or dichroic extinction
is predicted to be very low at high altitudes, and hence the extended
polarization pattern cannot be explained \citep[e.g.,][]{1996ApJ...465..127B,1997AJ....114.1405W,2017A&A...601A..92P}.
Therefore, the extended optical polarization indicates the existence
of a thick dust disk. In a seperate paper (Seon 2018, submitted),
we show that the extended optical polarization can be well explained
by the extraplanar dust layer which was inferred from the observations
of UV halos. Therefore, most of the extraplanar FUV emission measured
in our galaxies can be attributed to scattered light rather than to
direct starlight.

The simultaneous existence of the diffuse eDust and the extraplanar
H$\alpha$ emission suggests an interesting possibility that a large
fraction of the H$\alpha$ emission could originate from the galactic
plane and is scattered by the eDust into sightlines of the galactic
halo. \citet{1996ApJ...467L..69F} and \citet{2015MNRAS.447..559B}
investigated models for dust scattering of H$\alpha$ photons by assuming
only a thin dust disk, without taking into account eDust, and found
that less than $\sim20$\% of the total H\textgreek{a} intensity
can be attributed to dust scattering. Here, it should be emphasized
that the extraplanar FUV emission scattered by the eDust is more extended
than the extraplanar H$\alpha$ emission. The scattering cross-section
at H$\alpha$ is lower than that at FUV only by a factor of 1.9 for
the Milky Way dust \citep{2001ApJ...548..296W,2003ApJ...598.1017D}.
Therefore, there is no reason not to consider the possibility in which
a substantial portion of the total extraplanar H$\alpha$ emission
is attributed to dust scattering by eDust.

The most clear evidence on the existence of the ionized gas is provided 
by the pulsar dispersion measures. However, the pulsar dispersion measure 
alone provides only limited information (column density of electrons) on 
the ionized gas. The volume filling fraction and temperature of the DIG 
in the Milky Way were estimated from an ``implicit'' assumption that the 
pulsar dispersion measure and the H\textgreek{a} photons
probe the same ionized medium \citep{1989ApJ...339L..29R,2001ASPC..231..294H,2008PASA...25..184G}.
\citet{2001ASPC..231..294H} argued that the DIG probed by the diffuse
H$\alpha$ emission is needed to be distinguished from the ionized
gas that is traced by the pulsar dispersion measure. He showed that
the pulsar dispersion measures are highly likely to be produced mainly
by the warm ionized medium (WIM) predicted in the three phase model
of \citet{1977ApJ...218..148M}. In the thee phase model, the WIM
is predicted to occupy relatively a small fraction of the ISM.

It has been argued that the scattering effect does not seem to be
able to explain the observation that the ratios of forbidden lines
to Balmer line such as {[}\ion{N}{2}{]}/H$\alpha$ and {[}\ion{S}{2}{]}/H$\alpha$
increase with the altitude of a galaxy \citep{1985ApJ...294..256R,1987ApJ...323..118R,1994ApJ...431..156W}.
However, \citet{2012ApJ...758..109S} suggested a potential
solution to resolve this problem. The stellar continuum outside of
bright \ion{H}{2} regions is dominated by B- and A-type stars \citep{1992ApJ...388..310K,1992ApJS...79..255K}.
Balmer absorption lines in the underlying stellar continuum and its
scattered continuum background can give rise to underestimation of
the H$\alpha$ intensity and thus overestimation of the line ratios
of forbidden lines. However, in recent spectroscopic
studies of the DIG, the stellar continuum was fitted with stellar
sythesis models before the emission line ratios were measured \citep{2017A&A...599A.141J,2017MNRAS.466.3217Z}.
The resulting lines ratios in the DIG were found to be different from
those in \ion{H}{2} regions. Therefore, the present
results do not imply that most of the H$\alpha$ emission is caused
by the scattered light. Instead, it is suggested that a larger fraction
of the extraplanar H\textgreek{a} emission than that predicted by
\citet{1996ApJ...467L..69F} and \citet{2015MNRAS.447..559B} may
be caused by scattered H\textgreek{a} photons. We, therefore, need
to develop detailed models combining both photoionization and dust
scattering to investigate the importance of the dust scattering by
eDust. In a forthcoming paper, we will show how large fraction of
the extraplanar H$\alpha$ emission is attributable to the light scattered
by the eDust and discuss the effect of dust-scattered H$\alpha$ emission
on the line ratios.

\begin{figure}[t]
\begin{centering}
\medskip{}
\par\end{centering}
\begin{centering}
\includegraphics[clip,scale=0.48]{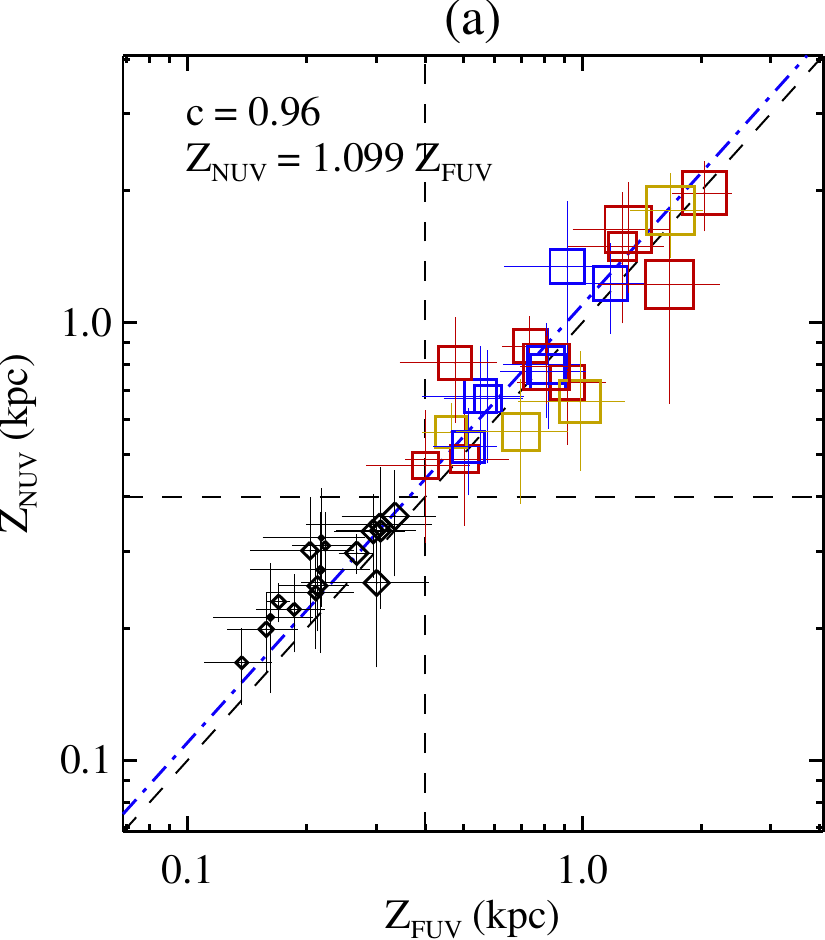}\ \ \includegraphics[clip,scale=0.48]{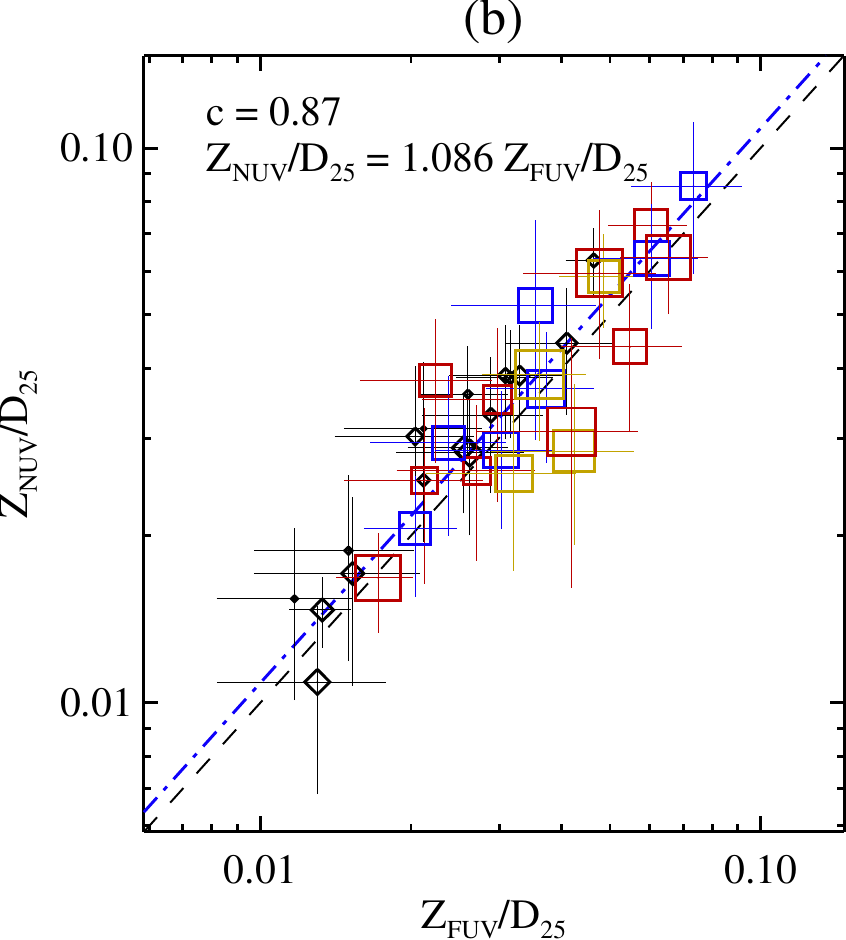}
\par\end{centering}
\begin{centering}
\medskip{}
\par\end{centering}
\begin{centering}
\includegraphics[clip,scale=0.48]{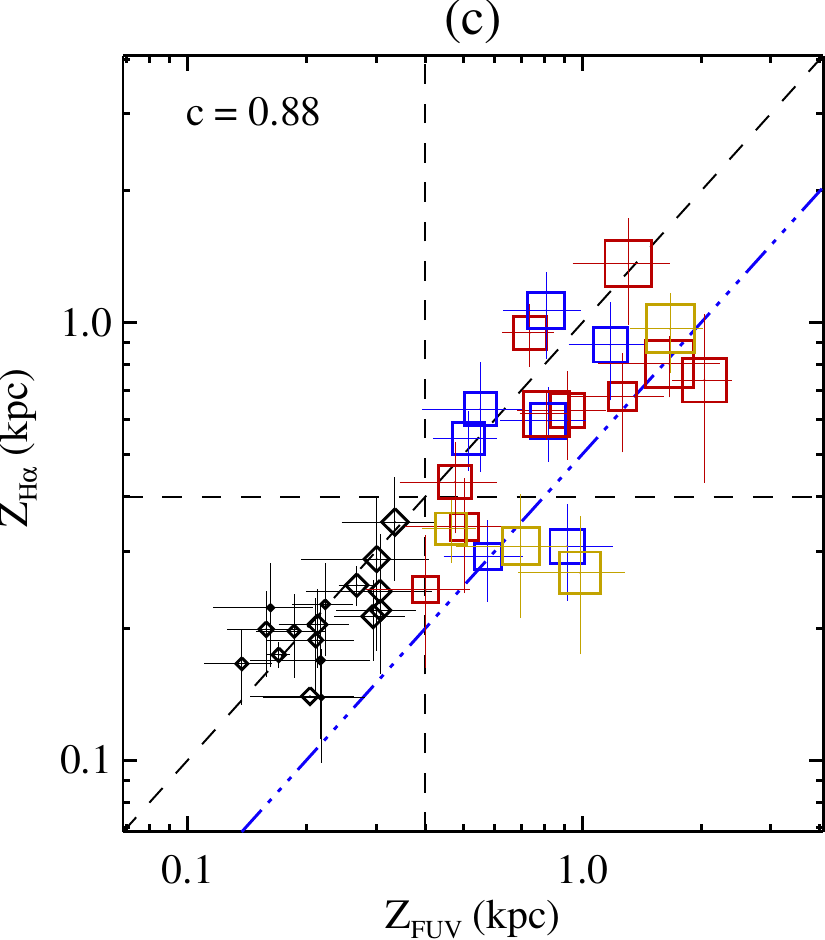}\ \ \includegraphics[clip,scale=0.48]{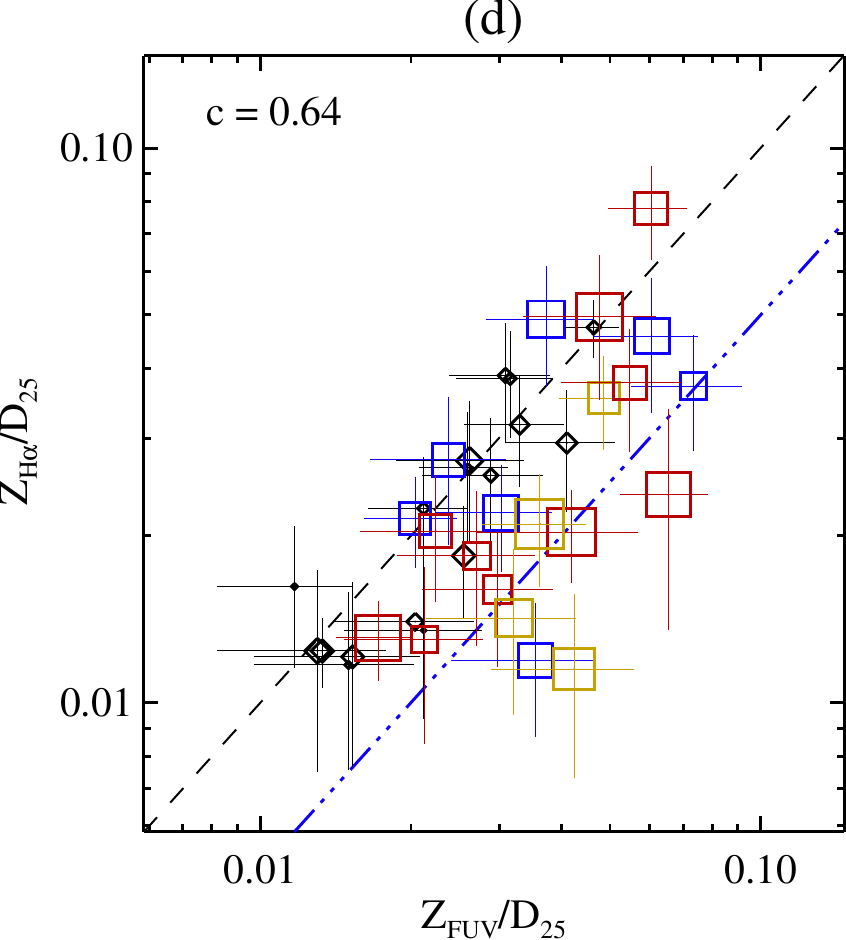}
\par\end{centering}
\begin{centering}
\medskip{}
\par\end{centering}
\caption{\label{fig2}Comparison of the scale heights of the H$\alpha$, FUV,
and NUV emissions for the 38 galaxies. Comparison of (a) scale heights
of the FUV and NUV emissions in kpc, (b) relative scale heights normalized
by D$_{25}$ of the host galaxies for the FUV and NUV emissions, (c)
scale heights measured at FUV and H$\alpha$ in kpc, and (d) relative
scale heights normalized by D$_{25}$ of the galaxies for the FUV
and H\textgreek{a} emissions. Black diamonds denote
Group A. Red, blue, and yellow squares denote Group B. Seven galaxies
with a relatively small inclination angle (NGC 5951, NGC 493, NGC
803, NGC 4020, NGC 3365, IC 2000, and NGC 5356) and four galaxies
showing disturbed disks (NGC 5107, NGC 4631, NGC 3432, and NGC 3628)
are denoted by the blue and yellow squares, respectively. The remaing
galaxies in Group B is denoted by the red squares. The symbol size
indicates the logarithm of the galaxy size. The blue, triple-dot dashed
lines in (c) and (d) denote a line corresponding to $Z_{{\rm H}\alpha}=0.5Z_{{\rm FUV}}$.}
\end{figure}

\begin{figure}[t]
\begin{centering}
\medskip{}
\par\end{centering}
\begin{centering}
\includegraphics[clip,scale=0.48]{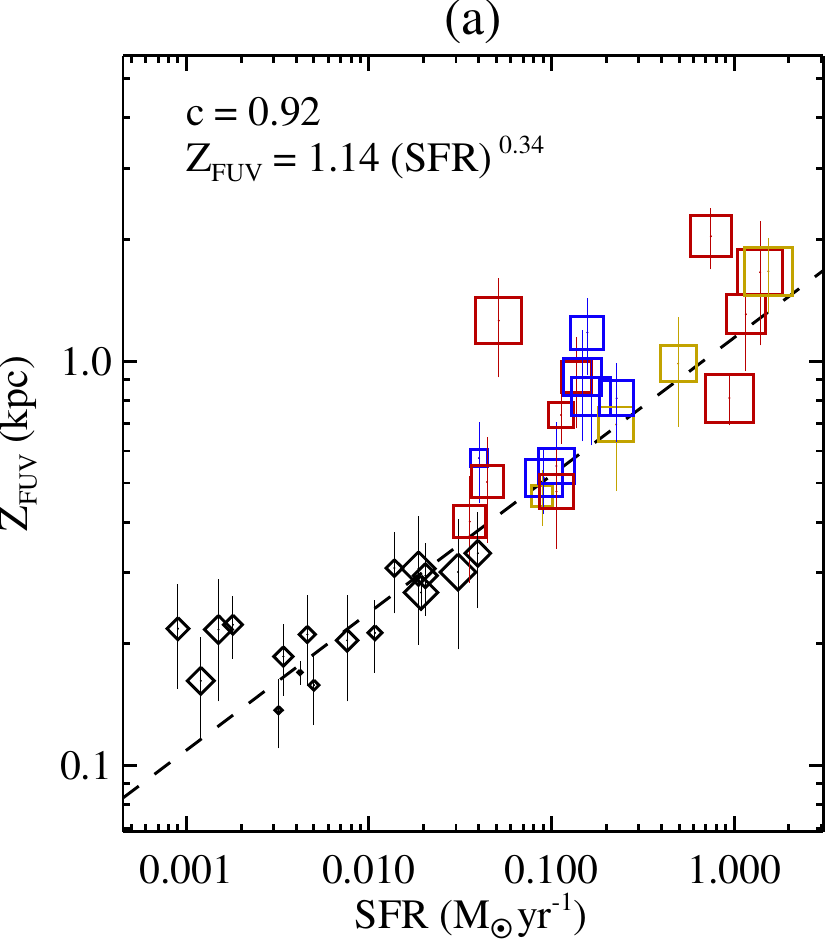}\ \ \includegraphics[clip,scale=0.48]{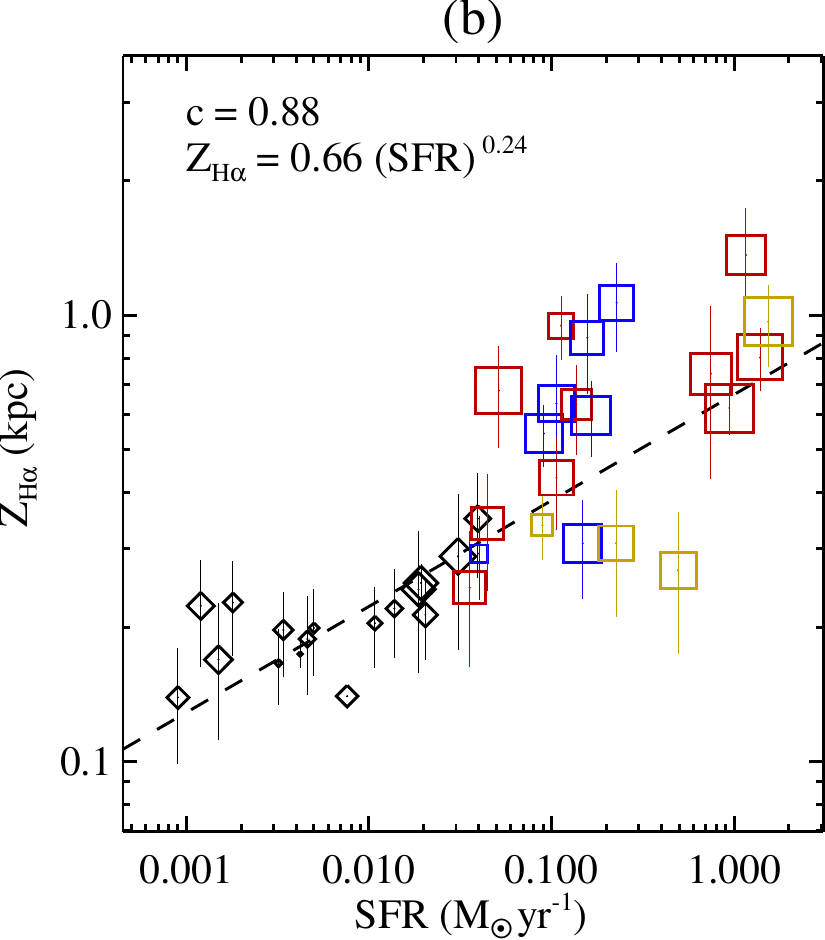}
\par\end{centering}
\begin{centering}
\medskip{}
\par\end{centering}
\caption{\label{fig3}Comparison of the scale heights of the FUV and H$\alpha$
emissions with the star formation rates (SFR$_{{\rm FIR}}$) of the
sample galaxies. The size of the symbol is proportional to the logarithm
of the galaxy size (D$_{25}$).}
\end{figure}

\begin{figure}[t]
\begin{centering}
\medskip{}
\par\end{centering}
\begin{centering}
\includegraphics[clip,scale=0.48]{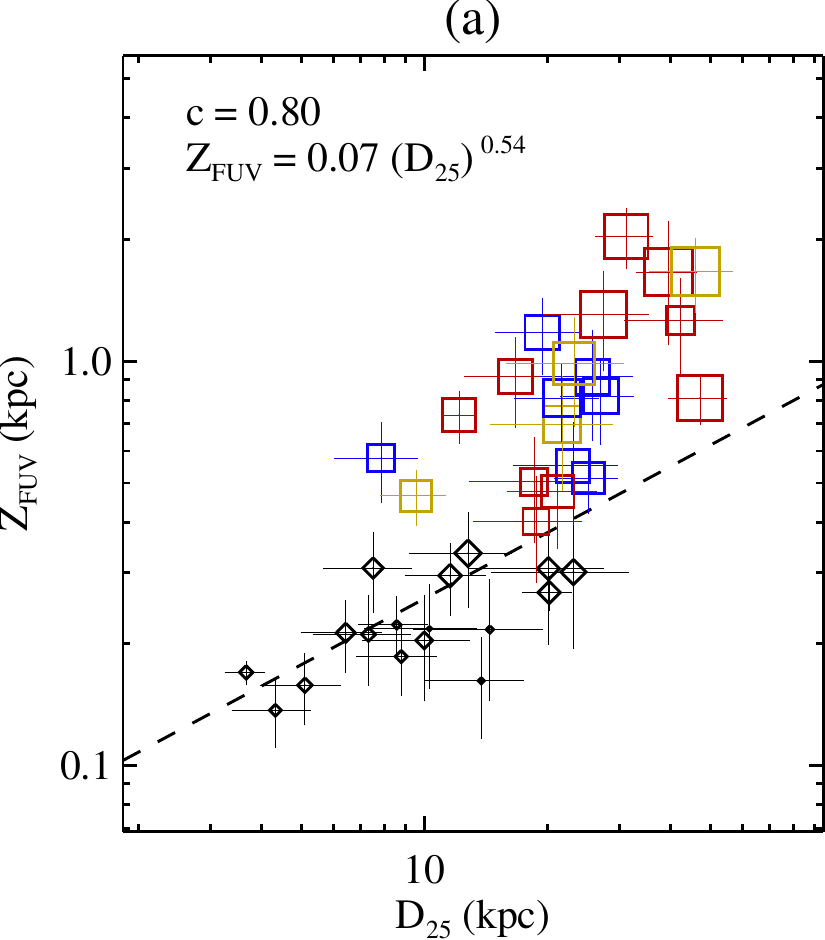}\ \ \includegraphics[clip,scale=0.48]{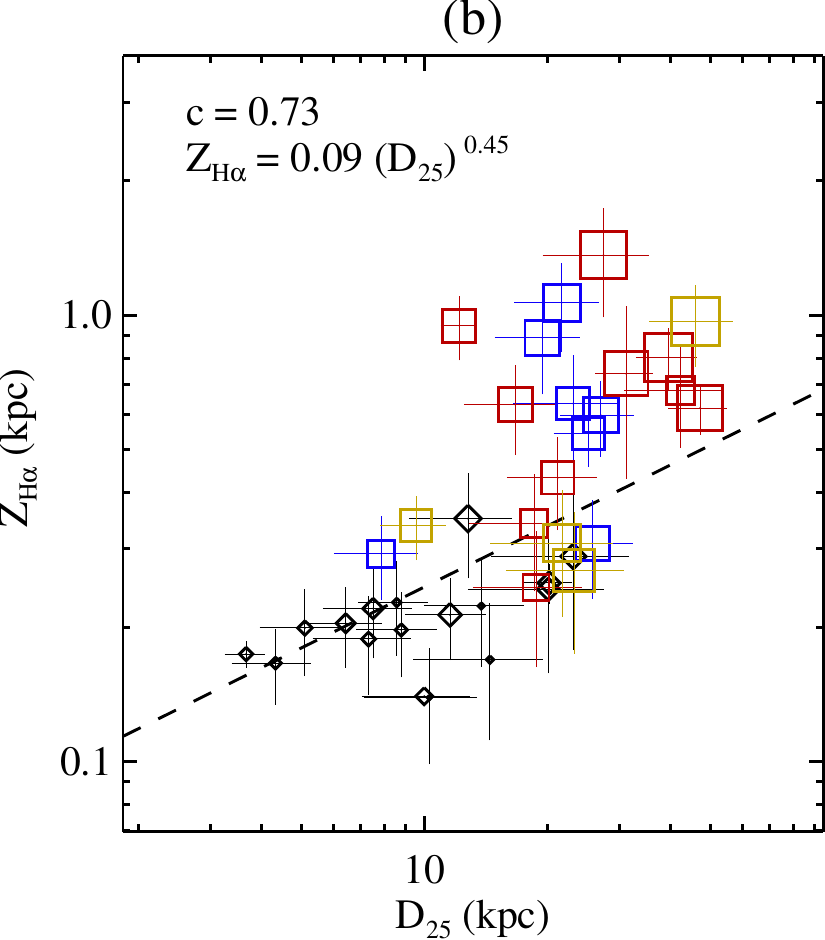}
\par\end{centering}
\begin{centering}
\medskip{}
\par\end{centering}
\caption{\label{fig4}Comparison of the scale heights of the FUV and H$\alpha$
emissions with the size of host galaxy (D$_{25}$). The size of the
symbol is proportional to the logarithmic scale of star formation
rates of the host galaxies (SFR$_{{\rm FIR}}$).}
\end{figure}

\begin{figure}[t]
\begin{centering}
\medskip{}
\par\end{centering}
\begin{centering}
\includegraphics[clip,scale=0.48]{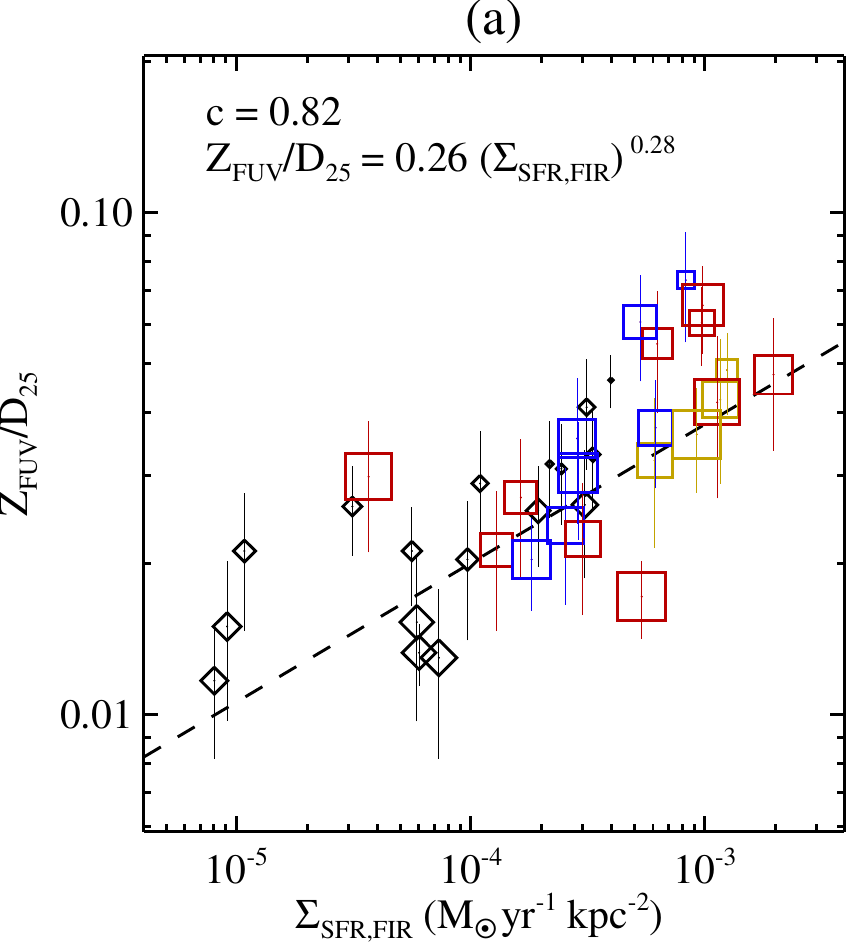}\ \ \includegraphics[clip,scale=0.48]{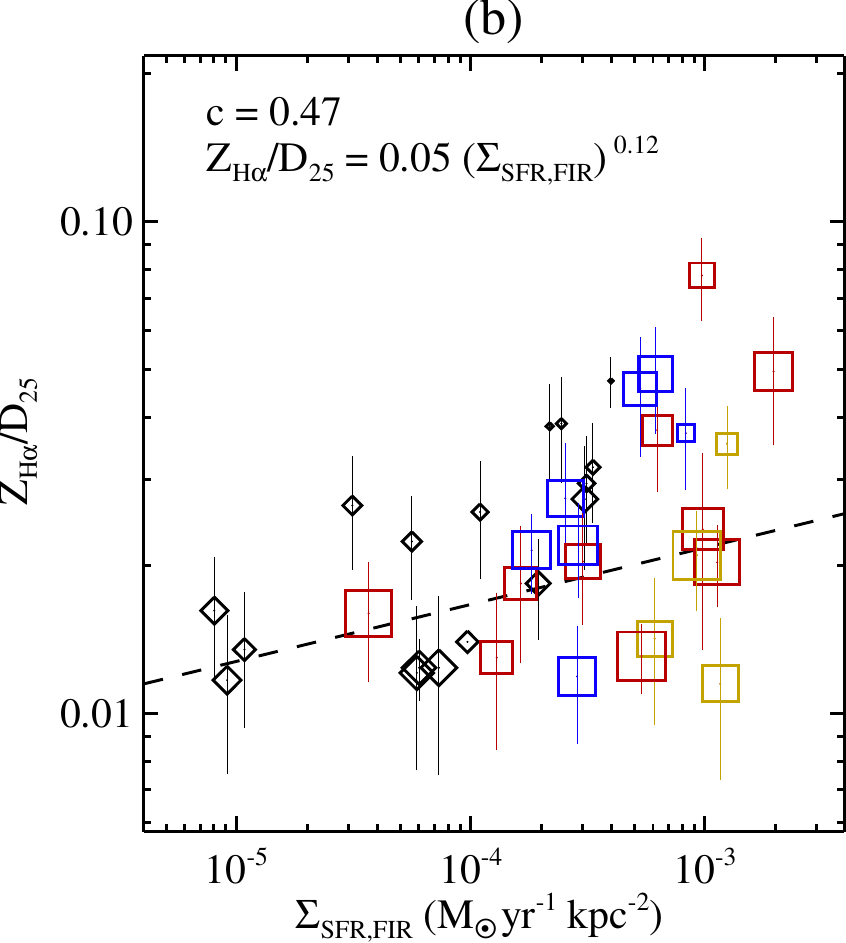}
\par\end{centering}
\begin{centering}
\medskip{}
\par\end{centering}
\begin{centering}
\includegraphics[clip,scale=0.48]{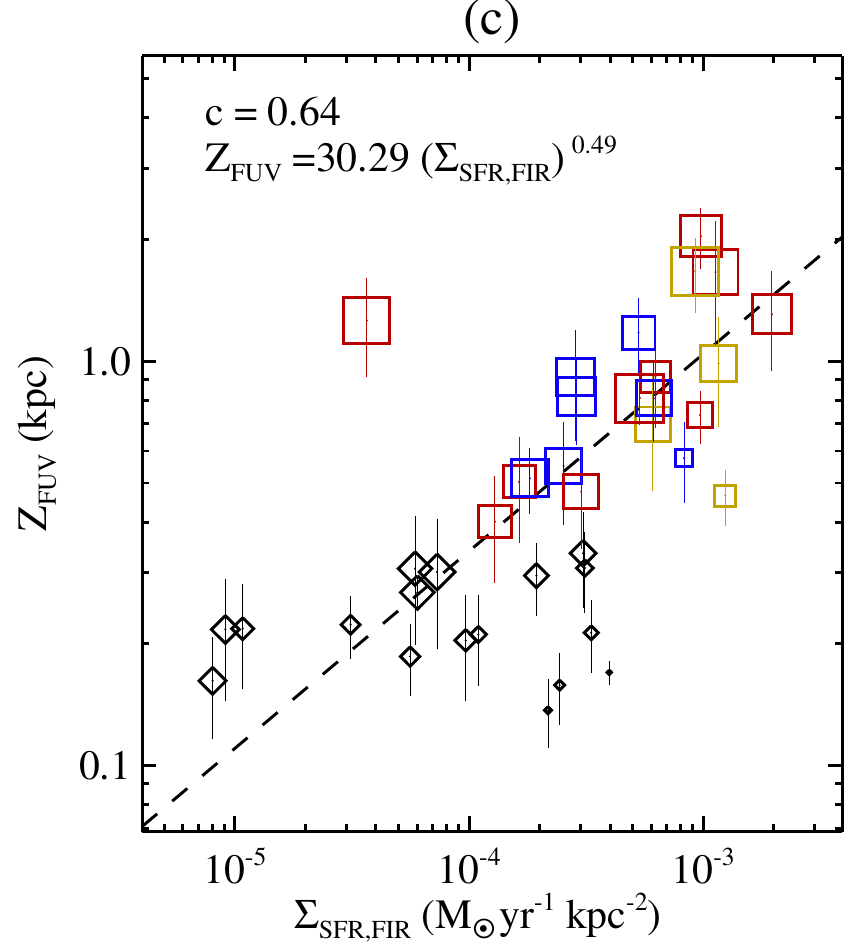}\ \ \includegraphics[clip,scale=0.48]{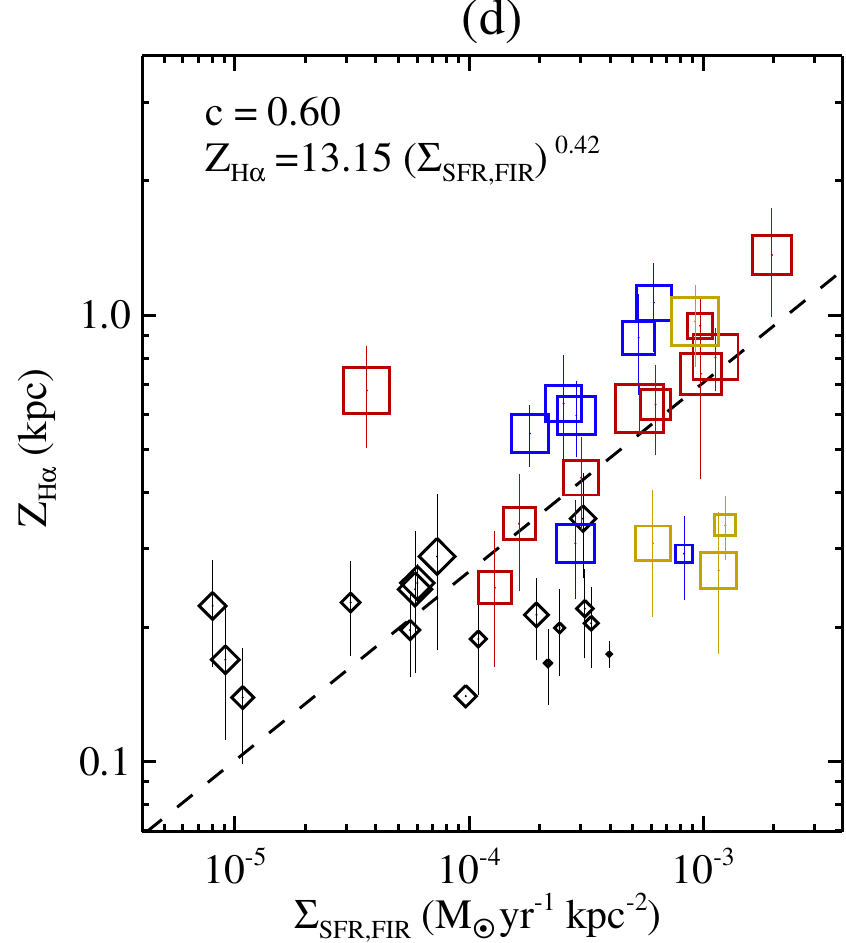}
\par\end{centering}
\begin{centering}
\medskip{}
\par\end{centering}
\caption{\label{fig5}Comparison of the normalized scale heights of the (a)
FUV and (b) H$\alpha$ emissions with star formation rate surface
densities ($\Sigma_{{\rm SFR,FIR}}$) of the host galaxies. Comparison
of the scale heights of the (c) FUV and (d) H$\alpha$ emissions with
$\Sigma_{{\rm SFR,FIR}}$. The size of the symbol is proportional
to the logarithmic scale of the size of the host galaxy (D$_{25}$).}
\end{figure}

\begin{figure}[t]
\begin{centering}
\medskip{}
\par\end{centering}
\begin{centering}
\includegraphics[clip,scale=0.48]{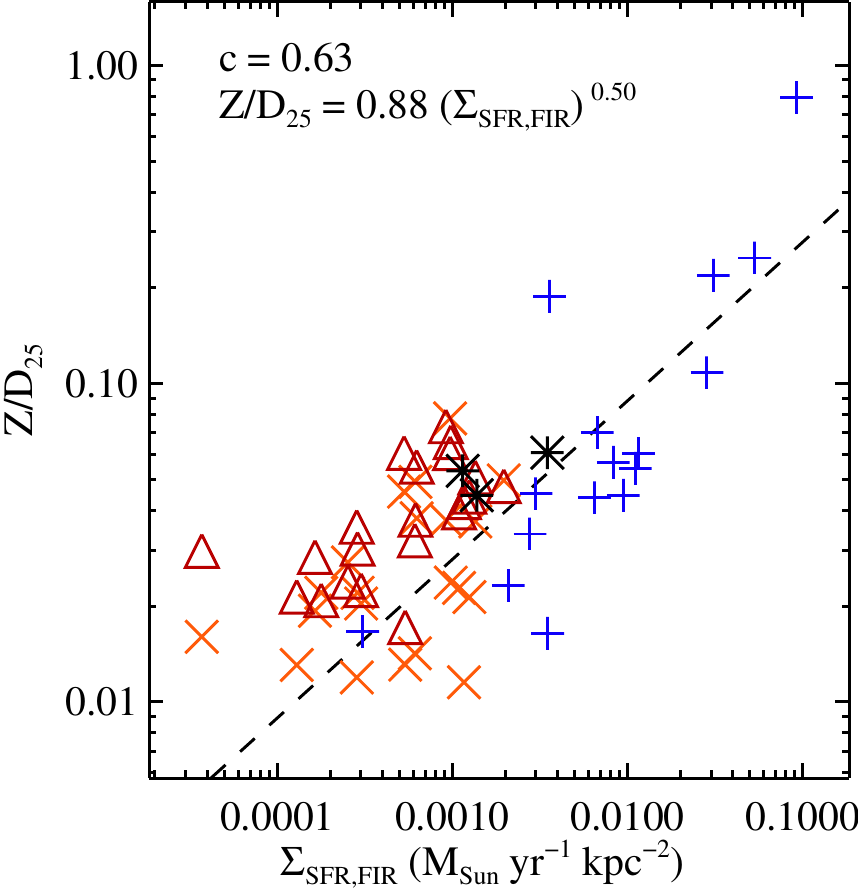}
\par\end{centering}
\begin{centering}
\medskip{}
\par\end{centering}
\caption{\label{fig6}Comparison of the normalized scale heights of the extraplanar
emissions with star formation rate surface densities ($\Sigma_{{\rm SFR,FIR}}$).
The blue pluses indicate 16 galaxies of \citet{2013ApJ...774..126M}
and the black asterisks denote three galaxies of \citet{2015ApJ...815..133S}.
The red triangles and orange crosses indicate the results obtained
from the extraplanar FUV and H\textgreek{a} emissions, respectively,
of 21 galaxies (Group B galaxies) in this study.}
 
\end{figure}

We now discuss the relationship between the FUV and H$\alpha$ scale
heights. If the total H$\alpha$ intensity in the galactic halo originates
from photoionized gas and the eDIG is uniformly mixed with the eDust,
which is exponentially distributed with a scale height of $Z_{{\rm eDust}}$,
then the scale height measured at H$\alpha$ will be given by ${\rm Z}_{{\rm H}\alpha}=0.5\ Z_{{\rm eDust}}$,
which is denoted by blue triple-dot-dashed lines in Figures \ref{fig2}(c)
and (d). This is because the emission measure is proportional to the
square of electron density. Therefore, the relation
between the two scale heights will provide a useful constraint in
understanding the properties of the extraplanar ISM. In this paper,
we found that $Z_{{\rm H}\alpha}\sim0.74\ Z_{{\rm FUV}}$, which appears
to be inconsistent with that expected from photoionized gas. In
the analyses of \citet{2014ApJ...785L..18S} and \citet{2015ApJ...815..133S},
we found that the scale height of dust-scattered light ($Z_{{\rm FUV}}$
or $Z_{{\rm NUV}}$) is similar to the intrinsic scale height of eDust
($Z_{{\rm eDust}}$), but not always the same as the intrinsic value.
The relation between the scale height of scattered light and the intrinsic
scale height of eDust is not clear at this moment. Therefore, the
relation $Z_{{\rm H}\alpha}\sim0.74\ Z_{{\rm FUV}}$ does not necessarily
indicate that the H$\alpha$ scale height is inconsistent with that
expected from photoionized gas.

It is necessary to investigate radiative transfer models to better
explain the present observations. The radiative transfer models could
also provide the amount of dust expelled by the SF activities for
our galaxies, as in \citet{2014ApJ...785L..18S} and \citet{2015ApJ...815..133S}.

\acknowledgements{This research was supported by the Korea Astronomy and Space Science
Institute under the R\&D program supervised by the Ministry of Science,
ICT, and Future Planning of Korea. This research was also supported
by the BK 21 plus program and Basic Science Research Program (2017R1D1A1B03031842)
through the National Research Foundation (NRF) funded by the Ministry
of Education of Korea. K.-I. Seon was supported by the National Research
Foundation of Korea (NRF) grant funded by the Korea government (MSIP)
(No. 2017R1A2B4008291). K.-I. Seon thanks Hyunjin Jung and Changhee
Rhee for helpful discussion on the star-formation rates and spectral
energy distribution of spiral galaxies.}

\pagebreak{}

\begin{table*}[tp]
\caption{\label{table1}Galaxy samples.}
\begin{centering}
\begin{tabular}{rllr@{\extracolsep{0pt}.}lr@{\extracolsep{0pt}.}lr@{\extracolsep{0pt}.}lr@{\extracolsep{0pt}.}lr@{\extracolsep{0pt}.}lr@{\extracolsep{0pt}.}lr@{\extracolsep{0pt}.}lr@{\extracolsep{0pt}.}ll}
\toprule 
\multirow{2}{*}{No.} & \multirow{2}{*}{Name} & \multirow{2}{*}{Morphology} & Gal& Lon. & Gal& Lat. & \multicolumn{2}{c}{Distance} & \multicolumn{2}{c}{D$_{{\rm major}}$} & \multicolumn{2}{c}{D$_{25}$} & \multicolumn{2}{c}{L$_{{\rm H}\alpha}$} & \multicolumn{2}{c}{SFR$_{{\rm H}\alpha}$} & \multicolumn{2}{c}{SFR$_{{\rm FIR}}$} & ref\tabularnewline
 &  &  & \multicolumn{2}{c}{(degree)} & \multicolumn{2}{c}{(degree)} & \multicolumn{2}{c}{(Mpc)} & \multicolumn{2}{c}{(arcmin)} & \multicolumn{2}{c}{(kpc)} & \multicolumn{2}{c}{($10^{40}$ erg s$^{-1}$)} & \multicolumn{2}{c}{(M$_{\odot}$ yr$^{-1}$)} & \multicolumn{2}{c}{(M$_{\odot}$ yr$^{-1}$)} & \tabularnewline
\midrule 
1 & UGCA 442 & SB(s)m? & 10&70 & -74&53 & 5&55 & 6&38 & 10&3 & 0&147 & 0&0117 & 0&0009 & SINGG\tabularnewline
2 & NGC 5951 & SBc? & 23&52 & 50&45 & 26&58 & 3&50 & 27&1 & 3&616 & 0&2870 & 0&1652 & H$\alpha$3\tabularnewline
3 & NGC 5107 & SB(s)d? & 96&01 & 76&98 & 18&54 & 1&70 & 9&2 & \multicolumn{2}{c}{} & \multicolumn{2}{c}{} & 0&0892 & H$\alpha$GS\tabularnewline
4 & NGC 5229 & SB(s)d? & 103&95 & 67&61 & 9&32 & 3&58 & 9&7 & 0&308 & 0&0244 & 0&0076 & LVL\tabularnewline
5 & UGC 08313 & SB(s)c? & 107&46 & 74&24 & 9&32 & 1&91 & 5&2 & 0&377 & 0&0299 & 0&0050 & LVL\tabularnewline
6 & NGC 5023 & Scd? & 110&38 & 72&58 & 9&36 & 7&28 & 19&8 & 0&798 & 0&0633 & 0&0204 & LVL\tabularnewline
7 & NGC 0493 & SAB(s)cd? & 138&91 & -60&97 & 21&94 & 3&40 & 21&7 & \multicolumn{2}{c}{} & \multicolumn{2}{c}{} & 0&2275 & H$\alpha$GS\tabularnewline
8 & NGC 0891 & SA(s)b? & 140&38 & -17&41 & 9&59 & 13&50 & 37&7 & 5&866 & 0&4655 & 1&3909 & Rossa\tabularnewline
9 & NGC 0784 & SBdm? & 140&90 & -31&59 & 4&21 & 6&60 & 8&1 & 0&408 & 0&0323 & 0&0034 & LVL\tabularnewline
10 & NGC 4631 & SB(s)d & 142&81 & 84&22 & 5&16 & 15&50 & 23&3 & 11&414 & 0&9059 & 0&4946 & SINGS\tabularnewline
11 & NGC 4144 & SAB(s)cd? & 143&17 & 69&01 & 6&14 & 6&00 & 10&7 & 0&962 & 0&0764 & 0&0193 & LVL\tabularnewline
12 & NGC 0803 & SA(s)c? & 147&17 & -43&41 & 22&29 & 3&00 & 19&5 & \multicolumn{2}{c}{} & \multicolumn{2}{c}{} & 0&1573 & H$\alpha$GS\tabularnewline
13 & NGC 4244 & SA(s)cd? & 154&57 & 77&16 & 4&11 & 19&38 & 23&2 & 1&023 & 0&0812 & 0&0308 & LVL\tabularnewline
14 & IC 2233 & SB(s)d? & 174&12 & 33&06 & 12&27 & 5&17 & 18&5 & 1&723 & 0&1368 & 0&0187 & LVL\tabularnewline
15 & NGC 3432 & SB(s)m & 184&77 & 63&16 & 10&98 & 6&80 & 21&7 & 6&732 & 0&5343 & 0&2260 & LVL\tabularnewline
16 & NGC 4020 & SBd? & 193&90 & 78&05 & 12&04 & 2&24 & 7&8 & 1&124 & 0&0892 & 0&0444 & LVL\tabularnewline
17 & NGC 3510 & SB(s)m & 202&36 & 66&21 & 13&95 & 4&35 & 17&7 & 2&380 & 0&1889 & 0&0401 & LVL\tabularnewline
18 & NGC 3190 & SA(s)a pec & 213&04 & 54&85 & 24&35 & 4&40 & 31&2 & \multicolumn{2}{c}{} & \multicolumn{2}{c}{} & 0&7462 & SINGS\tabularnewline
19 & NGC 3628 & Sb pec &  240&85 & 64&78 & 9&85 & 14&80 &  42&4 &  4&130 &  0&3278 &  1&5407 & H$\alpha$3\tabularnewline
20 & UGCA 193 & Sd? & 245&64 & 37&43 & 11&00 & 4&31 & 13&8 & 0&270 & 0&0215 & 0&0012 & SINGG\tabularnewline
21 & NGC 3365 & Scd? & 247&75 & 50&76 & 17&53 & 4&84 & 24&7 & 2&122 & 0&1684 & 0&0849 & SINGG\tabularnewline
22 & NGC 4455 & SB(s)d? & 251&64 & 83&29 & 9&13 & 2&80 & 7&4 & 0&888  & 0&0705  & 0&0138 & LVL\tabularnewline
23 & ESO 249- G 035 & SBcd? & 252&61 & -48&67 & 22&49 & 1&31 & 8&6 & 0&175 & 0&0139 & 0&0018 & SINGG\tabularnewline
24 & IC 2000 & SB(s)cd? & 257&65 & -49&60 & 19&39 & 4&10 & 23&1 & 2&980 & 0&2365 & 0&1064 & SINGG\tabularnewline
25 & IC 1959 & SB(s)m? & 261&28 & -51&54 & 7&90 & 2&80 & 6&4 & 0&667 & 0&0530 & 0&0108 & SINGG\tabularnewline
26 & NGC 1311 & SB(s)m? & 265&29 & -52&66 & 4&96 & 3&00 & 4&3 & 0&214 & 0&0170 & 0&0032 & SINGG\tabularnewline
27 & NGC 4313 & SA(rs)ab? & 277&74 & 73&25 & 14&62 & 4&99 & 21&2 & 0&811 & 0&0644 & 0&1065 & H$\alpha$3\tabularnewline
28 & NGC 4388 & SA(s)b? & 279&12 & 74&34 & 19&50 & 4&84 & 27&5 & \multicolumn{2}{c}{} & \multicolumn{2}{c}{} & 1&1619 & Rossa\tabularnewline
29 & NGC 4469 & SB0/a?(s) & 286&13 & 70&90 & 16&75 & 2&50 & 12&2 & \multicolumn{2}{c}{} & \multicolumn{2}{c}{} & 0&1130 & H$\alpha$GS\tabularnewline
30 & NGC 4866 & SA0\textasciicircum{}+(r)? & 311&54 & 76&91 & 23&09 & 6&30 & 42&3 & 2&792 & 0&2216 & 0&0515 & H$\alpha$3\tabularnewline
31 & IC 5176 & SAB(s)bc? & 323&00 & -43&69 & 26&86 & 6&05 & 47&3 & 4&240 & 0&3365 & 0&9423 & SINGG\tabularnewline
32 & IC 5052 & SBd? & 325&18 & -35&81 & 7&46 & 5&90 & 12&8 & 2&263 & 0&1796  & 0&0395 & SINGG\tabularnewline
33 & IC 4951 & SB(s)dm? & 334&89 & -32&85 & 8&97 & 2&80 & 7&3 & 0&278 & 0&0221 & 0&0046 & SINGG\tabularnewline
34 & NGC 5348 & SBbc? & 340&03 & 63&49 & 18&44 & 3&50 & 18&8 & 1&481 & 0&1176 & 0&0357 & H$\alpha$3\tabularnewline
35 & NGC 5356 & SABbc? & 340&53 & 63&47 & 23&89 & 3&71 & 25&8 & 1&930 & 0&1532 & 0&1485 & H$\alpha$3\tabularnewline
36 & NGC 7090 & SBc? & 341&30 & -45&39 & 7&76 & 7&40 & 16&7 & 2&831 & 0&2247 & 0&1370 & LVL\tabularnewline
37 & NGC 7412A & SBdm? & 351&39 & -62&04 & 9&74 & 5&11 & 14&5 & 0&212 & 0&0168 & 0&0025  & SINGG\tabularnewline
38 & ESO 347- G 017 & SB(s)m? & 357&78 & -69&49 & 7&89 & 1&60 & 3&7 & 0&175 & 0&0139 & 0&0042 & SINGG\tabularnewline
\bottomrule
\end{tabular}
\par\end{centering}

Column (1): the running index number. Column (2): galaxy name taken
from the NED\textquoteright s preferred object name. Column (3): galaxy
morphology. Column (4): Galactic longitude. Column (5): Galactic latitude.
Column (6): average value of the redshift-independent distances. Column
(7): major axis diameter in arcmin. Column (8): major axis diameter
in kpc calculated using Column (6) and (7): Column (9): integrated
H$\alpha$ luminosity taken from the reference papers. Column (10):
star formation rate based on the H$\alpha$ luminosity. Column (11):
star formation rate estimated using the FIR luminosity.
\centering{}\medskip{}
\end{table*}

\begin{table*}[tp]
\caption{\label{table2}Scale heights of the sample galaxies.}
\begin{centering}
\begin{tabular}{rlcr@{\extracolsep{0pt}.}lr@{\extracolsep{0pt}.}lr@{\extracolsep{0pt}.}lr@{\extracolsep{0pt}.}lr@{\extracolsep{0pt}.}lr@{\extracolsep{0pt}.}lr@{\extracolsep{0pt}.}l}
\hline 
\multirow{2}{*}{No.} & \multirow{2}{*}{Name} & \multirow{2}{*}{Group} & \multicolumn{2}{c}{SFR$_{{\rm FIR}}$} & \multicolumn{12}{c}{scale height (kpc)}\tabularnewline
\cline{6-17} 
 &  &  & \multicolumn{2}{c}{(M$_{\odot}$ yr$^{-1}$)} & \multicolumn{2}{c}{H$\alpha$} & \multicolumn{2}{c}{H$\alpha$ err} & \multicolumn{2}{c}{FUV} & \multicolumn{2}{c}{FUV err} & \multicolumn{2}{c}{NUV} & \multicolumn{2}{c}{NUV err}\tabularnewline
\hline 
1 & UGCA 442 & A & 0&0009 & 0&138 & 0&005 & 0&216 & 0&007 & 0&318 & 0&029\tabularnewline
2 & NGC 5951 & B & 0&1652 & 0&595 & 0&043 & 0&818 & 0&134 & 0&770 &  0&141\tabularnewline
3 & NGC 5107 & B & 0&0892 & 0&319 & 0&018 & 0&442 & 0&013 & 0&538 & 0&029\tabularnewline
4 & NGC 5229 & A & 0&0076 & 0&136 & 0&293 & 0&198 & 0&018 & 0&293 & 0&048\tabularnewline
5 & UGC 08313 & A & 0&0050 & 0&194 & 0&017 & 0&160 & 0&006 &  0&201 & 0&014\tabularnewline
6 & NGC 5023 & A & 0&0204 & 0&238 & 0&012 & 0&302 & 0&012 & 0&340 &  0&027\tabularnewline
7 & NGC 0493 & B & 0&2275 & 1&068 & 0&077 & 0&811 & 0&050 & 0&801 & 0&095\tabularnewline
8 & NGC 0891 & B & 1&3909 & 0&821 & 0&052 & 1&778 & 0&462 & 1&340 & 0&497\tabularnewline
9 & NGC 0784 & A & 0&0034 & 0&180 & 0&016 & 0&171 & 0&005 & 0&203 & 0&008\tabularnewline
10 & NGC 4631 & B & 0&4946 & 0&266 & 0&045 & 0&979 & 0&018 & 0&655 & 0&025\tabularnewline
11 & NGC 4144 & A & 0&0193 & 0&197 & 0&011 & 0&272 & 0&010 & 0&308 & 0&024\tabularnewline
12 & NGC 0803 & B & 0&1573 & 0&890 & 0&121 & 1&178 & 0&047 & 1&229 & 0&117\tabularnewline
13 & NGC 4244 & A & 0&0308 & 0&288 & 0&040 & 0&302 & 0&008 & 0&254 & 0&012\tabularnewline
14 & IC 2233 & A & 0&0187 & 0&229 & 0&008 & 0&245 & 0&005 & 0&271 & 0&010\tabularnewline
15 & NGC 3432 & B & 0&2260 & 0&308 & 0&003 & 0&697 & 0&013 & 0&562 & 0&020\tabularnewline
16 & NGC 4020 & B & 0&0444 & 0&293 & 0&010 & 0&578 & 0&049 & 0&672 & 0&131\tabularnewline
17 & NGC 3510 & B & 0&0401 & 0&322 & 0&014 & 0&478 & 0&011 & 0&462 & 0&022\tabularnewline
18 & NGC 3190 & B & 0&7462 & 0&737 & 0&293 & 2&035 & 0&239 & 1&973 & 0&263\tabularnewline
19 & NGC 3628 & B &  1&5407 &  0&862 &  0&015 &  1&529 &  0&064 &  1&654 &  0&122\tabularnewline
20 & UGCA 193 & A & 0&0012 & 0&222 & 0&018 & 0&162 & 0&020 & 0&211 & 0&043\tabularnewline
21 & NGC 3365 & B & 0&0849 & 0&533 & 0&028 & 0&500 & 0&055 & 0&509 & 0&086\tabularnewline
22 & NGC 4455 & A & 0&0138 & 0&218 & 0&008 & 0&303 & 0&013 & 0&331 & 0&025\tabularnewline
23 & ESO 249- G 035 & A & 0&0018 & 0&225 & 0&039 & 0&220 & 0&014 & 0&308 & 0&032\tabularnewline
24 & IC 2000 & B & 0&1064 & 0&634 & 0&054 & 0&551 & 0&055 & 0&679 & 0&095\tabularnewline
25 & IC 1959 & A & 0&0108 & 0&203 & 0&009 & 0&212 & 0&007 & 0&249 & 0&018\tabularnewline
26 & NGC 1311 & A & 0&0032 & 0&166 & 0&002 & 0&136 & 0&004 & 0&167 & 0&007\tabularnewline
27 & NGC 4313 & B & 0&1065 & 0&431 & 0&027 & 0&473 & 0&073 &  0&808 & 0&117\tabularnewline
28 & NGC 4388 & B & 1&1619 & 1&361 & 0&052 & 1&307 & 0&072 & 1&633 & 0&126\tabularnewline
29 & NGC 4469 & B & 0&1130 & 0&944 & 0&155 & 0&734 & 0&109 & 0&881 & 0&152\tabularnewline
30 & NGC 4866 & B & 0&0515 & 0&677 & 0&028 & 1&260 & 0&137 & 1&490 & 0&313\tabularnewline
31 & IC 5176 & B & 0&9423 & 0&618 & 0&008 & 0&802 & 0&052 & 0&794 & 0&096\tabularnewline
32 & IC 5052 & A & 0&0395 & 0&347 & 0&008 & 0&333 & 0&016 & 0&360 & 0&022\tabularnewline
33 & IC 4951 & A & 0&0046 & 0&188 & 0&005 & 0&211 & 0&008 & 0&241 & 0&014\tabularnewline
34 & NGC 5348 & B & 0&0357 & 0&245  & 0&044 & 0&402 & 0&034 & 0&473 & 0&084\tabularnewline
35 & NGC 5356 & B & 0&1485 & 0&307 & 0&027 & 0&915 & 0&177 & 1&341 & 0&449\tabularnewline
36 & NGC 7090 & B & 0&1370 & 0&627 & 0&006 & 0&911 & 0&097 & 0&727 & 0&114\tabularnewline
37 & NGC 7412A & A & 0&0025  & 0&169 & 0&010 & 0&216 & 0&007 & 0&273 & 0&030\tabularnewline
38 & ESO 347- G 017 & A & 0&0042 & 0&175 & 0&007 & 0&170 & 0&007 & 0&230 & 0&020\tabularnewline
\hline 
\end{tabular}
\par\end{centering}
\centering{}\medskip{}
\end{table*}

\pagebreak{}

\begin{continuedfigure*}[tp]
\begin{centering}
\medskip{}
\par\end{centering}
\begin{centering}
\includegraphics[clip,scale=0.7]{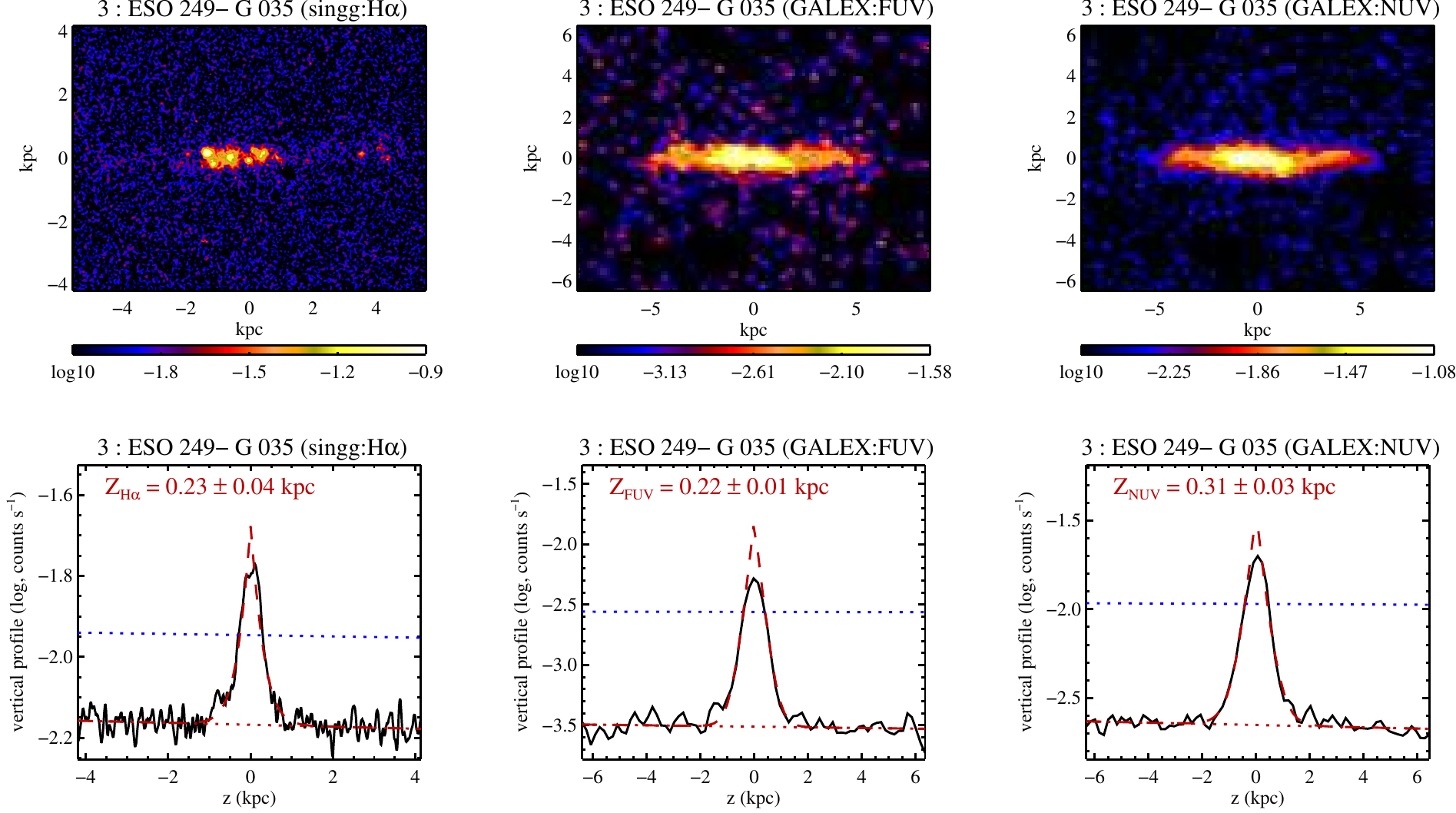}
\par\end{centering}
\begin{centering}
\includegraphics[clip,scale=0.7]{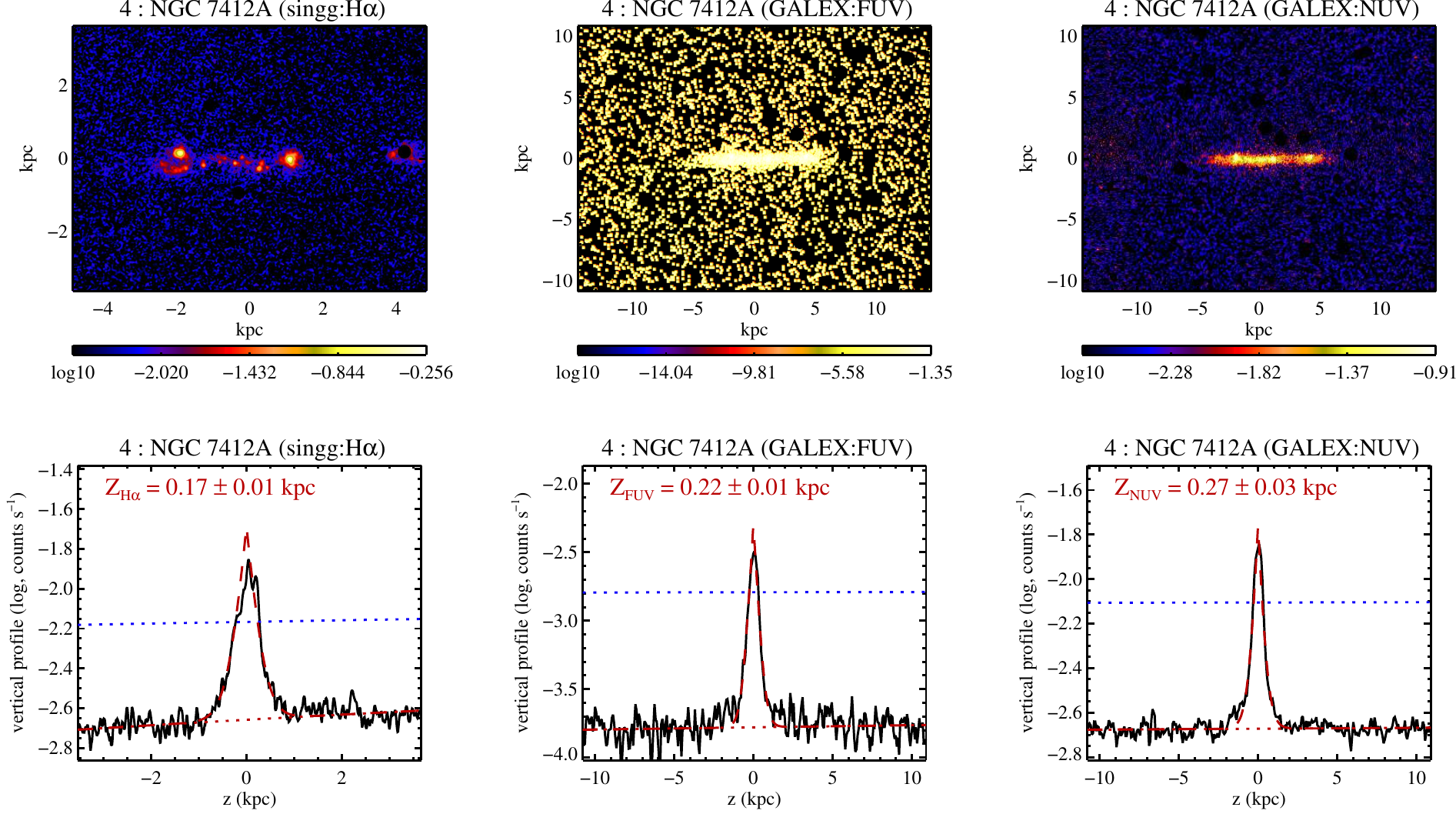}
\par\end{centering}
\begin{centering}
\includegraphics[clip,scale=0.7]{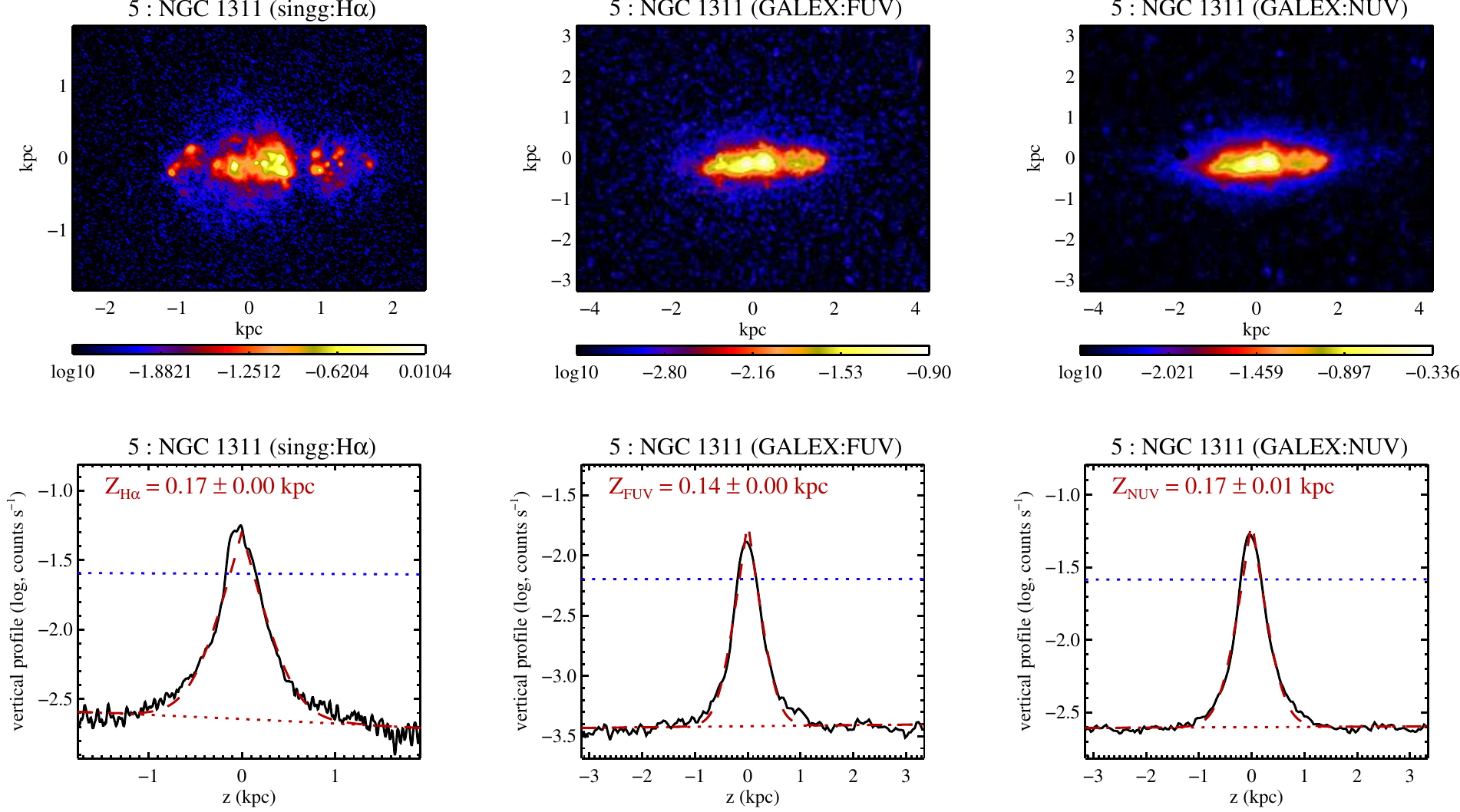}
\par\end{centering}
\begin{centering}
\medskip{}
\par\end{centering}
\caption{Continued.}
\end{continuedfigure*}

\begin{continuedfigure*}[tp]
\begin{centering}
\medskip{}
\par\end{centering}
\begin{centering}
\includegraphics[clip,scale=0.7]{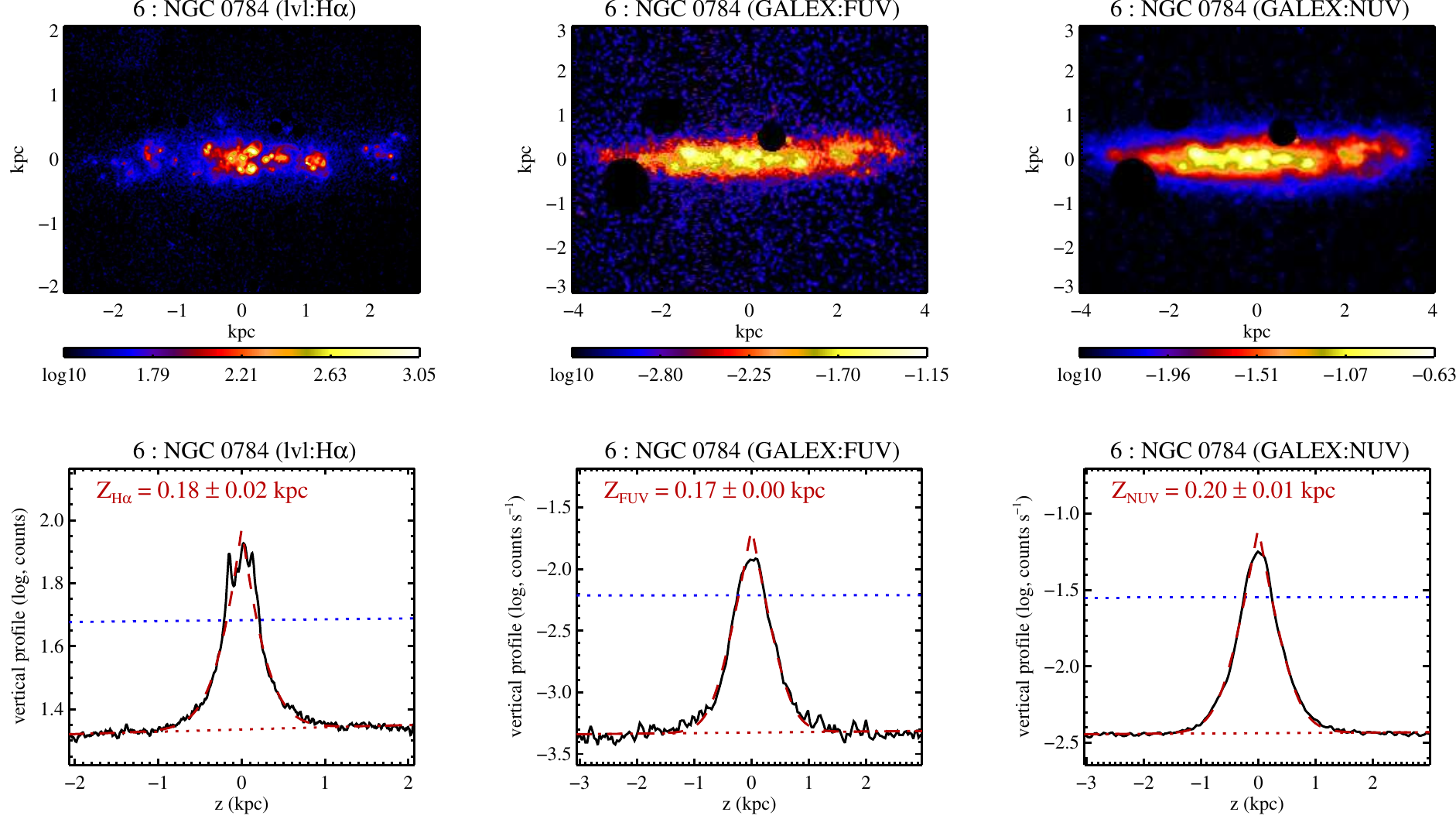}
\par\end{centering}
\begin{centering}
\includegraphics[clip,scale=0.7]{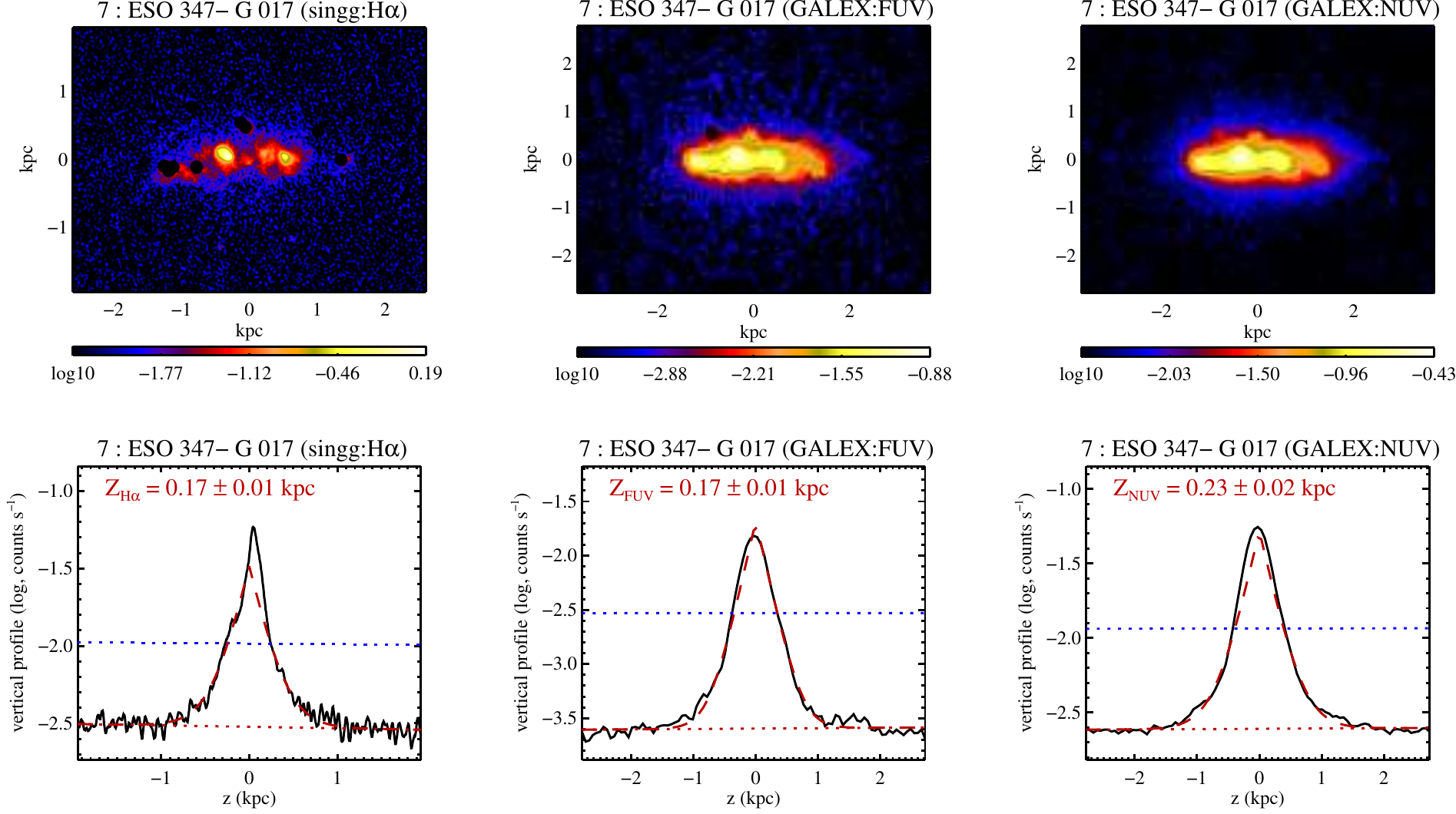}
\par\end{centering}
\begin{centering}
\includegraphics[clip,scale=0.7]{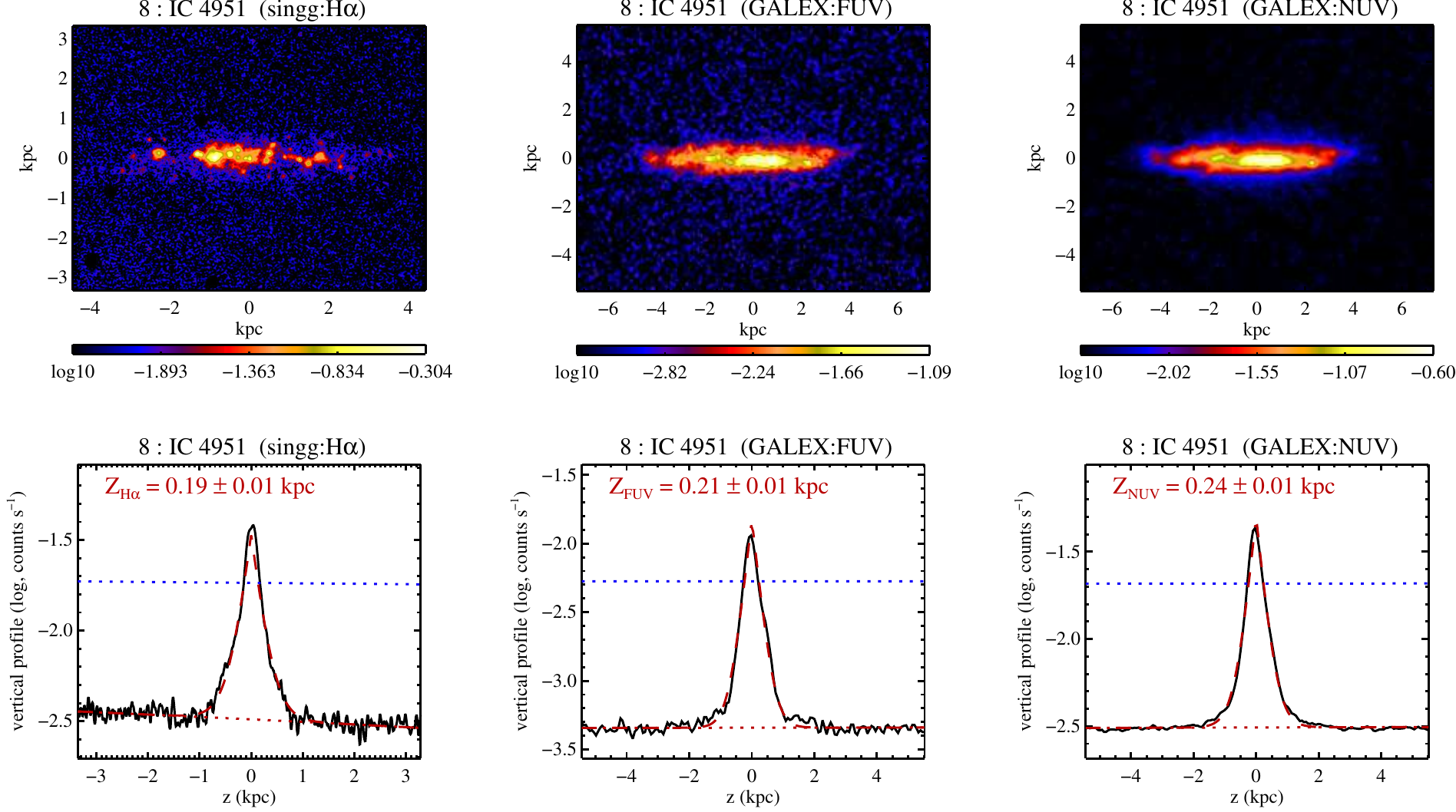}
\par\end{centering}
\begin{centering}
\medskip{}
\par\end{centering}
\caption{Continued.}
\end{continuedfigure*}

\begin{continuedfigure*}[tp]
\begin{centering}
\medskip{}
\par\end{centering}
\begin{centering}
\includegraphics[clip,scale=0.7]{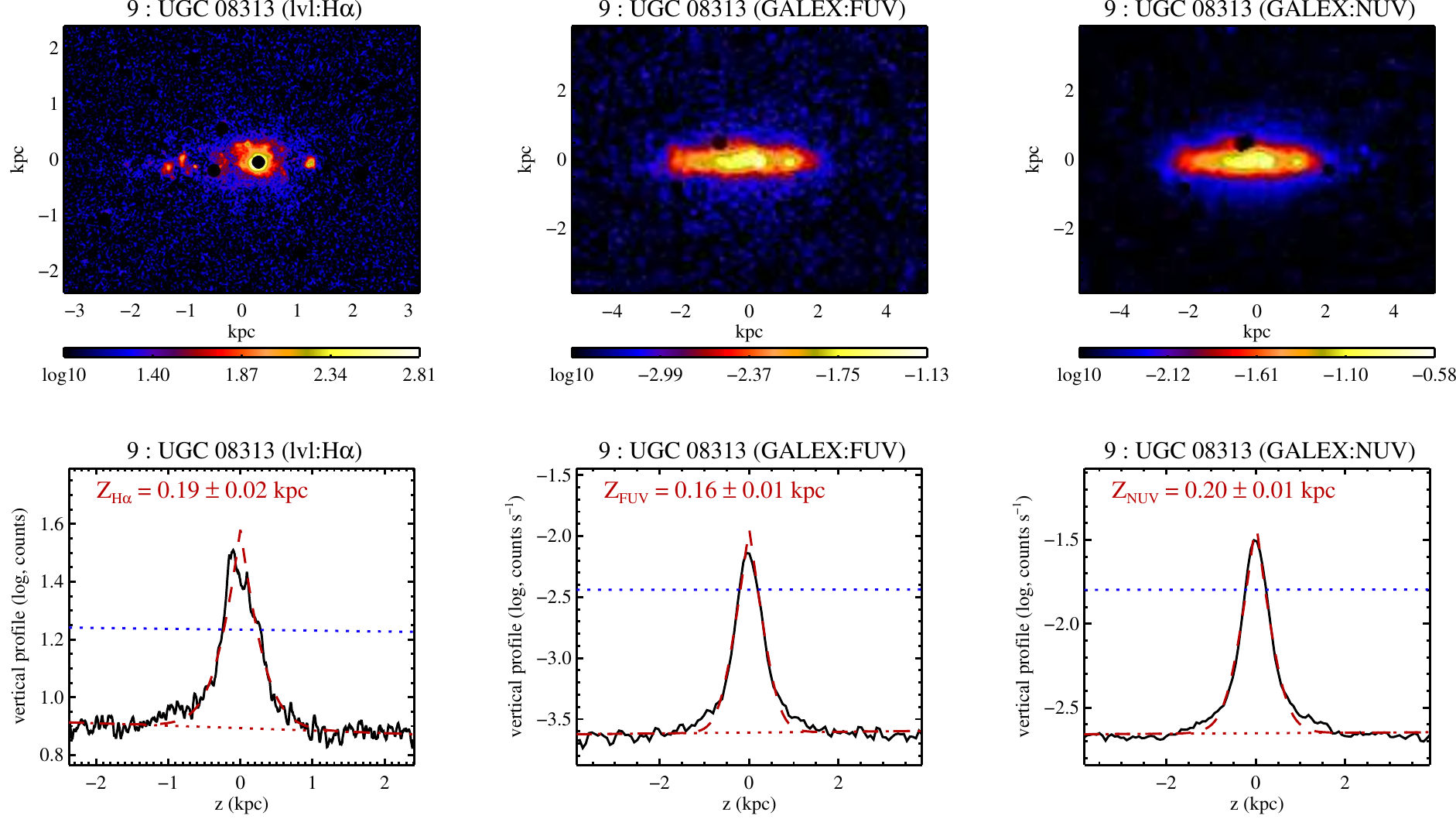}
\par\end{centering}
\begin{centering}
\includegraphics[clip,scale=0.7]{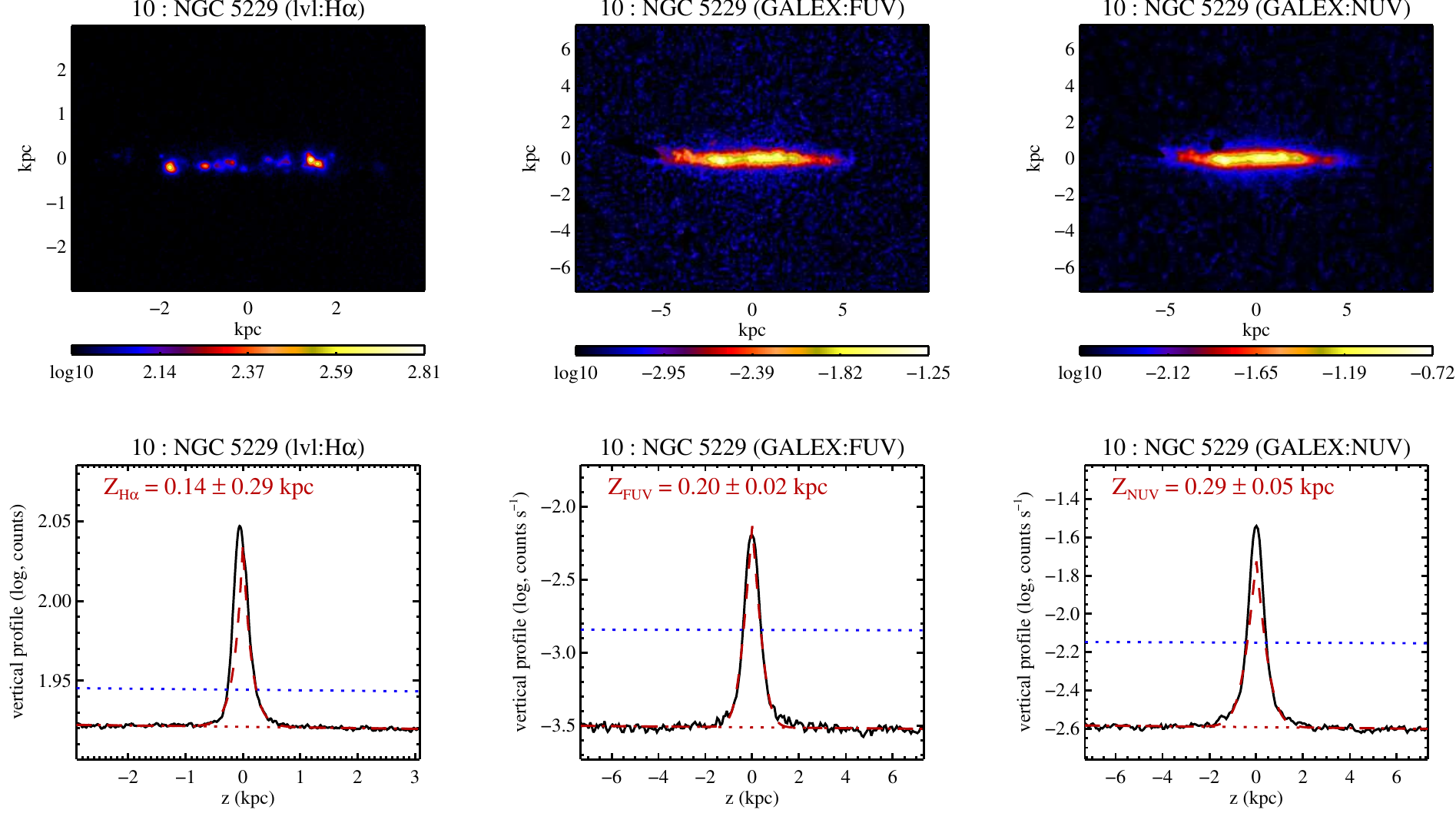}
\par\end{centering}
\begin{centering}
\includegraphics[clip,scale=0.7]{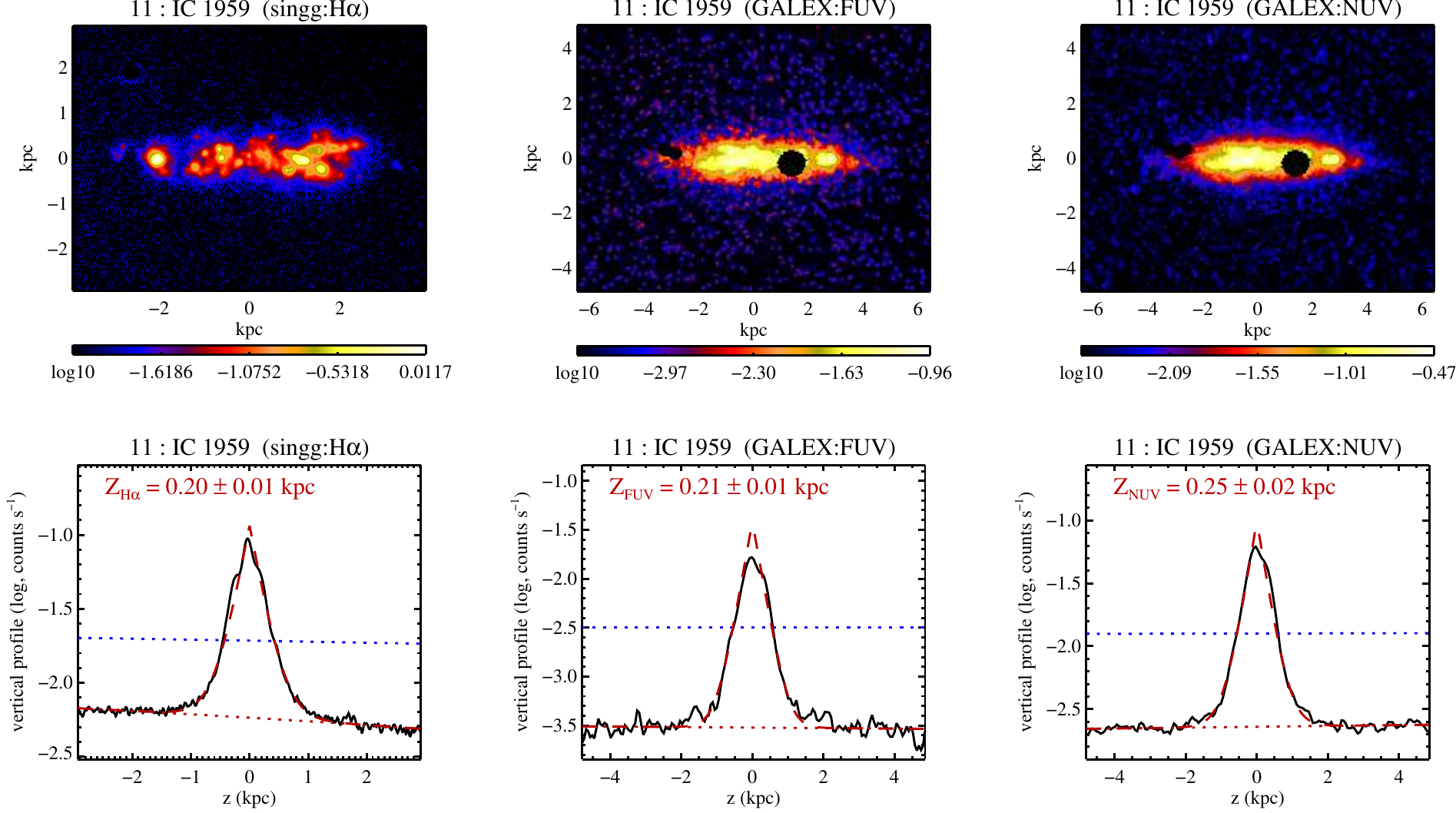}
\par\end{centering}
\begin{centering}
\medskip{}
\par\end{centering}
\caption{Continued.}
\end{continuedfigure*}

\begin{continuedfigure*}[tp]
\begin{centering}
\medskip{}
\par\end{centering}
\begin{centering}
\includegraphics[clip,scale=0.7]{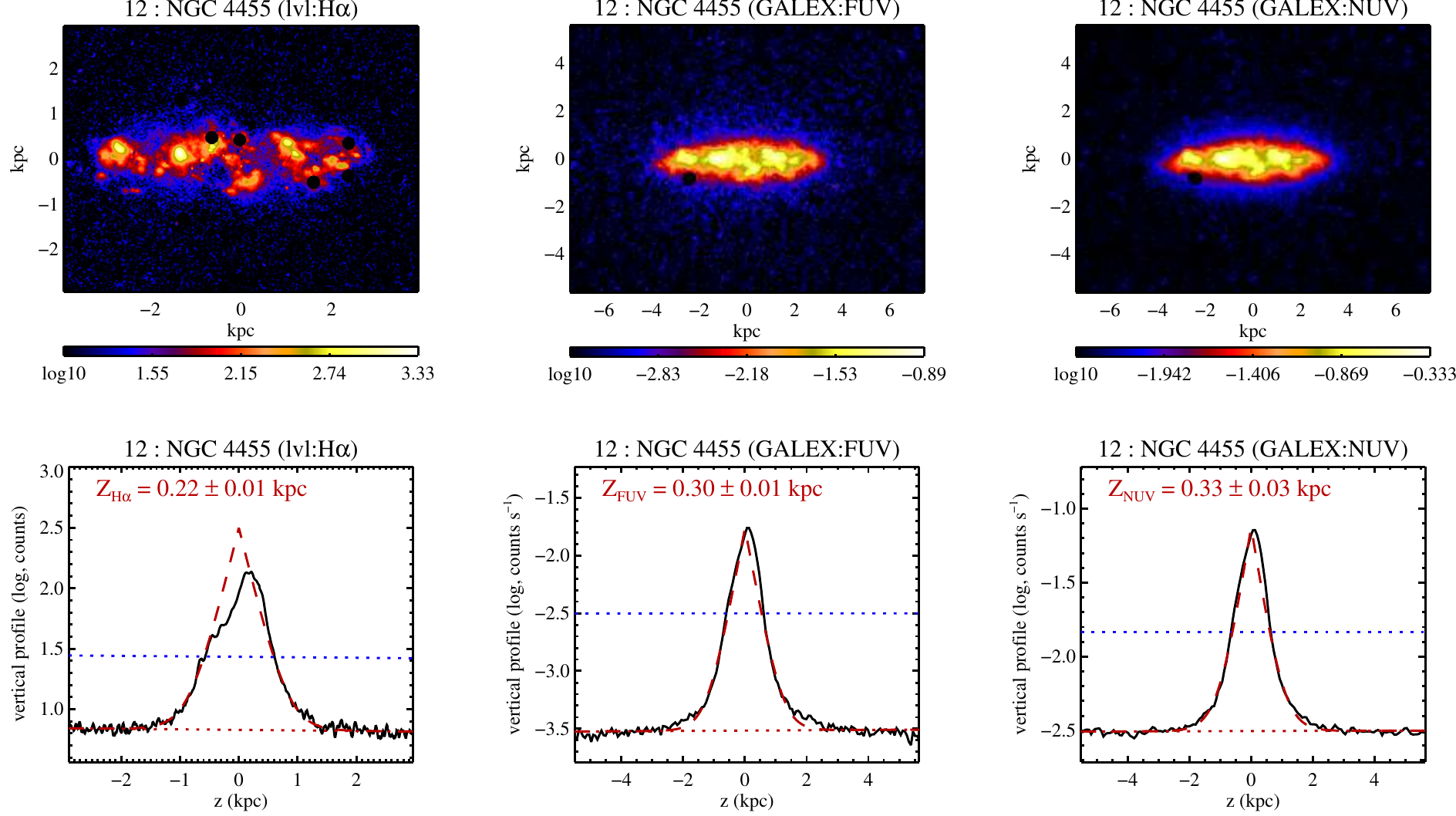}
\par\end{centering}
\begin{centering}
\includegraphics[clip,scale=0.7]{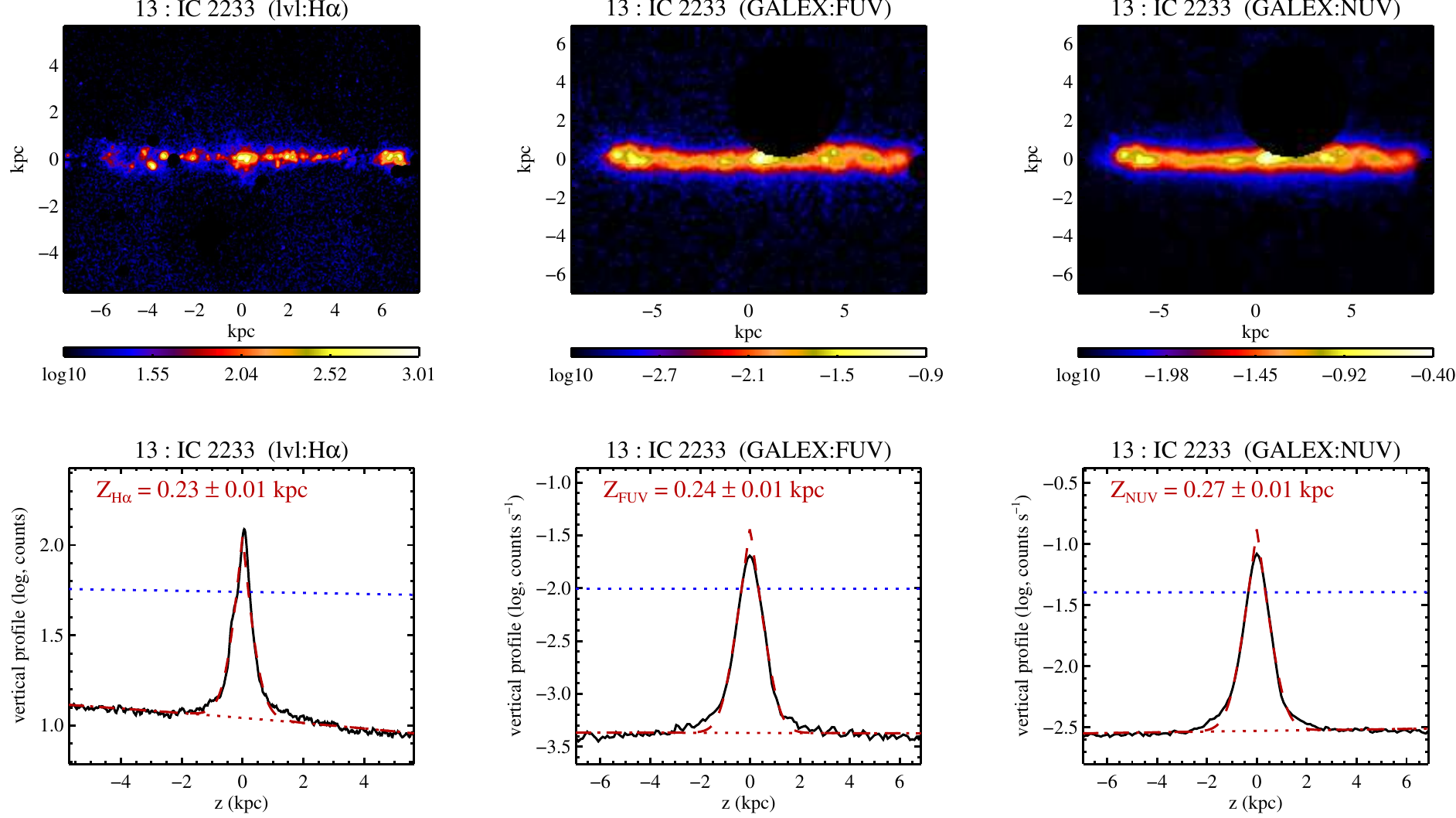}
\par\end{centering}
\begin{centering}
\includegraphics[clip,scale=0.7]{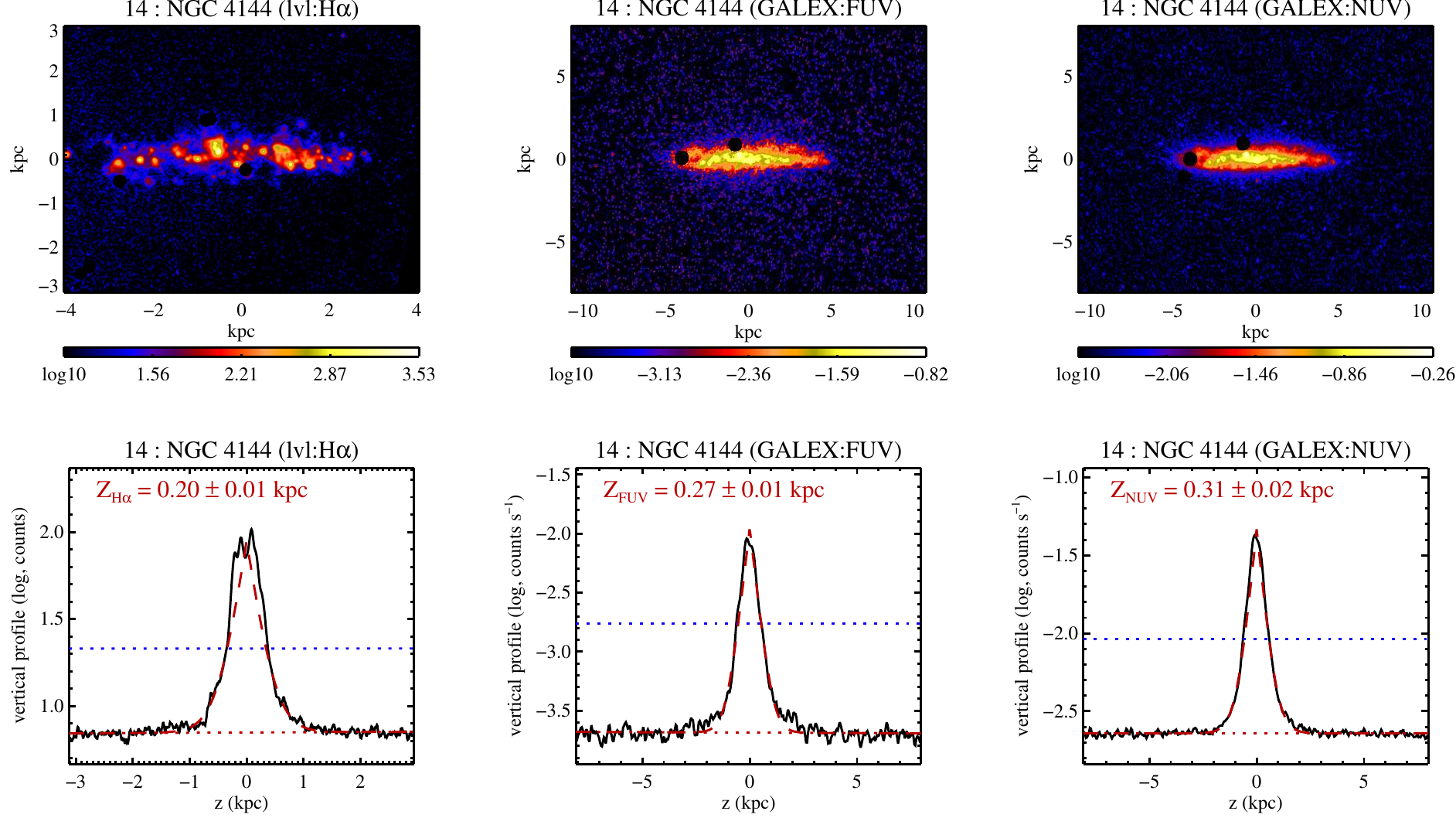}
\par\end{centering}
\begin{centering}
\medskip{}
\par\end{centering}
\caption{Continued.}
\end{continuedfigure*}

\begin{continuedfigure*}[tp]
\begin{centering}
\medskip{}
\par\end{centering}
\begin{centering}
\includegraphics[clip,scale=0.7]{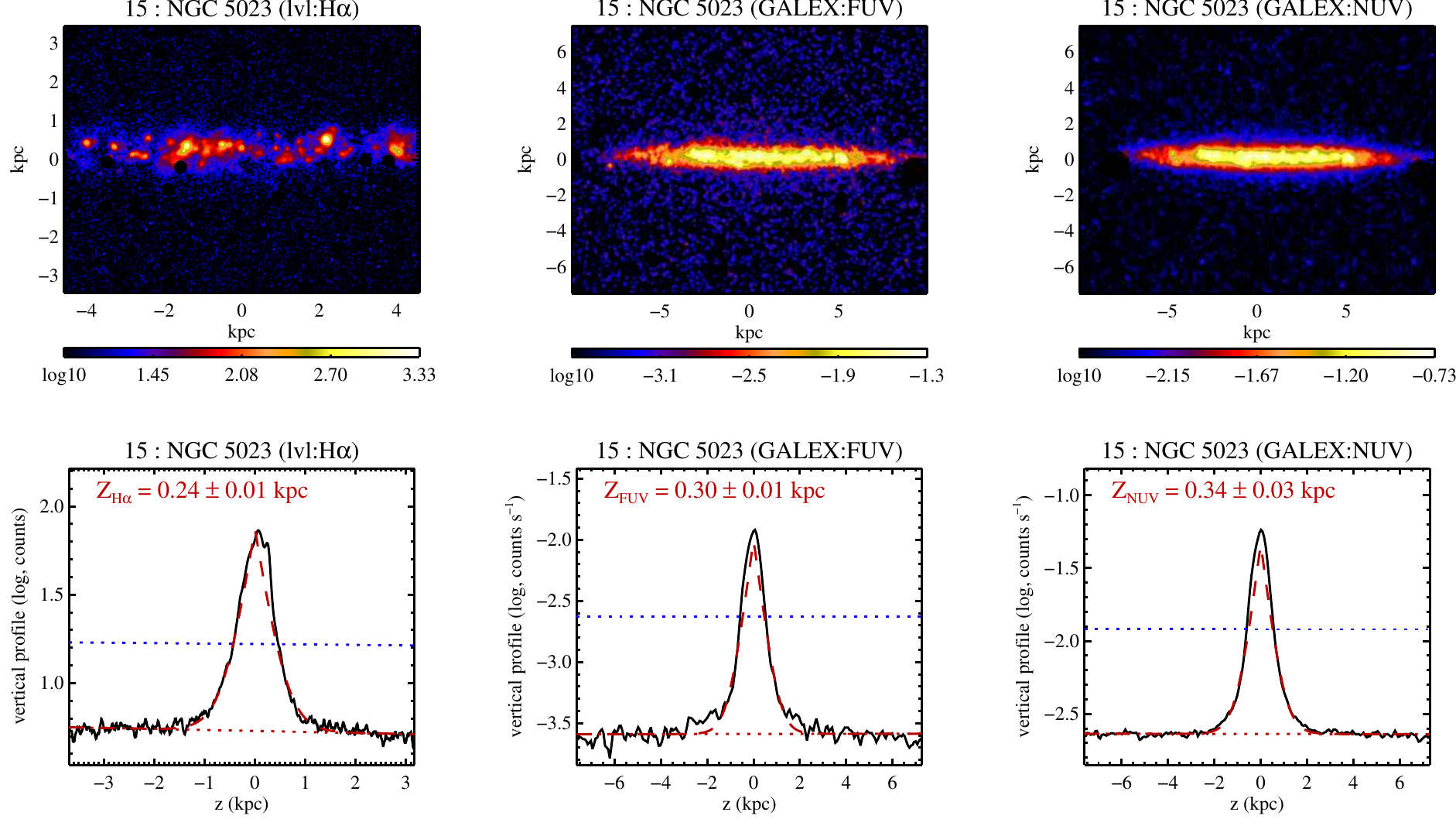}
\par\end{centering}
\begin{centering}
\includegraphics[clip,scale=0.7]{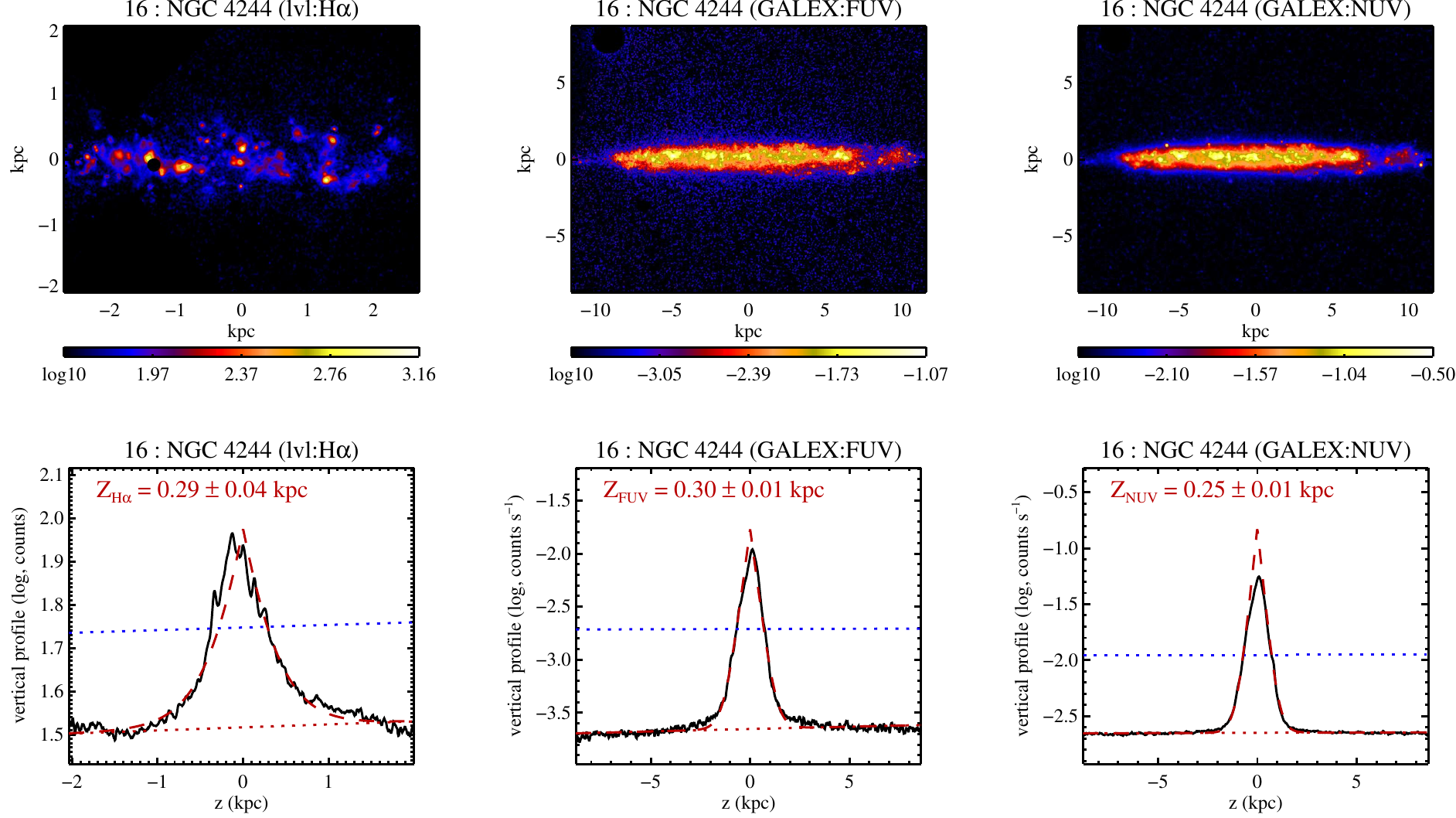}
\par\end{centering}
\begin{centering}
\includegraphics[clip,scale=0.7]{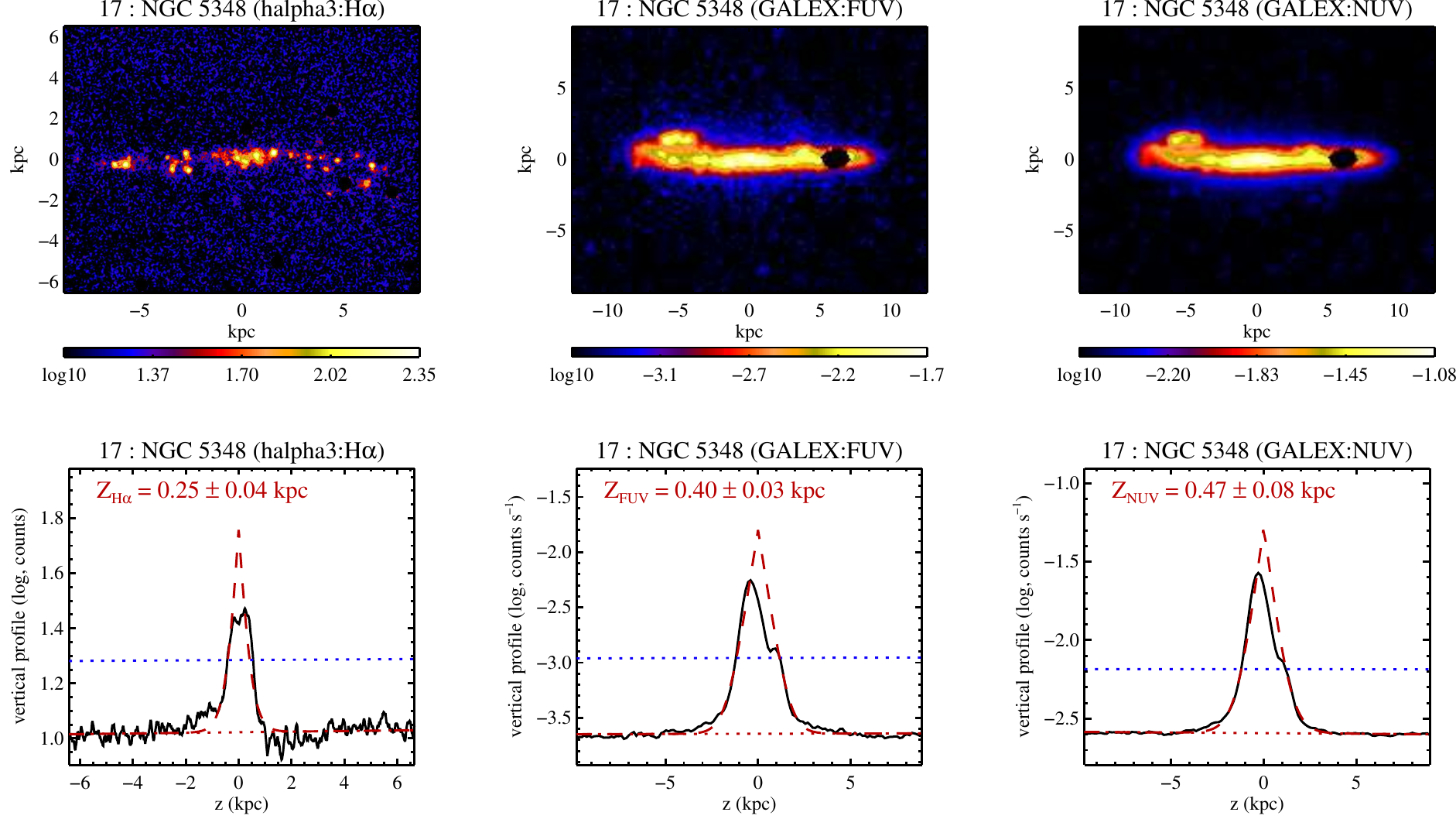}
\par\end{centering}
\begin{centering}
\medskip{}
\par\end{centering}
\caption{Continued.}
\end{continuedfigure*}

\begin{continuedfigure*}[tp]
\begin{centering}
\medskip{}
\par\end{centering}
\begin{centering}
\includegraphics[clip,scale=0.7]{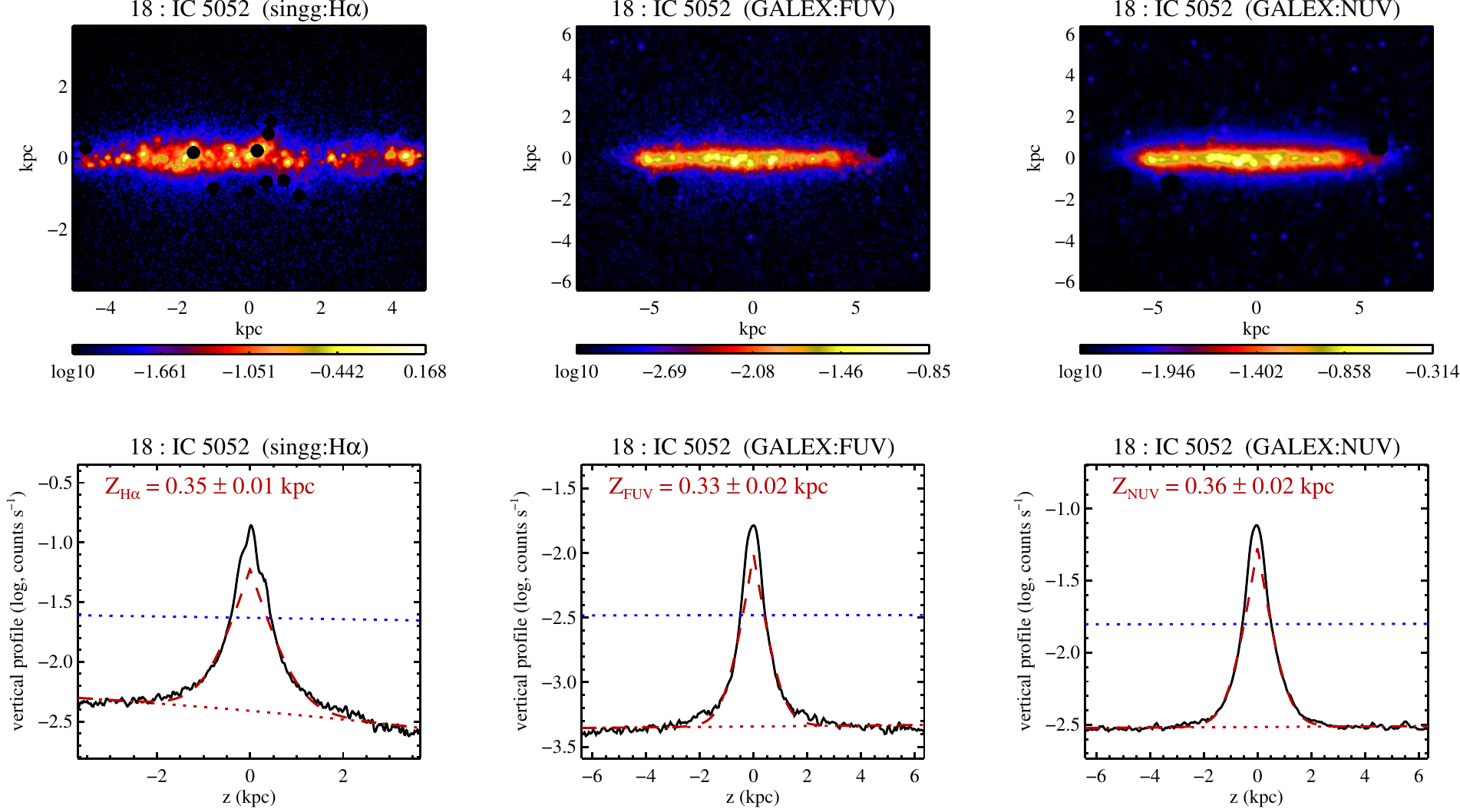}
\par\end{centering}
\begin{centering}
\includegraphics[clip,scale=0.7]{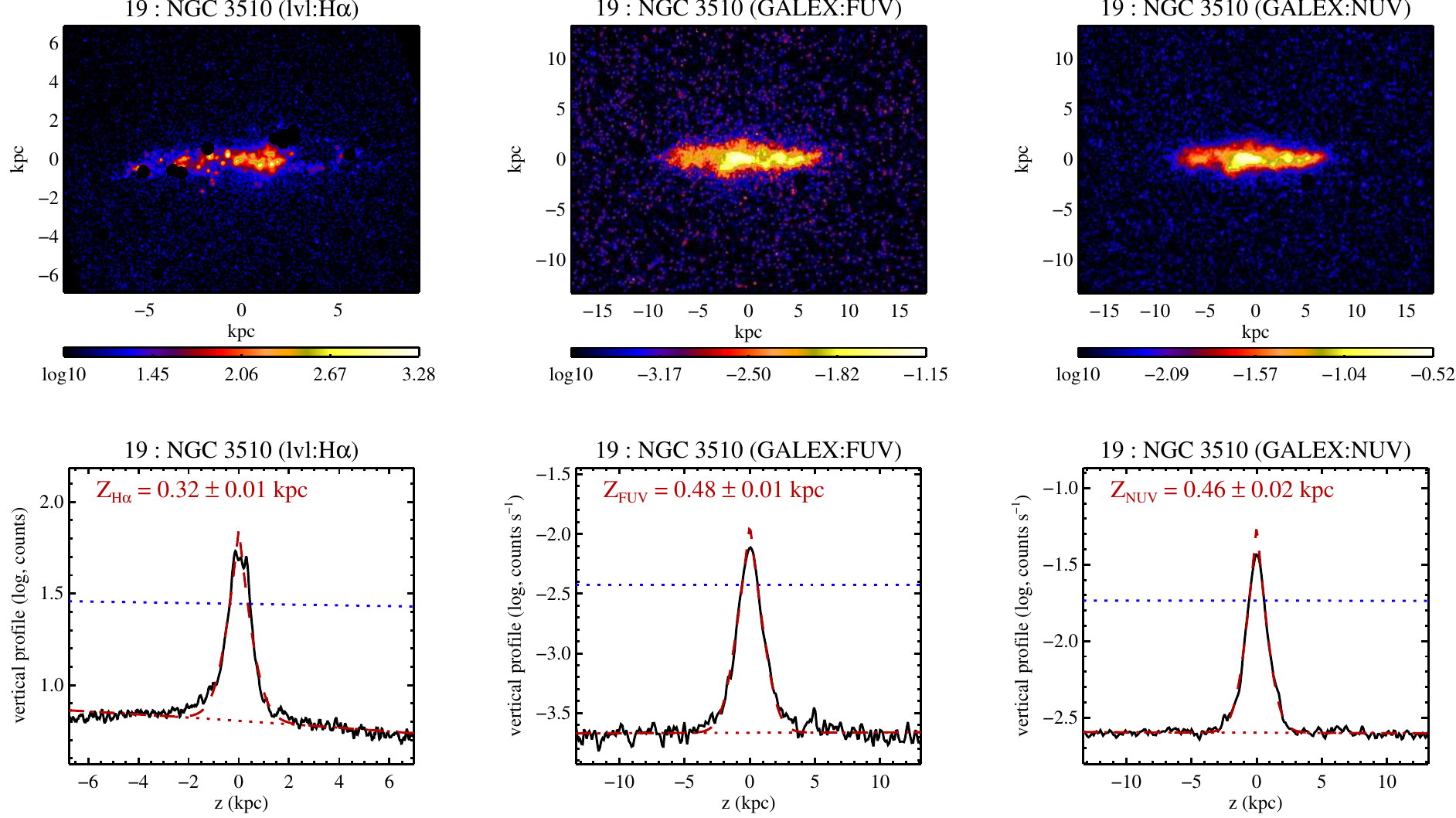}
\par\end{centering}
\begin{centering}
\includegraphics[clip,scale=0.7]{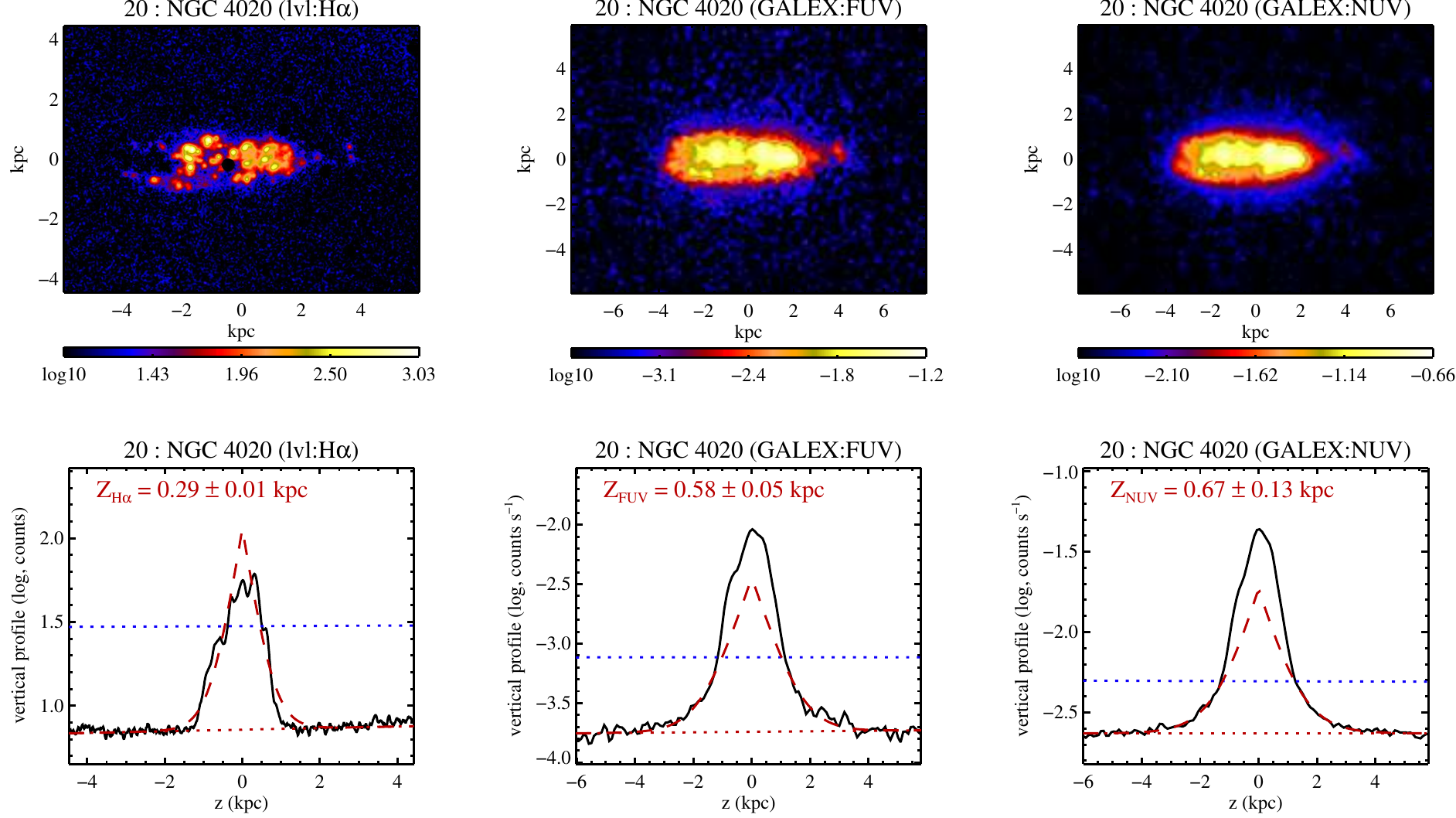}
\par\end{centering}
\begin{centering}
\medskip{}
\par\end{centering}
\caption{Continued.}
\end{continuedfigure*}

\begin{continuedfigure*}[tp]
\begin{centering}
\medskip{}
\par\end{centering}
\begin{centering}
\includegraphics[clip,scale=0.7]{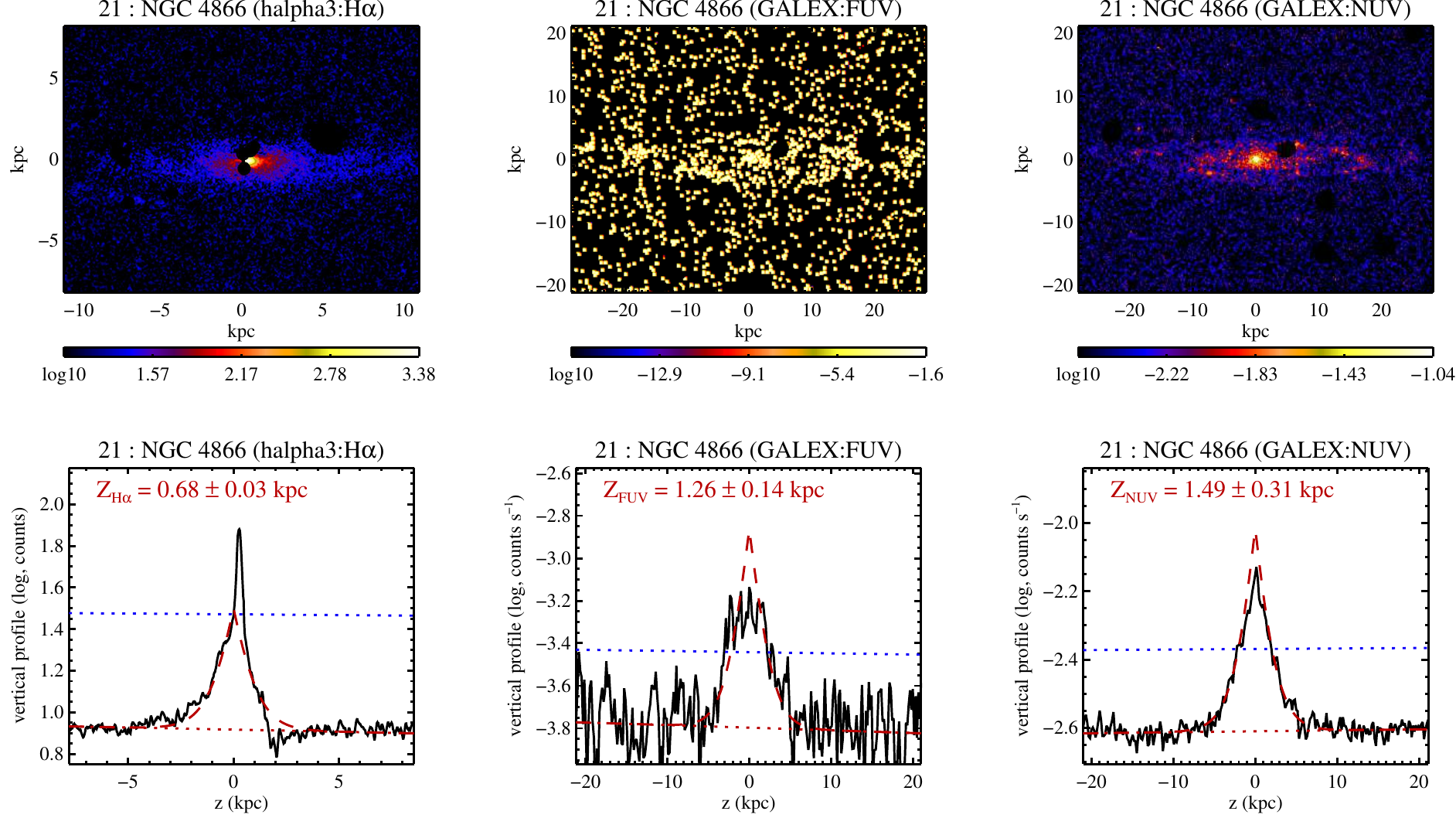}
\par\end{centering}
\begin{centering}
\includegraphics[clip,scale=0.7]{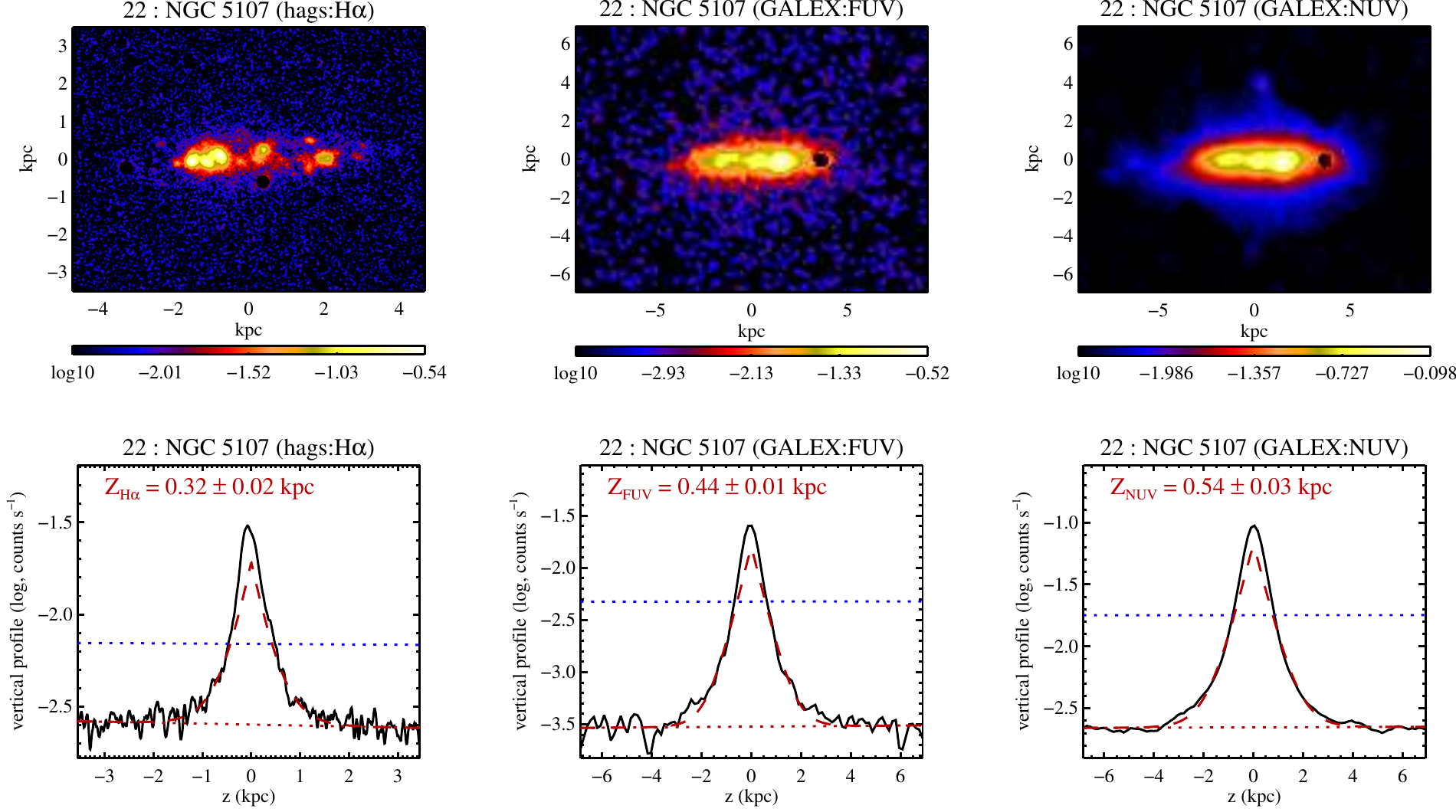}
\par\end{centering}
\begin{centering}
\includegraphics[clip,scale=0.7]{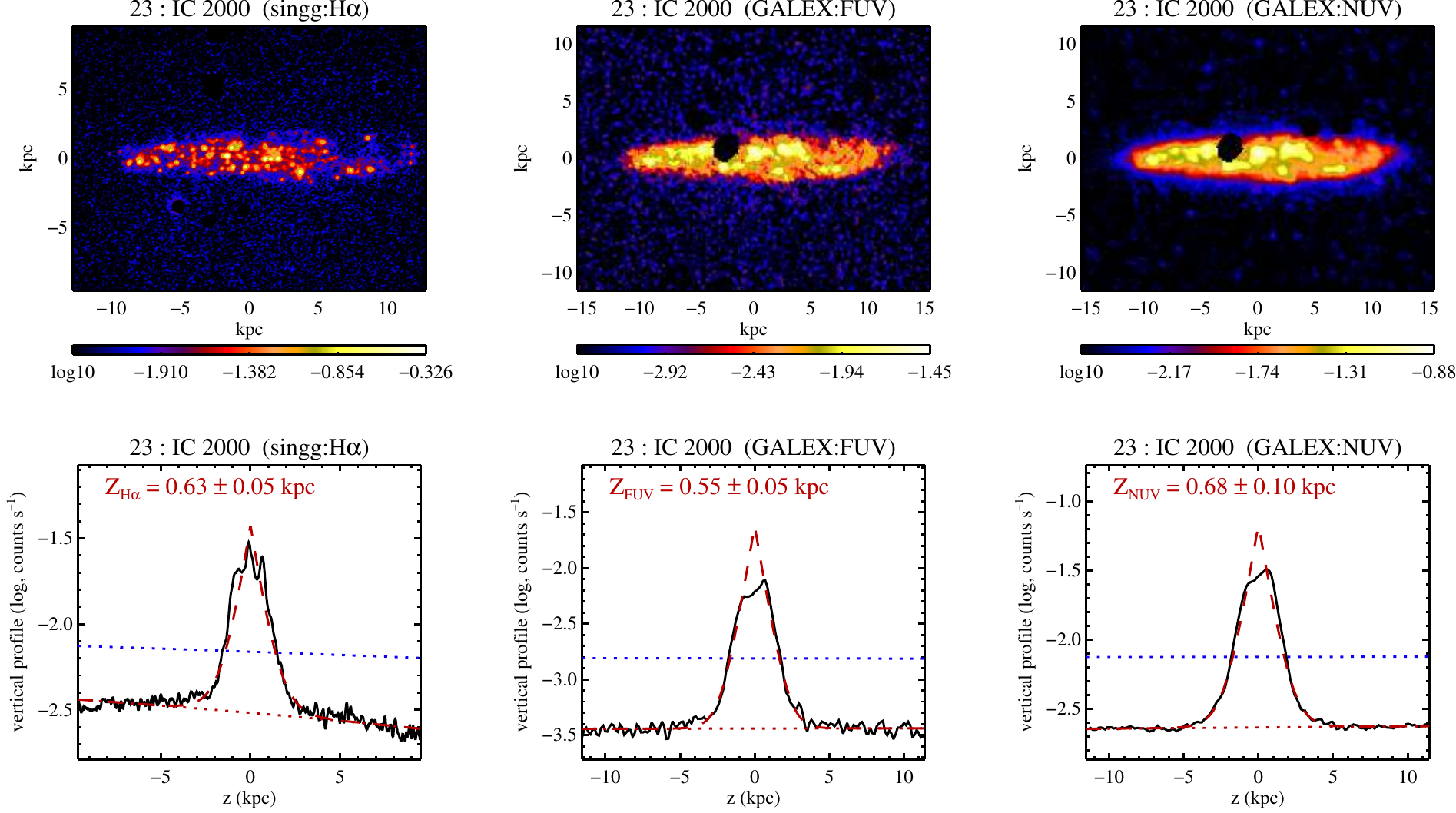}
\par\end{centering}
\begin{centering}
\medskip{}
\par\end{centering}
\caption{Continued.}
\end{continuedfigure*}

\begin{continuedfigure*}[tp]
\begin{centering}
\medskip{}
\par\end{centering}
\begin{centering}
\includegraphics[clip,scale=0.7]{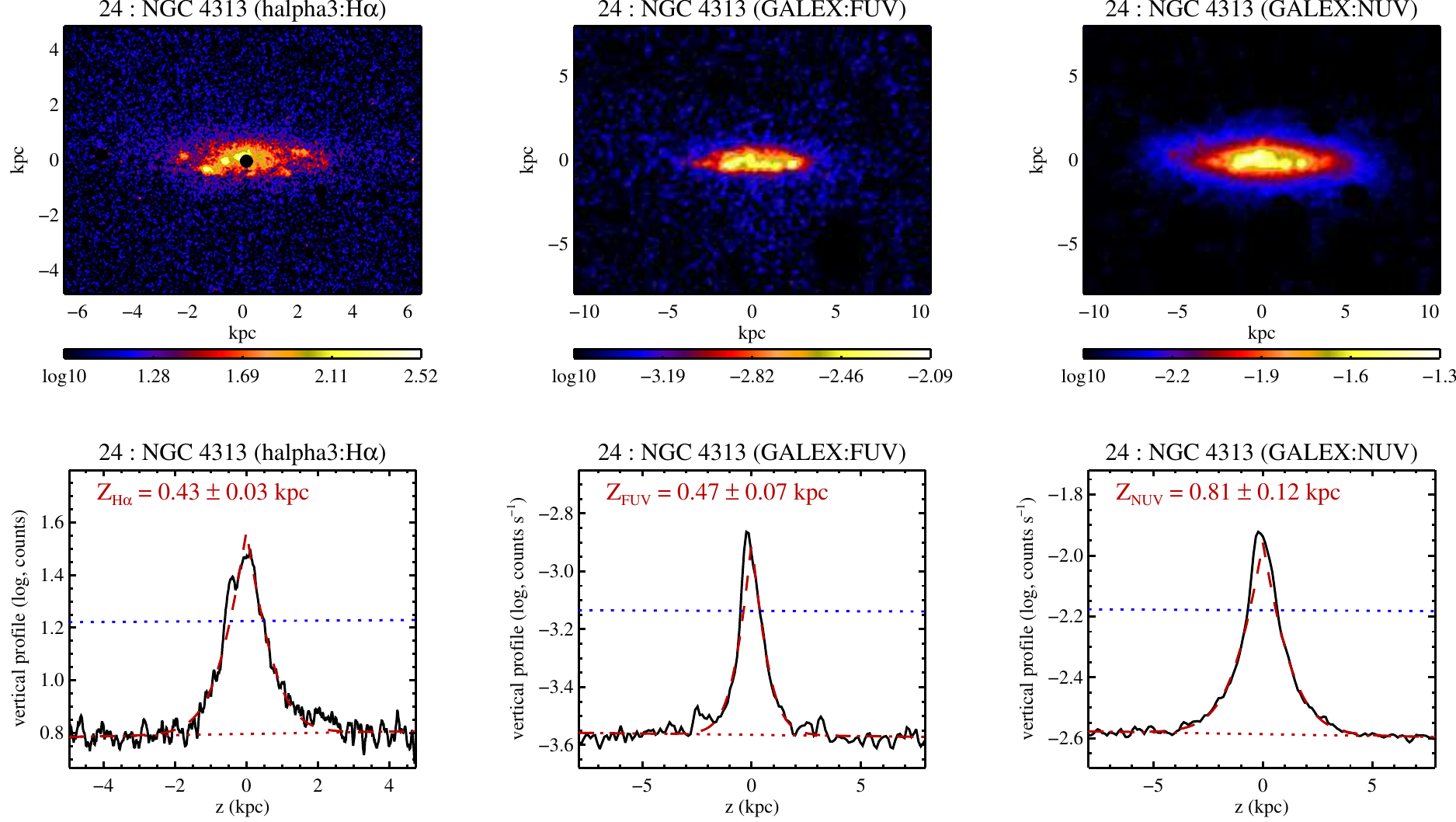}
\par\end{centering}
\begin{centering}
\includegraphics[clip,scale=0.7]{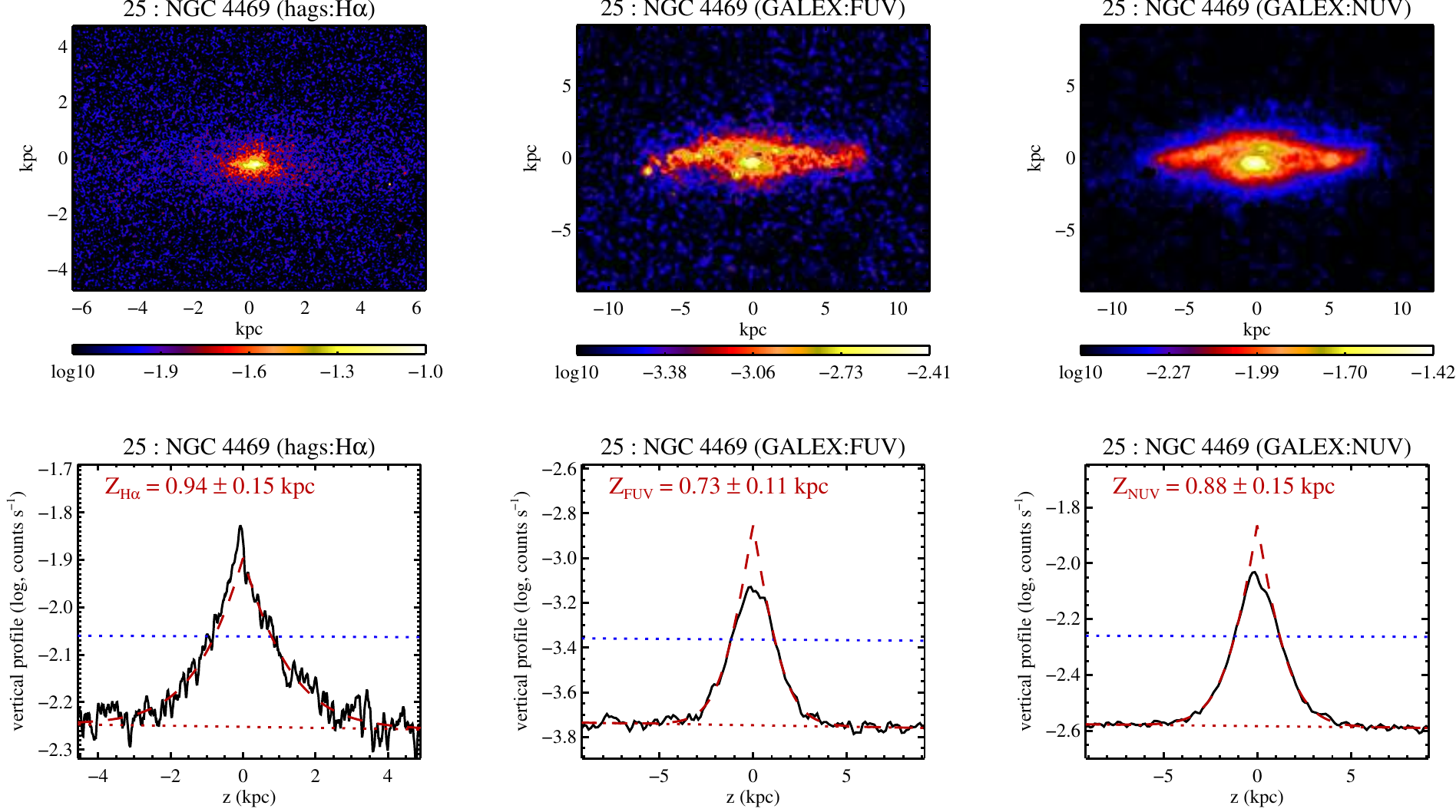}
\par\end{centering}
\begin{centering}
\includegraphics[clip,scale=0.7]{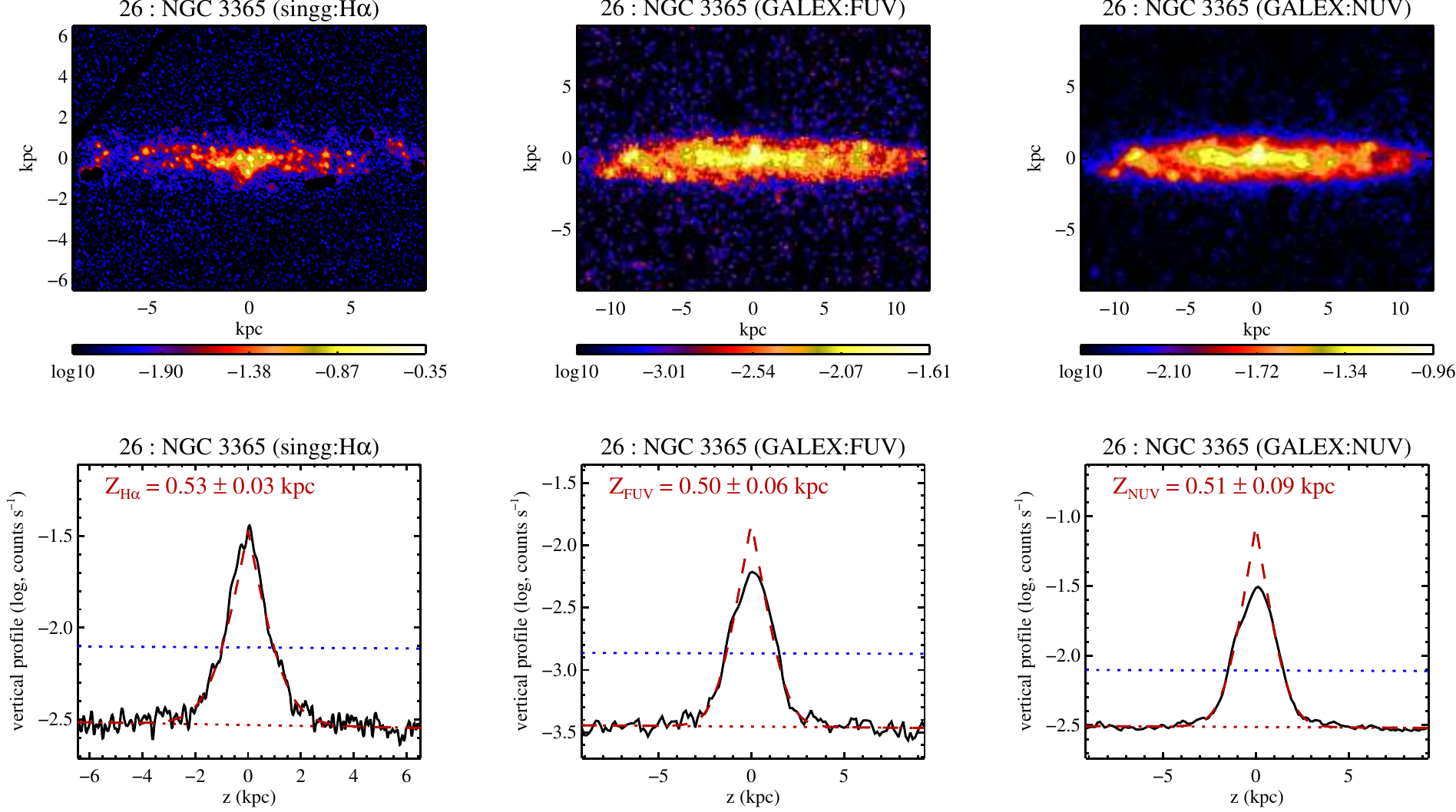}
\par\end{centering}
\begin{centering}
\medskip{}
\par\end{centering}
\caption{Continued.}
\end{continuedfigure*}

\begin{continuedfigure*}[tp]
\begin{centering}
\medskip{}
\par\end{centering}
\begin{centering}
\includegraphics[clip,scale=0.7]{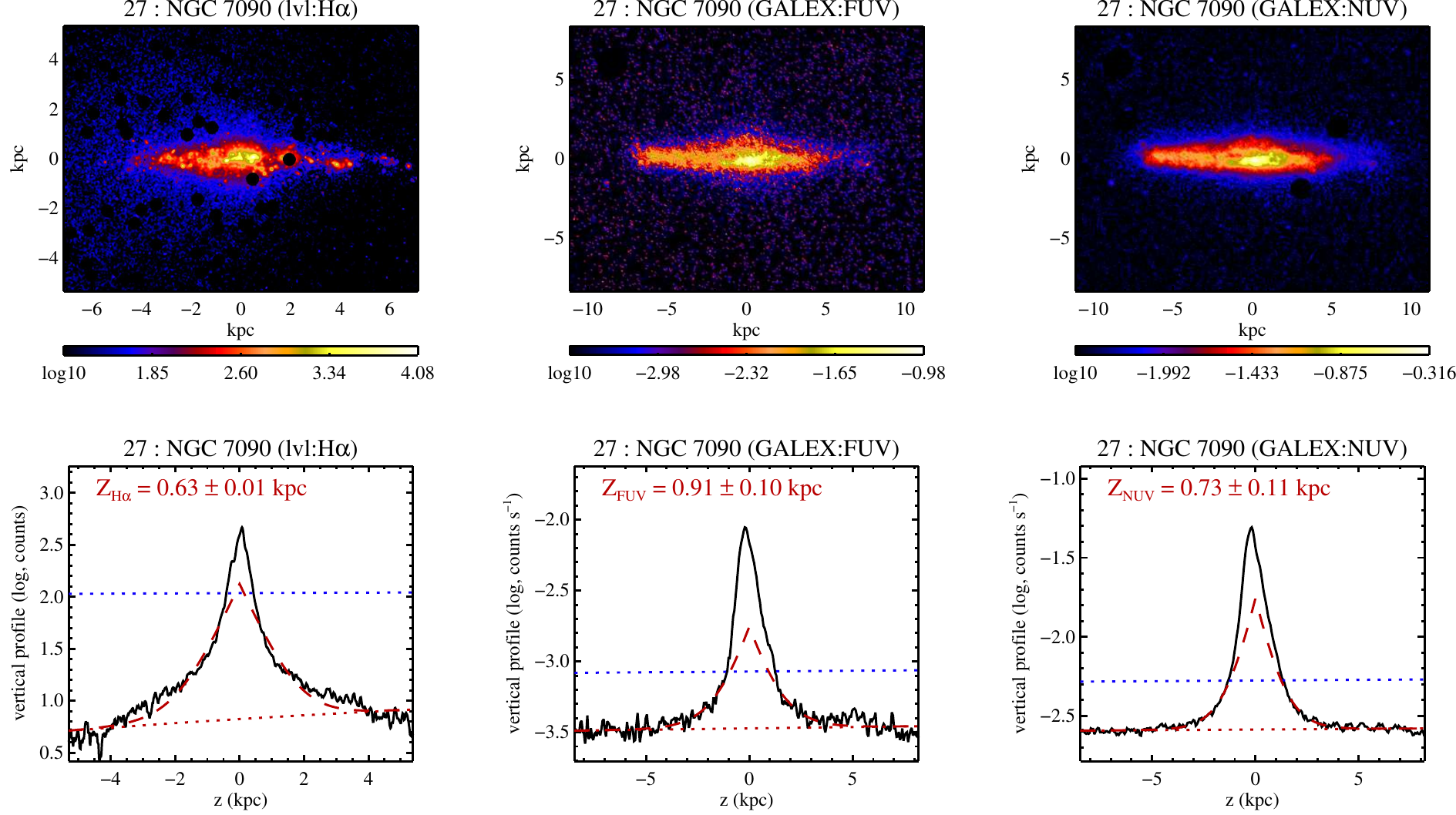}
\par\end{centering}
\begin{centering}
\includegraphics[clip,scale=0.7]{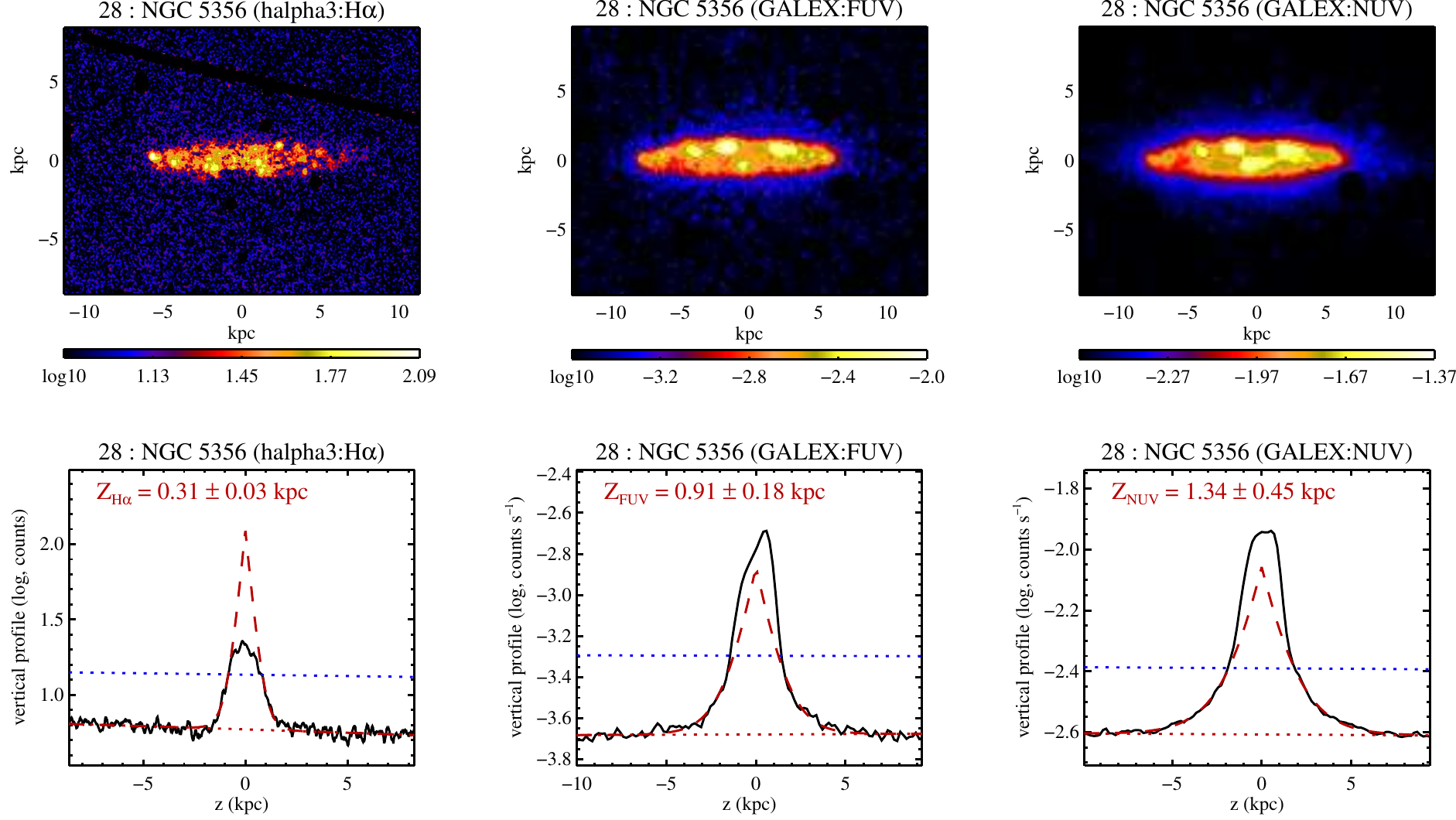}
\par\end{centering}
\begin{centering}
\includegraphics[clip,scale=0.7]{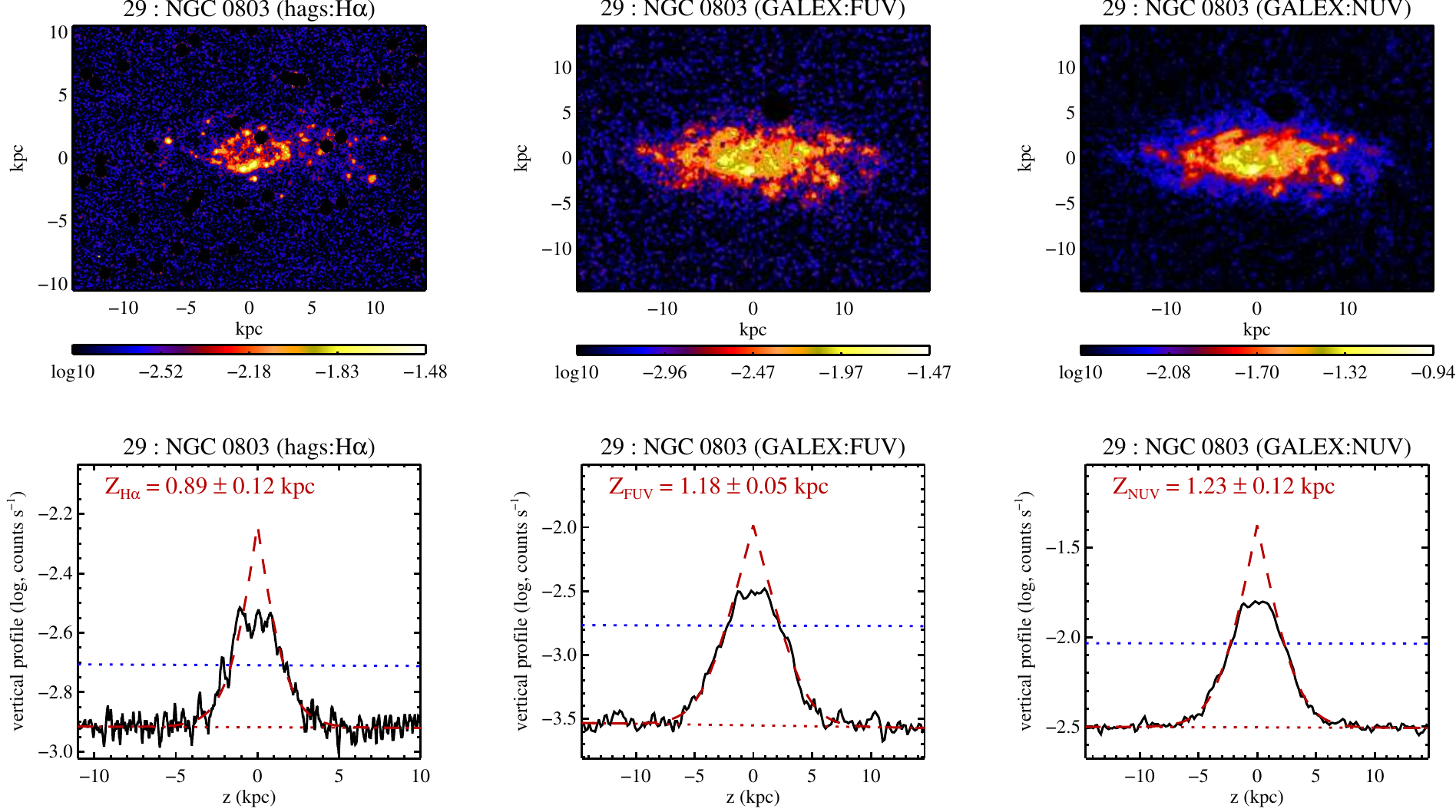}
\par\end{centering}
\begin{centering}
\medskip{}
\par\end{centering}
\caption{Continued.}
\end{continuedfigure*}

\begin{continuedfigure*}[tp]
\begin{centering}
\medskip{}
\par\end{centering}
\begin{centering}
\includegraphics[clip,scale=0.7]{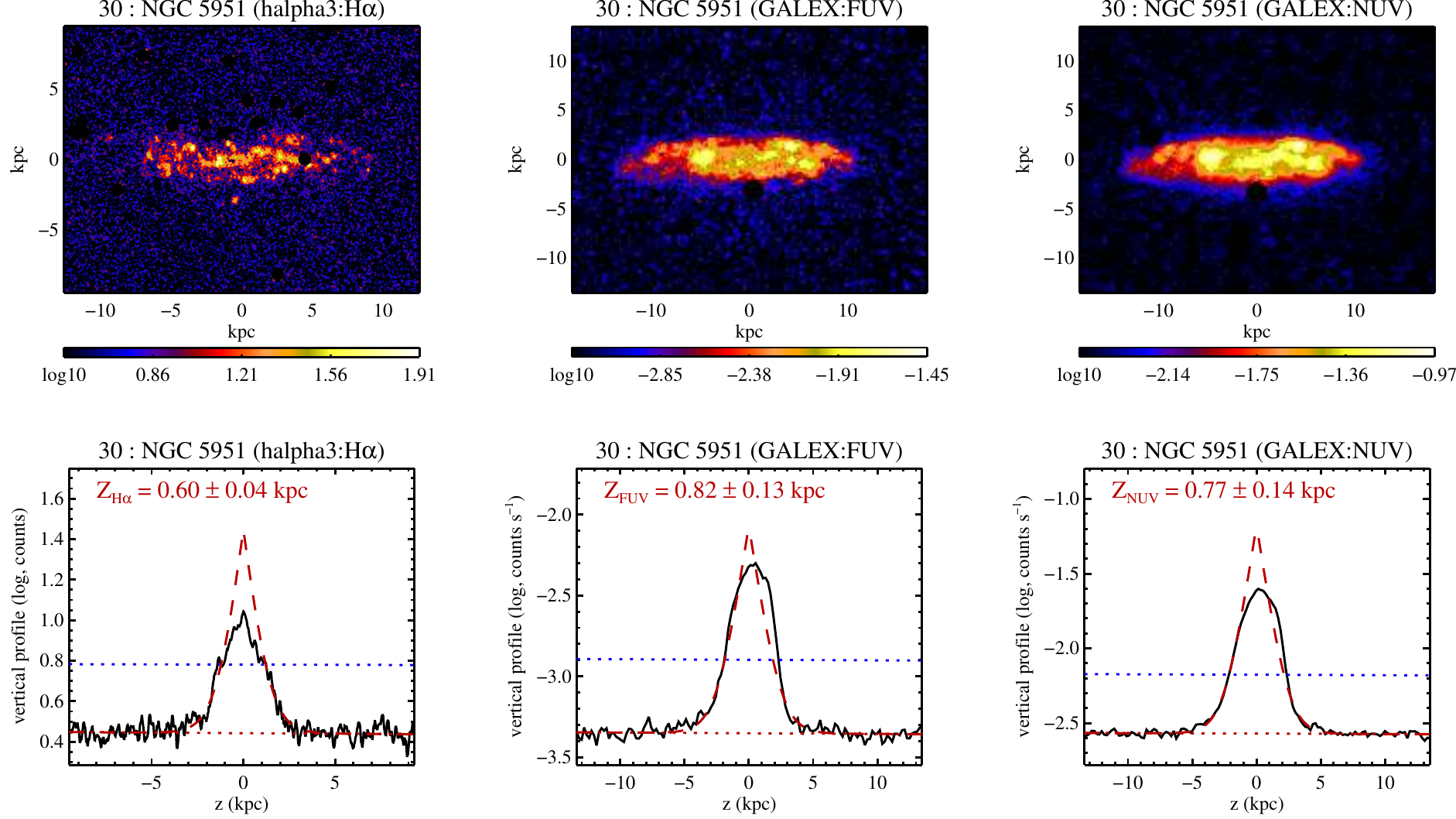}
\par\end{centering}
\begin{centering}
\includegraphics[clip,scale=0.7]{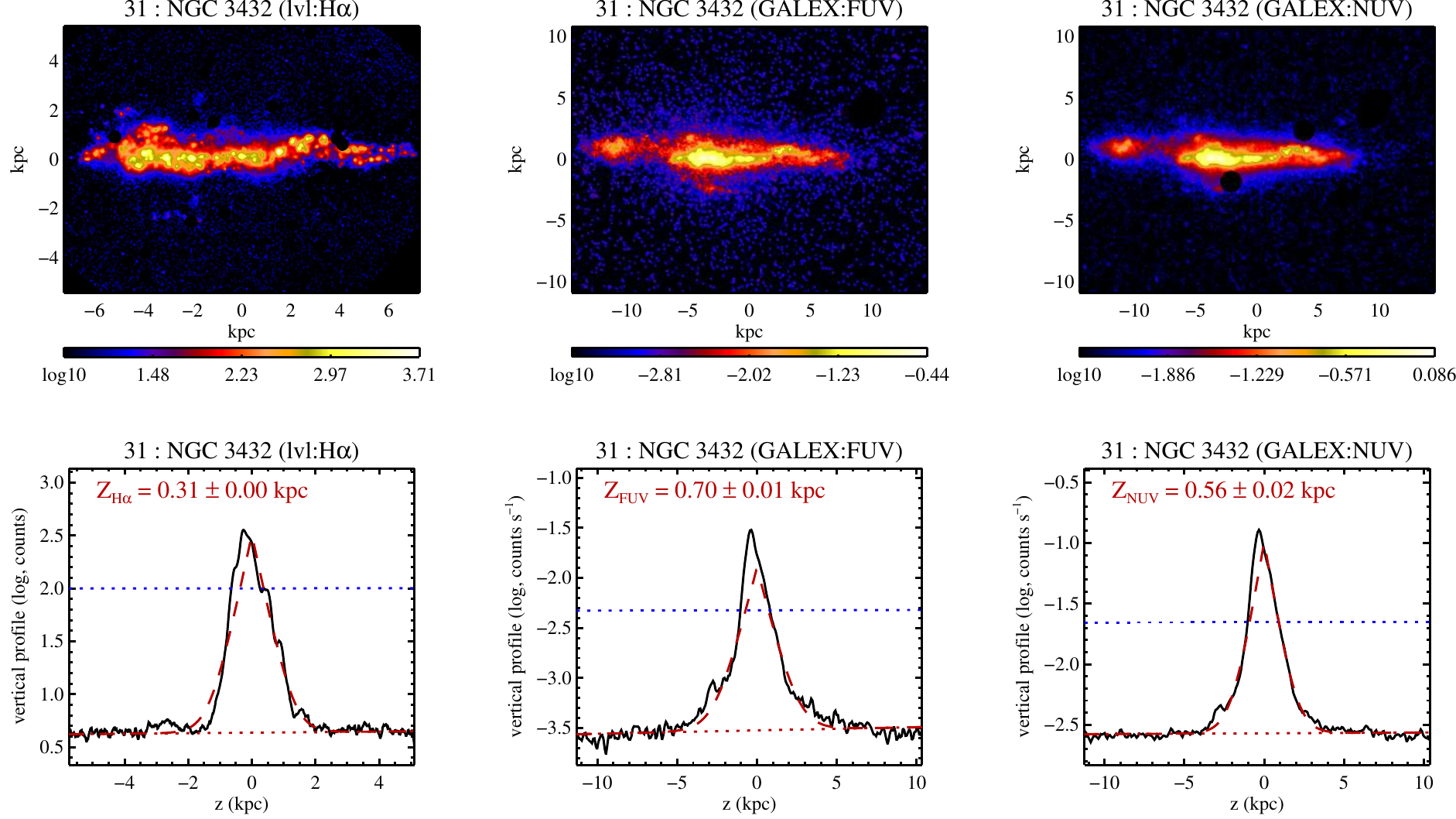}
\par\end{centering}
\begin{centering}
\includegraphics[clip,scale=0.7]{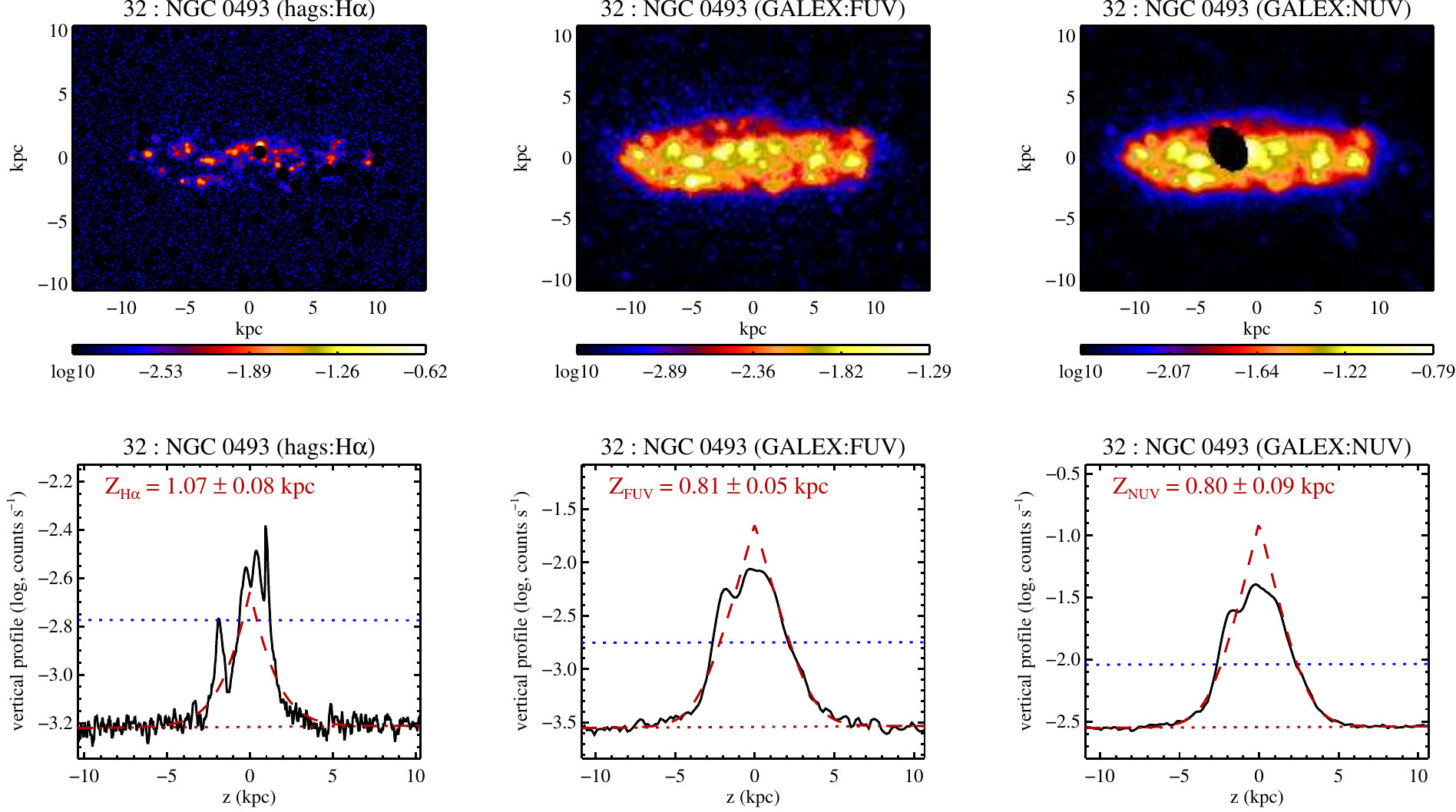}
\par\end{centering}
\begin{centering}
\medskip{}
\par\end{centering}
\caption{Continued.}
\end{continuedfigure*}

\begin{continuedfigure*}[tp]
\begin{centering}
\medskip{}
\par\end{centering}
\begin{centering}
\includegraphics[clip,scale=0.7]{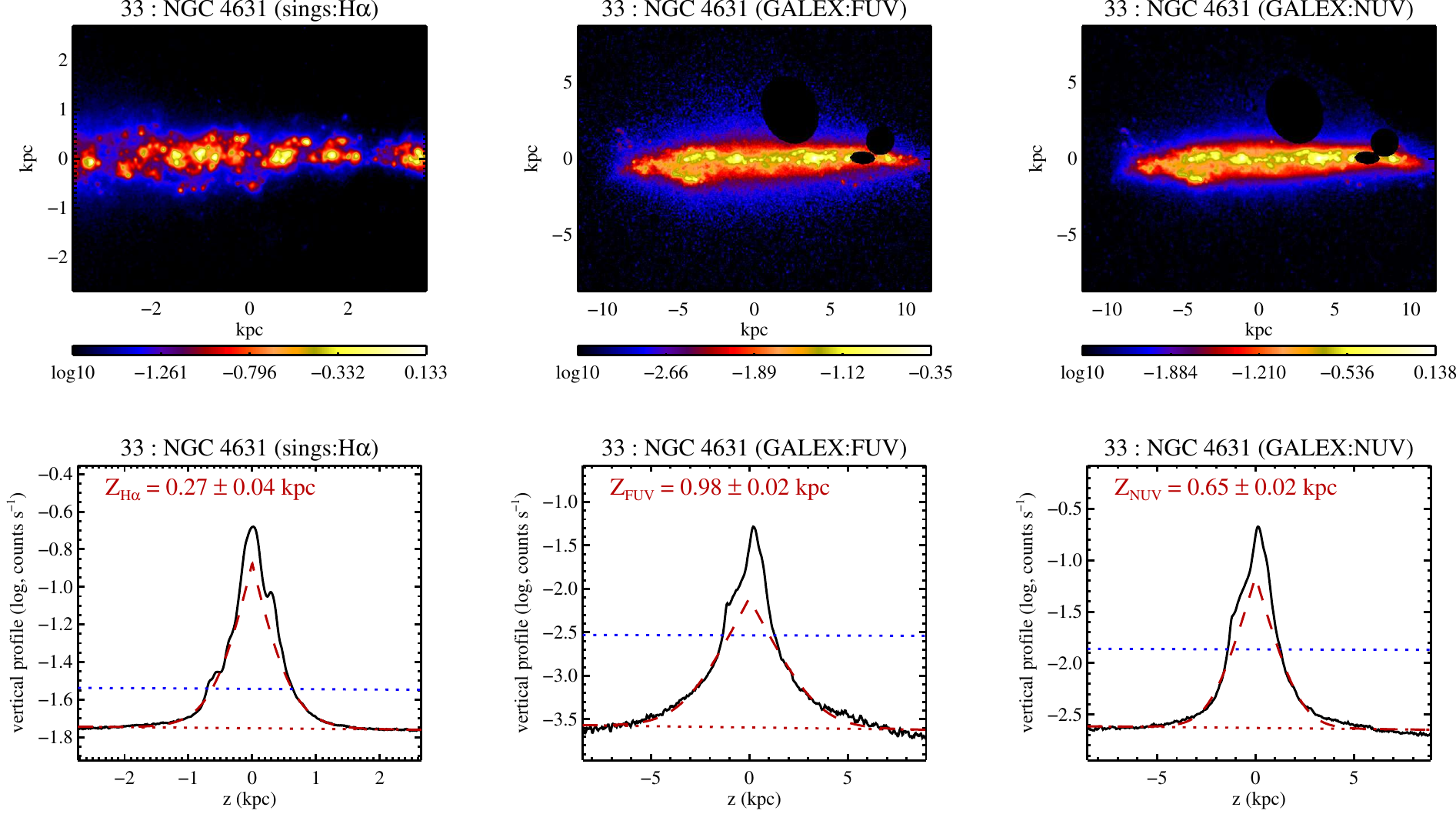}
\par\end{centering}
\begin{centering}
\includegraphics[clip,scale=0.7]{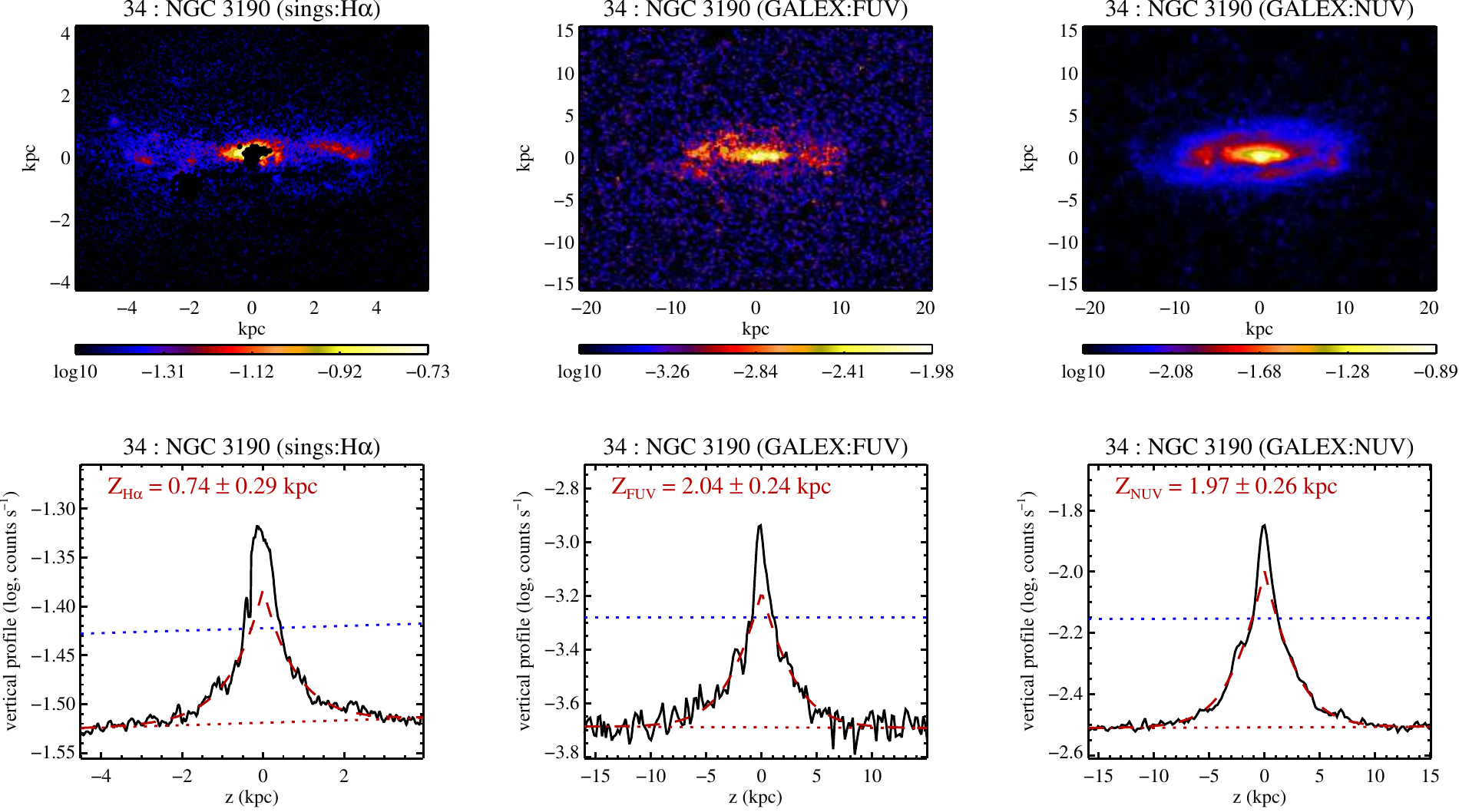}
\par\end{centering}
\begin{centering}
\includegraphics[clip,scale=0.7]{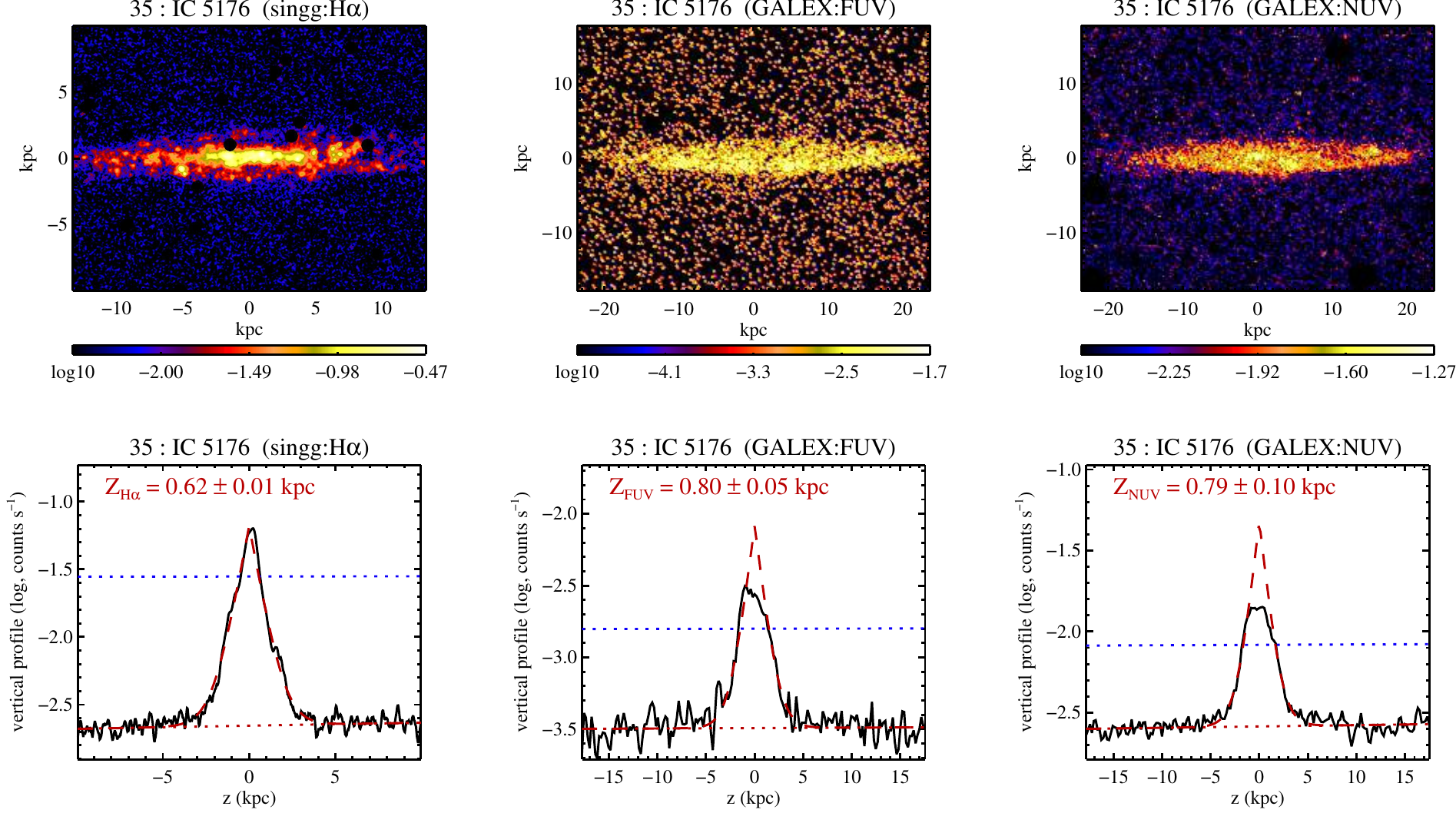}
\par\end{centering}
\begin{centering}
\medskip{}
\par\end{centering}
\caption{Continued.}
\end{continuedfigure*}

\begin{continuedfigure*}[tp]
\begin{centering}
\medskip{}
\par\end{centering}
\begin{centering}
\includegraphics[clip,scale=0.7]{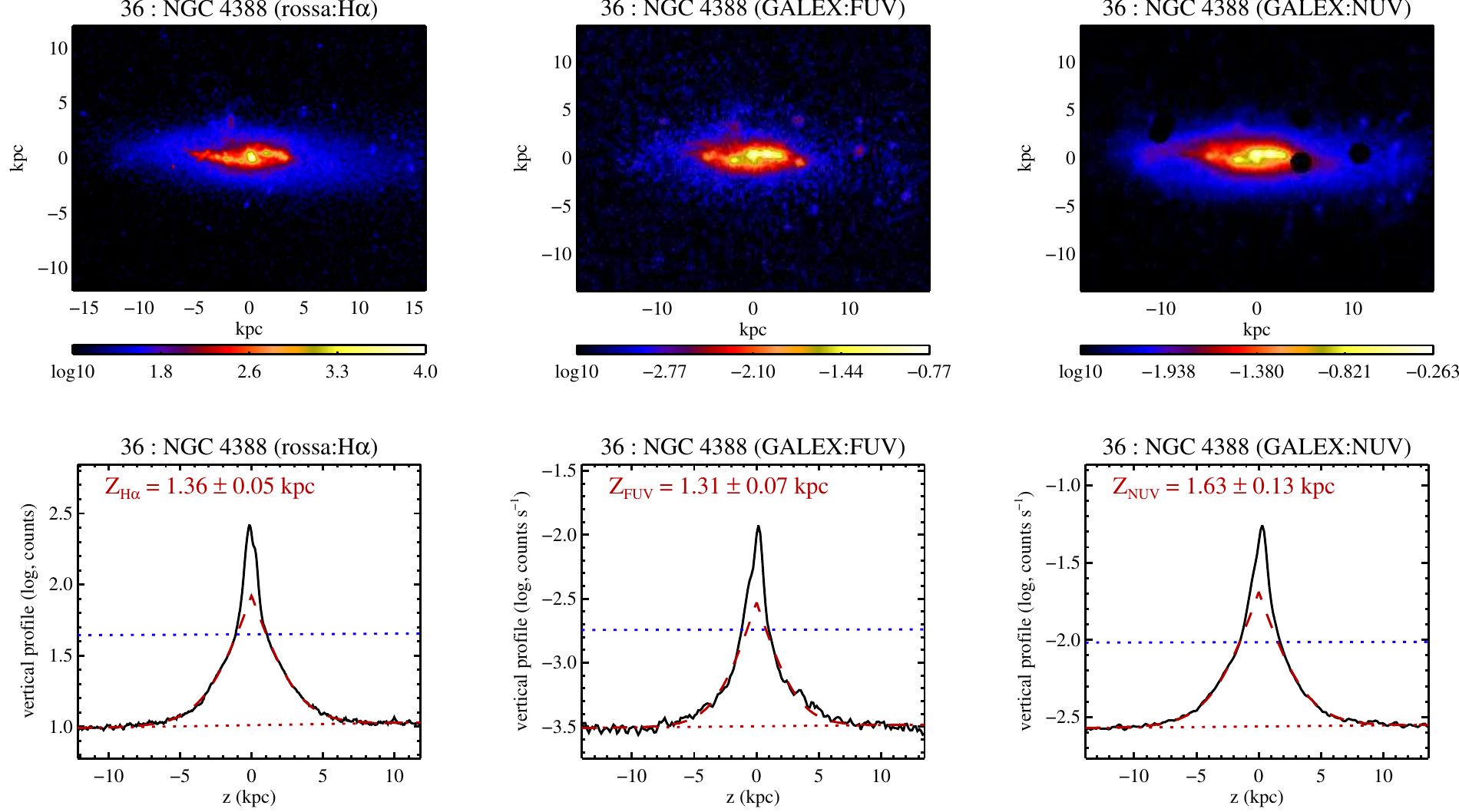}
\par\end{centering}
\begin{centering}
\includegraphics[clip,scale=0.7]{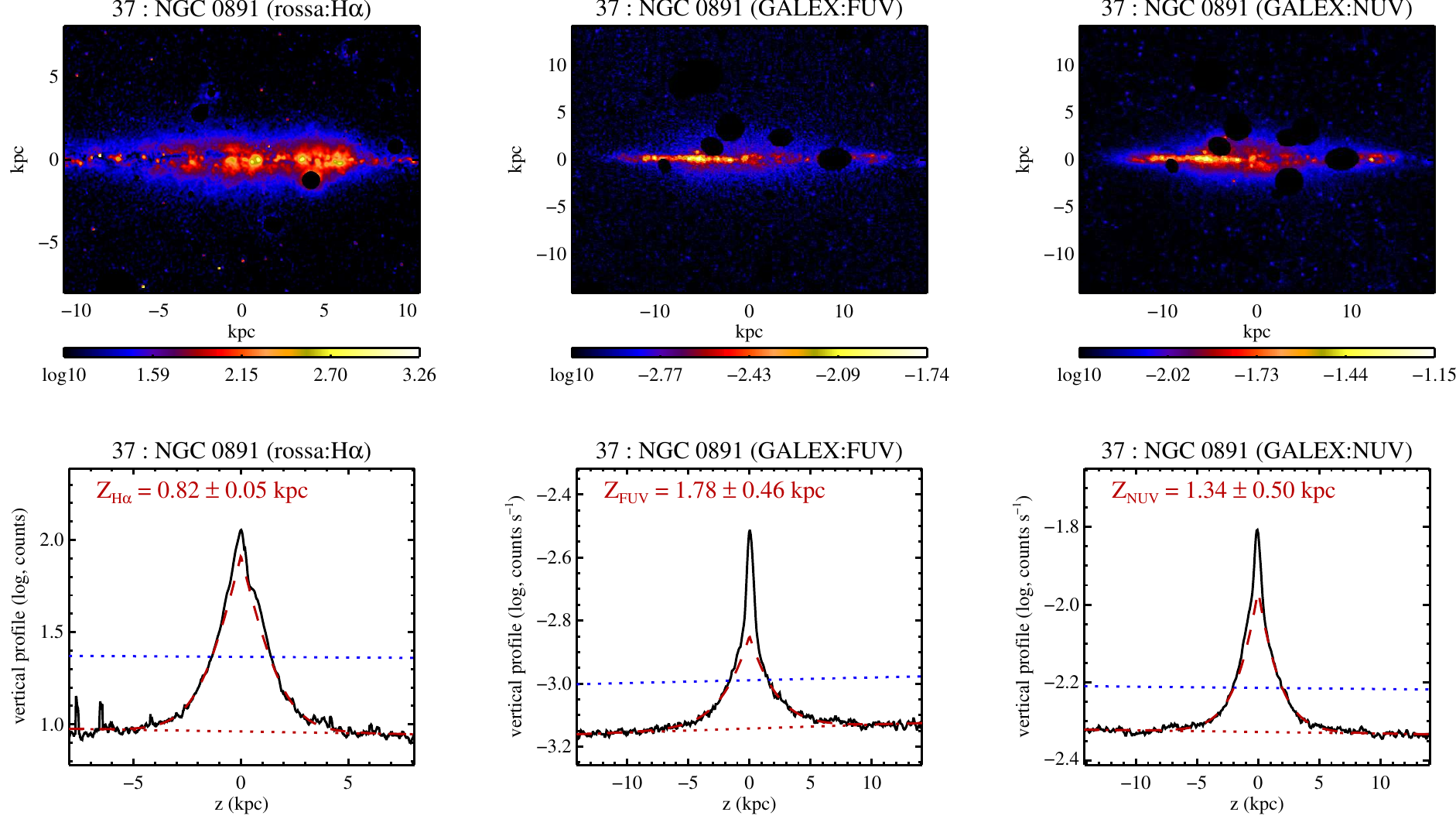}
\par\end{centering}
\begin{centering}
\includegraphics[clip,scale=0.7]{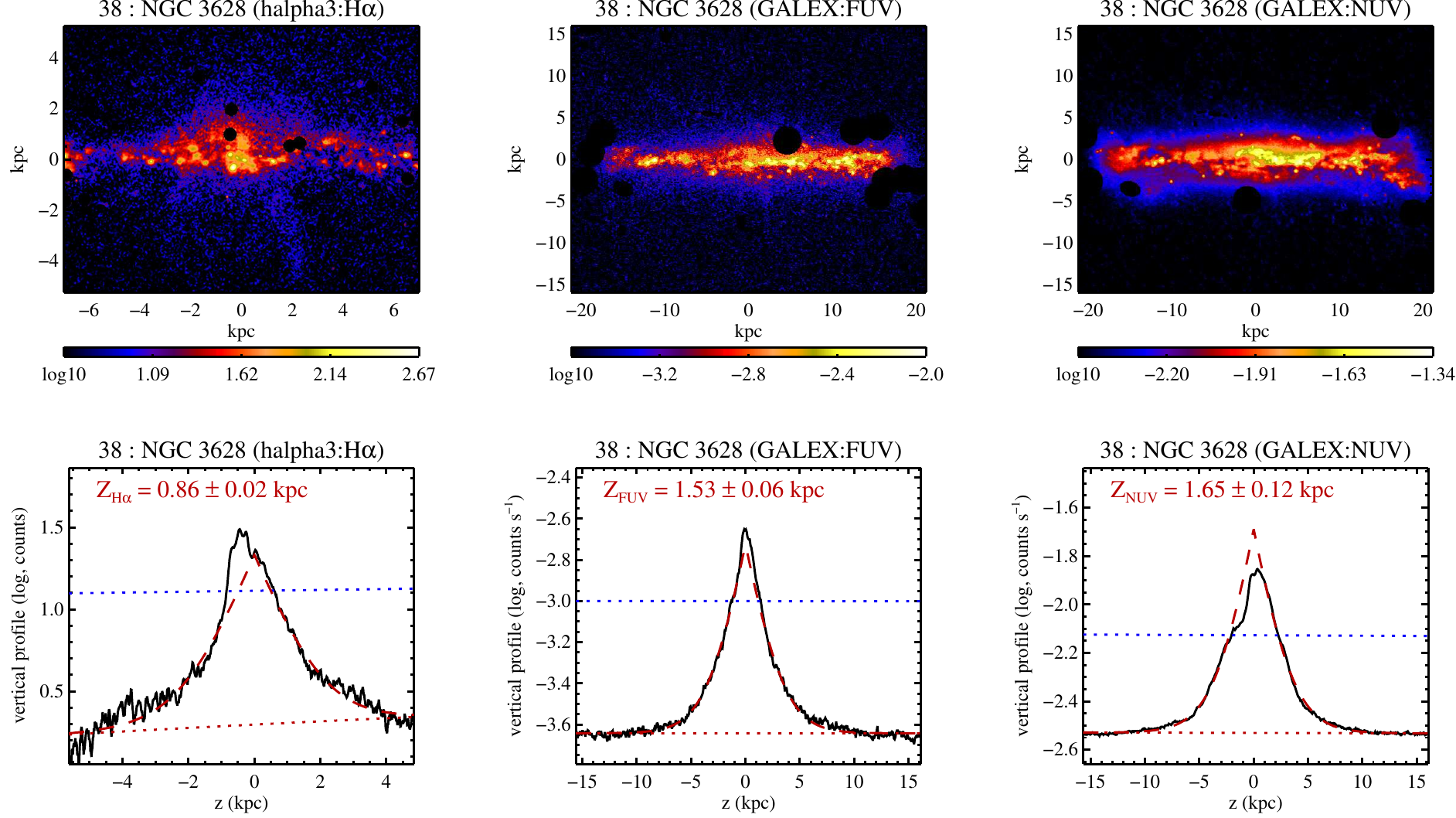}
\par\end{centering}
\begin{centering}
\medskip{}
\par\end{centering}
\caption{Continued.}
\end{continuedfigure*}

\end{document}